\documentclass[AMA,STIX1COL]{WileyNJD-v2}

\articletype{RESEARCH ARTICLE}%

\received{DD MONTH YEAR}
\revised{DD MONTH YEAR}
\accepted{DD MONTH YEAR}

\usepackage{mathtools}
\usepackage{subfigure}
\usepackage{epsfig}
\usepackage{siunitx}
\DeclareSIUnit\Molar{M}     %

\usepackage{textcomp}    %

\usepackage{color}

\usepackage{hyperref}   %

\makeatletter %

\def\input@path{{figures/}}

\newcommand{\defeq}{\overset{!}{=}}
\newcommand{\defvariable}{:=}%

\newcommand{\bigO}{\mathcal{O}}%

\newcommand{\secref}[1]{Sec.~\ref{#1}}%
\newcommand{\appref}[1]{App.~\ref{#1}}%
\newcommand{\figref}[1]{Fig.~\ref{#1}}%
\renewcommand{\eqref}[1]{Eq.~(\ref{#1})}%

\newcommand\DeclareBoldMathCommand[2]{%
  \protected@edef\@tempb{%
    \noexpand\DeclareRobustCommand{\csname #1\endcsname}{\boldsymbol{\ensuremath{#2}}}}
  \@tempb}
\@tfor\AM@letter:=abcdefghijklmnopqrstuvwxyz%
\do{\DeclareBoldMathCommand{v\AM@letter}{\AM@letter}}
\@tfor\AM@letter:=ABCDEFGHIJKLMNOPQRSTUVWXYZ%
\do{\DeclareBoldMathCommand{v\AM@letter}{\AM@letter}}
\DeclareBoldMathCommand{vnull}{0}
\DeclareBoldMathCommand{vone}{1}
\DeclareBoldMathCommand{valpha}{\alpha}
\DeclareBoldMathCommand{vbeta}{\beta}
\DeclareBoldMathCommand{vgamma}{\gamma}
\DeclareBoldMathCommand{vdelta}{\delta}
\DeclareBoldMathCommand{vepsilon}{\epsilon}
\DeclareBoldMathCommand{vvarepsilon}{\varepsilon}
\DeclareBoldMathCommand{vzeta}{\zeta}
\DeclareBoldMathCommand{veta}{\eta}
\DeclareBoldMathCommand{vtheta}{\theta}
\DeclareBoldMathCommand{vvartheta}{\vartheta}
\DeclareBoldMathCommand{viota}{\iota}
\DeclareBoldMathCommand{vkappa}{\kappa}
\DeclareBoldMathCommand{vlambda}{\lambda}
\DeclareBoldMathCommand{vmu}{\mu}
\DeclareBoldMathCommand{vnu}{\nu}
\DeclareBoldMathCommand{vpi}{\pi}
\DeclareBoldMathCommand{vxi}{\xi}
\DeclareBoldMathCommand{vrho}{\varrho}
\DeclareBoldMathCommand{vsigma}{\sigma}
\DeclareBoldMathCommand{vtau}{\tau}
\DeclareBoldMathCommand{vupsilon}{\upsilon}
\DeclareBoldMathCommand{vphi}{\varphi}
\DeclareBoldMathCommand{vchi}{\chi}
\DeclareBoldMathCommand{vpsi}{\psi}
\DeclareBoldMathCommand{vomega}{\omega}
\DeclareBoldMathCommand{vGamma}{\Gamma}
\DeclareBoldMathCommand{vDelta}{\Delta}
\DeclareBoldMathCommand{vTheta}{\Theta}
\DeclareBoldMathCommand{vLambda}{\Lambda}
\DeclareBoldMathCommand{vXi}{\Xi}
\DeclareBoldMathCommand{vPi}{\Pi}
\DeclareBoldMathCommand{vSigma}{\Sigma}
\DeclareBoldMathCommand{vUpsilon}{\Upsilon}
\DeclareBoldMathCommand{vPhi}{\Phi}
\DeclareBoldMathCommand{vPsi}{\Psi}
\DeclareBoldMathCommand{vOmega}{\Omega}

\newcommand\DeclareDiscreteBoldMathCommand[2]{%
  \protected@edef\@tempc{%
    \noexpand\DeclareRobustCommand{\csname #1\endcsname}{\boldsymbol{\mathrm{#2}}}}
  \@tempc}
\@tfor\AM@letter:=abcdefghijklmnopqrstuvwxyz%
\do{\DeclareDiscreteBoldMathCommand{vd\AM@letter}{\AM@letter}}
\@tfor\AM@letter:=ABCDEFGHIJKLMNOPQRSTUVWXYZ%
\do{\DeclareDiscreteBoldMathCommand{vd\AM@letter}{\AM@letter}}
\DeclareDiscreteBoldMathCommand{vdnull}{0}
\DeclareDiscreteBoldMathCommand{vdone}{1}
\DeclareDiscreteBoldMathCommand{vdalpha}{\alpha}
\DeclareDiscreteBoldMathCommand{vdbeta}{\beta}
\DeclareDiscreteBoldMathCommand{vdgamma}{\gamma}
\DeclareDiscreteBoldMathCommand{vddelta}{\delta}
\DeclareDiscreteBoldMathCommand{vdepsilon}{\epsilon}
\DeclareDiscreteBoldMathCommand{vdvarepsilon}{\varepsilon}
\DeclareDiscreteBoldMathCommand{vdzeta}{\zeta}
\DeclareDiscreteBoldMathCommand{vdeta}{\eta}
\DeclareDiscreteBoldMathCommand{vdtheta}{\theta}
\DeclareDiscreteBoldMathCommand{vdvartheta}{\vartheta}
\DeclareDiscreteBoldMathCommand{vdiota}{\iota}
\DeclareDiscreteBoldMathCommand{vdkappa}{\kappa}
\DeclareDiscreteBoldMathCommand{vdlambda}{\lambda}
\DeclareDiscreteBoldMathCommand{vdmu}{\mu}
\DeclareDiscreteBoldMathCommand{vdnu}{\nu}
\DeclareDiscreteBoldMathCommand{vdpi}{\pi}
\DeclareDiscreteBoldMathCommand{vdxi}{\xi}
\DeclareDiscreteBoldMathCommand{vdrho}{\varrho}
\DeclareDiscreteBoldMathCommand{vdsigma}{\sigma}
\DeclareDiscreteBoldMathCommand{vdtau}{\tau}
\DeclareDiscreteBoldMathCommand{vdupsilon}{\upsilon}
\DeclareDiscreteBoldMathCommand{vdphi}{\varphi}
\DeclareDiscreteBoldMathCommand{vdchi}{\chi}
\DeclareDiscreteBoldMathCommand{vdpsi}{\psi}
\DeclareDiscreteBoldMathCommand{vdomega}{\omega}
\DeclareDiscreteBoldMathCommand{vdGamma}{\Gamma}
\DeclareDiscreteBoldMathCommand{vdDelta}{\Delta}
\DeclareDiscreteBoldMathCommand{vdTheta}{\Theta}
\DeclareDiscreteBoldMathCommand{vdLambda}{\Lambda}
\DeclareDiscreteBoldMathCommand{vdXi}{\Xi}
\DeclareDiscreteBoldMathCommand{vdPi}{\Pi}
\DeclareDiscreteBoldMathCommand{vdSigma}{\Sigma}
\DeclareDiscreteBoldMathCommand{vdUpsilon}{\Upsilon}
\DeclareDiscreteBoldMathCommand{vdPhi}{\Phi}
\DeclareDiscreteBoldMathCommand{vdPsi}{\Psi}
\DeclareDiscreteBoldMathCommand{vdOmega}{\Omega}

\providecommand*{\dd}{%
  \@ifnextchar^{\@dd}{\@dd^{}}}
\def\@dd^#1{%
  \mathop{\mathrm{\mathstrut d}}%
  \nolimits^{#1}\dd@gobblespace}
\def\dd@gobblespace{%
  \futurelet\diffarg\dd@opspace}
\def\dd@opspace{%
  \let\dd@space\!%
  \ifx\diffarg(%
\let\dd@space\relax%
\else%
\ifx\diffarg[%
\let\dd@space\relax%
\else%
\ifx\diffarg\{%
\let\dd@space\relax%
\fi%
\fi%
\fi%
\dd@space}

\newcommand{\Frac}{%
  \@ifnextchar[%
  {\Frac@i}
  {\Frac@ii}}
\newcommand{\Frac@i}{}
\def\Frac@i[#1]#2#3{%
  \genfrac{}{}{#1}{}{\displaystyle{#2}}{\displaystyle{#3}}}
\newcommand{\Frac@ii}[2]{\frac{\displaystyle{#1}}{\displaystyle{#2}}}
      \newcommand{\diff@diffspace}{\,}
\newcommand{\diff@mathfrac}[2]{\frac{#1}{#2}}
\newcommand{\diff@mathFrac}[2]{\Frac{#1}{#2}}
\newcommand{\diff@textfrac}[2]{%
  \bgroup #1\egroup\mkern-1mu/\mkern-1mu\bgroup #2\egroup}
\newcommand{\diff}{%
  \global\let\diff@diffop\dd
  \global\let\diff@frac\diff@mathfrac
  \@ifnextchar[%
  {\diff@i}
  {\diff@ii}}
\newcommand{\Diff}{%
  \global\let\diff@diffop\dd
  \global\let\diff@frac\diff@mathFrac
  \@ifnextchar[%
  {\diff@i}
  {\diff@ii}}
\newcommand{\tdiff}{%
  \global\let\diff@diffop\dd
  \global\let\diff@frac\diff@textfrac
  \@ifnextchar[%
  {\diff@i}
  {\diff@ii}}
\newcommand{\pdiff}{%
  \global\let\diff@diffop\partial
  \global\let\diff@frac\diff@mathfrac
  \@ifnextchar[%
  {\diff@i}
  {\diff@ii}}
\newcommand{\Pdiff}{%
  \global\let\diff@diffop\partial
  \global\let\diff@frac\diff@mathFrac
  \@ifnextchar[%
  {\diff@i}
  {\diff@ii}}
\newcommand{\tpdiff}{%
  \global\let\diff@diffop\partial
  \global\let\diff@frac\diff@textfrac
  \@ifnextchar[%
  {\diff@i}
  {\diff@ii}}

\newcommand*{\diff@i}{}
\def\diff@i[#1]#2#3{\eval{\diff@ii{#2}{#3}}_{#1}}

\newcommand*{\diff@ii}[2]{%
  \begingroup
  \toks0={}\count0=0
  \diff@degree #2\diff@degree
  \diff@frac{\diff@diffop\ifnum\count0>1^{\the\count0}\fi\diff@diffspace#1}%
  {\the\toks0}%
  \endgroup}

\newcommand*{\diff@degree}[1]{%
  \ifx #1\diff@degree \expandafter\diff@stopd
  \else \expandafter\diff@addd \fi #1^1$#1\diff@addd}
\newcommand{\diff@stopd}{}
\def\diff@stopd #1\diff@addd{}
\newcommand*{\diff@addd}{}
\def\diff@addd #1^#2#3$#4\diff@addd{%
  \advance\count0 #2
  \toks0=\expandafter{\the\toks0%
    {\diff@diffop\diff@diffspace #4}%
    \diff@diffspace}\diff@degree}

\def\rs#1{\@ifnextchar[%
  {\@rs{#1}}{\@@rs{#1}}}
\def\@rs#1[#2]#3{\mathinner{%
    \setbox\@ne\hbox{$\displaystyle{\vphantom{#3}}#1{#3}\m@th$}%
    \setbox\tw@\hbox{$\displaystyle{#3}#2\m@th$}%
    \hskip\wd\@ne\hskip-\wd\tw@\mathord{\hskip\wd\tw@\hskip-\wd\@ne%
      {\vphantom{#3}}#1{#3}#2}}}
\def\@@rs#1#2{\mathinner{%
    \setbox\@ne\hbox{$\displaystyle{\vphantom{#2}}#1{#2}\m@th$}%
    \hskip\wd\@ne\mathord{\hskip-\wd\@ne%
      {\vphantom{#2}}#1{#2}}}}

\newcommand{\MR}{\mathalpha{\mathbb{R}}}

\newcommand*{\abs}[1]{\mathinner{\vert#1\vert}}

\newcommand*{\norm}[1]{\mathinner{\Vert#1\Vert}}

\DeclareMathOperator{\sgn}{sgn}

\definecolor{notecolor}{cmyk}{0,1,1,.2}
\newcommand*\AM@notesname{Notes}

\makeatother

\graphicspath{{figures/}}

\begin{document}

\title{Asymptotically consistent and computationally efficient modeling of short-ranged molecular interactions between curved slender fibers undergoing large 3D deformations}

\author{Maximilian J.~Grill*}
\author{Wolfgang A.~Wall}
\author{Christoph Meier}

\corres{* \email{maximilian.grill@tum.de}}

\address{Institute for Computational Mechanics, Technical University of Munich, Boltzmannstr.~15, 85748 Garching b.~M\"unchen, Germany}

\abstract[Summary]{%

This article proposes a novel computational modeling approach for short-ranged molecular interactions between curved slender fibers undergoing large 3D deformations, and gives a detailed overview how it fits into the framework of existing fiber or beam interaction models, either considering microscale molecular or macroscale contact effects. The direct evaluation of a molecular interaction potential between two general bodies in 3D space would require to integrate molecule densities over two 3D volumes, leading to a sixfold integral to be solved numerically. By exploiting the short-range nature of the considered class of interaction potentials as well as the fundamental kinematic assumption of undeformable fiber cross-sections, as typically applied in mechanical beam theories, a recently derived, closed-form analytical solution is applied for the interaction potential between a given section of the first fiber (slave beam) and the entire second fiber (master beam), whose geometry is linearly expanded at the point with smallest distance to the given slave beam section. This novel approach based on a pre-defined section-beam interaction potential (SBIP) requires only one single integration step along the slave beam length to be performed numerically. In addition to significant gains in computational efficiency, the total beam-beam interaction potential resulting from this approach is shown to exhibit an asymptotically consistent angular and distance scaling behavior. Critically for the numerical solution scheme, a regularization of the interaction potential in the zero-separation limit as well as the finite element discretization of the interacting fibers, modeled by the geometrically exact beam theory, are presented. In addition to elementary two-fiber systems, carefully chosen to verify accuracy and asymptotic consistence of the proposed SBIP approach, a potential practical application in form of adhesive nanofiber-grafted surfaces is studied. Involving a large number of helicoidal fibers undergoing large 3D deformations, arbitrary mutual fiber orientations as well as frequent local fiber pull-off and snap-into-contact events, this example demonstrates the robustness and computational efficiency of the new approach.

}


\maketitle

\section{Introduction}

This work is motivated by the abundance and manifoldness of biological, fiber-like structures on the nano- and microscale, including filamentous actin, collagen, and DNA, among others.
These slender, deformable fibers form a variety of complex, hierarchical assemblies such as networks (e.g.~cytoskeleton, extracellular matrix, mucus) or bundles (e.g.~muscle, tendon, ligament), which are crucial for numerous essential processes in the human body. Mainly due to the involved length and time scales and the complex composition of these systems, the design and working principles on the mesoscale often remain poorly understood and their significance regarding human physiology and pathophysiology can only be estimated so far.
In this context, computational modeling and simulation is expected to complement theoretical and experimental approaches and significantly contribute to the scientific progress in this field~\cite{lindstrom2010biopolymer,Castro2011,Gautieri2012,Sauer2009,Mueller2014rheology,Mueller2015interpolatedcrosslinks,Negi2018,Goodrich2018,GrillParticleMobilityHydrogels,Eichinger2021,BundlesPNAS}.
Moreover, the application of the accurate, efficient and versatile computational models and methods developed in this context may well be extended towards (future) technologies using, e.g.,~synthetic polymer, glass or carbon nanofibers in form of meshes, webbings, bundles or fibers embedded in a matrix material~\cite{durville2010,Kulachenko2012,Durville2015,Weeger2016,Meier2017b,pattinson2019additive,steinbrecher2020mortar,Khristenko2021,steinbrecher2021}.

The targeted class of problems is characterized by involving length and time scales that are inaccessible for simulations with atomistic resolution, but require a level of detail, which precludes homogenized continuum models.
Focusing on the efficient and accurate modeling of (short-ranged) molecular interactions such as van der Waals adhesion and steric repulsion, this article tackles one of the major challenges of the simulation-based investigation of these complex biophysical systems:
In many cases, these interactions between atoms or charges are the key to the functionality and behavior on the system level.
Yet, the inherent high dimensionality of interactions within an ensemble of objects in 3D poses a great challenge in terms of computational efficiency.
Our strategy hereby is to rigorously derive models from the first principles of molecular interactions, which are formulated as interaction potentials of atoms or unit charges, and at the same time exploit the dimensionally reduced, slender structure of the fibers, which is inspired by mechanical beam theories, as well as the short-range nature of the considered class of interaction potentials. This leads to accurate, efficient and versatile beam-beam interaction formulations that allow for the computational study of so far intractable problems in complex biophysical systems of slender fibers.

There is a rich body of literature dealing with the analytical and computational modeling of molecular interactions between arbitrarily shaped, solid bodies in 3D space~\cite{Argento1997,Sauer2007a,Sauer2009a,Sauer2013,Fan2015,Du2019,Mergel2019}. A direct evaluation of the interaction potential between two general bodies in 3D space would require to integrate molecule densities over their volumes, leading to a sixfold integral (two nested 3D integrals) that has to be solved numerically. Even though dimensionally reduced models have been derived that are tailored for short-ranged interactions of such general-shaped 3D bodies and, thus, only require a numerical integration across the interacting surfaces~\cite{Sauer2009a}, the solution of the remaining fourfold integral in combination with the sharp gradients characterizing these short-ranged interaction forces is still too demanding from a computational point of view to simulate representative 3D systems of curved slender fibers, which necessitates the development of reduced-order models for slender fibers that consistently account for their molecular interactions. While there is a large number of articles~\cite{wriggers1997,litewka2005,durville2010,Kulachenko2012,Chamekh2014,GayNeto2016a,Konyukhov2016,Weeger2017,meier2016,Meier2017a,bosten2022mortar} focusing on macroscale contact interaction between slender fibers respectively beams, comparable formulations for microscale molecular interactions are still missing. Important steps into this direction have been made by the works~\cite{Sauer2009,Sauer2014,Schmidt2015}, however limited to the interaction of fibers respectively beams with a rigid half-space.

Based on the fundamental kinematic assumption of undeformable fiber cross-sections, as typically applied in mechanical beam theories, the authors recently proposed a generalized formalism to postulate section-section interaction potentials (SSIP) integrated into the framework of Cosserat beam theories, which allows to consider general interaction potentials between curved fibers in 3D space characterized by large deformations, arbitrary mutual orientations, initial curvatures and cross-section shapes as well as inhomogeneous molecule/charge distributions within the cross-sections~\cite{Meier2021CosseratPotential}. In a previous contribution of the authors, where this SSIP approach has originally been proposed, exemplary closed-form analytical solutions for the required SSIP laws could be derived for different long-ranged (e.g., electrostatic) and short-ranged (e.g., van der Waals adhesion and steric repulsion) interactions, relying on the additional assumptions of circular cross-section shapes and homogeneous molecule distributions and considering the asymptotic limits of either large distances for long-ranged or small distances for short-ranged interactions~\cite{GrillSSIP}. Due to the pre-calculated analytical representation of section-section interaction potentials, this SSIP approach only requires the twofold integration along the fiber length directions to be performed numerically. While this approach is general in the sense that tailored SSIP laws could be derived for either long- or short-ranged interactions, the latter class turned out to be critical in terms of accuracy and algorithmic complexity, i.e., the derived SSIP laws for short-ranged interactions did not exhibit a consistent asymptotic scaling behavior and the required numerical solution of a twofold integral was still dominating the overall computational costs when simulating fiber systems. Following up these conclusions from our previous contribution~\cite{GrillSSIP}, we aim to develop an enhanced approach for the specifically important and challenging case of very short-ranged interactions.

By exploiting the short-range nature of the considered class of interaction potentials and the kinematic assumption of undeformable fiber cross-sections, a recently derived, closed-form analytical solution~\cite{Grilldiskcylpot} is applied for the interaction potential between a given section of the first fiber (slave beam) and the entire second fiber (master beam), whose geometry is linearly expanded at the point with smallest distance to the given slave beam section. This novel approach based on a pre-defined section-beam interaction potential (SBIP) requires only one single integration step along the slave beam length to be performed numerically. To formulate a specific computational model for fiber systems, this interaction model is combined with the geometrically exact beam theory, representing the mechanics of individual fibers, and discretized on basis of the finite element method.

Based on elementary two-fiber test cases considering pairs of either straight\&straight, straight\&curved or curved\&curved beams, the total beam-beam interaction potential resulting from this approach is shown to exhibit an asymptotically consistent angular and distance scaling behavior in the decisive regime of small separations. Thus, remarkably, the newly proposed SBIP approach turns out to be superior to the previously derived SSIP approach, when applied to short-ranged interactions, not only in terms of computational efficiency but also in terms of model accuracy. Based on conservative estimates for the algorithmic complexity and by considering practically relevant parameter settings, it is demonstrated that the SBIP approach has the potential to reduce the computational costs by at least one order of magnitude as compared to the SSIP approach, and by at least five orders of magnitude as compared to the direct approach of sixfold numerical integration.

The remainder of this article is structured as follows.
First, fundamentals required for the modeling of molecular interactions and of slender fibers will be briefly summarized in \secref{sec::fundamentals_SBIP}.
In the following, the novel SBIP approach will be presented in \secref{sec::method_single_length_specific_integral} and the applied closed-form analytical interaction law as well as the verification of its consistent scaling behavior by means of elementary straight-fiber-pair test cases with analytical solutions will be presented in~\secref{sec::ia_pot_single_length_specific_evaluation_vdW}.
These aspects will be combined to derive the resulting virtual work contribution in \secref{sec::virtual_work_disk-cyl-pot_m}, its discretization based on finite elements as well as the consistent linearization required for tangent-based solution schemes. Eventually, Sec.~\secref{sec::beam_interaction_formulations_miscellaneous} gives a detailed overview on how the resulting computational models according to the novel SBIP approach and the previously proposed SSIP approach fit into the framework of existing fiber or beam interaction models, either considering microscale molecular or macroscale contact effects.

Important numerical and algorithmic aspects, such as the regularization of the interaction potential in the (singular) zero separation limit, search schemes for relevant pairs of interaction partners and overall algorithmic complexity are presented in~\secref{sec::discretization_algorithmic_aspects}. This discussion finally paves the way for the set of numerical examples in \secref{sec::numerical_examples_SBIP}. As further elementary test case to verify the model accuracy of the proposed SBIP approach, the peeling and pull-off behavior of a pair consisting either of a straight and a curved beam or of two curved beams will be studied in detail. Critically, by switching the master-slave assignment, the variant with one straight and one curved beam allows to verify the fundamental modeling assumption of approximating the geometry of one beam (master) as straight cylinder following a first-order expansion. Most notably, this section includes an application-oriented example mimicking adhesive nanofiber-grafted surfaces, which confirms the effectiveness, computational efficiency and robustness of the novel formulation in combination with implicit time integration for large-scale, complex 3D systems with arbitrary mutual configurations, frequent local pull-off and snap-into-contact events as well as large deformations of the interacting, strongly curved fibers. Finally, this article will be concluded by a summary and outlook in \secref{sec::SBIP_conclusions_outlook}.

\section{Fundamentals}\label{sec::fundamentals_SBIP}
This section briefly summarizes essential aspects from the fields of molecular interactions and beam theory that will be referred to in the remainder of this article and are thus crucial for the overall understanding.

\subsection{Two-body interaction from a molecular perspective}\label{sec::two-body_molecular_interaction_SBIP}
\begin{figure}[htpb]%
  \centering
  \def\svgwidth{0.35\textwidth}
  \input{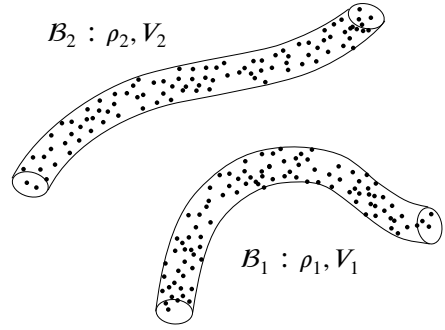}
  \vspace{-0.8cm}
  \caption{Two arbitrarily shaped, deformable bodies~$\mathcal{B}_1$ and~$\mathcal{B}_2$ with volumes~$V_1,V_2$ and continuous particle densities $\rho_1,\rho_2$.}
  \label{fig::beam_to_beam_interaction_particle_clouds}
\end{figure}

This section briefly recapitulates how the interaction of two extended, macromolecular or macroscopic bodies can be described starting from the first principles of atom-atom, i.e., point-pair interaction.
The following summary is reproduced from our previous work~\cite{GrillSSIP} for the reader's convenience.
Refer e.g.~to the textbook~\cite{israel2011} for further details.
\figref{fig::beam_to_beam_interaction_particle_clouds} schematically visualizes the distribution of elementary interaction partners, i.e., atoms or molecules, within two macromolecular or macroscopic bodies.

Consider a point pair interaction potential $\Phi(r)$ as a function of the mutual distance~$r$.
Popular examples include the inverse-sixth power law valid for van der Waals (vdW) interactions
\begin{align}\label{eq::pot_ia_vdW_pointpair}
  \Phi_\text{vdW}(r) = k_6 r^{-6} = - C_\text{vdW} r^{-6}
\end{align}
and the LJ interaction law, which extends the attractive vdW part by a repulsive steric contribution:
\begin{align}\label{eq::pot_ia_LJ_pointpair}
  \Phi_\text{LJ}(r) = k_{12} r^{-12} + k_6 r^{-6} = -\Phi_\text{LJ,eq} \left( \left( \frac{r_\text{LJ,eq}}{r} \right)^{12} - 2 \,\left( \frac{r_\text{LJ,eq}}{r} \right)^{6} \right)
\end{align}
Both are typical examples for short-ranged interactions and the generalized form of an inverse power law with high exponent $m>3$
\begin{align}\label{eq::pot_ia_pointpair_generic_inverse_powerlaw}
  \Phi_\text{m}(r) = k_\text{m} r^{-m}
\end{align}
will serve as the prime example to be used throughout this article.
To ease the comparison with the literature, a few equivalent forms using the most popular definitions and notations of the constant parameters~$k_\text{m},C_\text{vdW},r_\text{LJ,eq},\Phi_\text{LJ,eq}$ are stated above.

Assuming additivity, we apply \textit{pairwise summation} to arrive at the two-body interaction potential
\begin{equation}
 \Pi_\text{ia} = \sum_{i\in \mathcal{B}_1} \sum_{j\in \mathcal{B}_2} \Phi(r_{ij}).
\end{equation}
Note that the assumption of additivity is known to not hold unconditionally for instance in the important case of vdW interaction.
However, the distance dependency obtained from pairwise summation is still valid and only the prefactor called Hamaker ``constant'' needs to be obtained from advanced Lifshitz theory, which yields Hamaker-Lifshitz hybrid forms~\cite{parsegian2005,israel2011} and motivates us to still apply pairwise summation here.
Further assuming a continuous atomic density $\rho_i$, $i={1,2}$, the total interaction potential can alternatively be rewritten as nested integrals over the volumes $V_1, V_2$ of both bodies~$\mathcal{B}_1$ and~$\mathcal{B}_2$:
\begin{equation}\label{eq::pot_fullvolint}
 \Pi_\text{ia} = \iint_{V_1,V_2} \rho_1(\vx_1) \rho_2(\vx_2) \Phi(r) \dd V_2 \dd V_1 \qquad \text{with} \quad r = \norm{\vx_1-\vx_2}
\end{equation}
It can be shown that this continuum approach is the result of \textit{coarse-graining}, i.e., smearing out the discrete positions of atoms in a system into a smooth atomic density function $\rho(\vx)$.\cite{Sauer2007a}

\subsection{General strategy to account for molecular interactions}\label{sec::problem_statement_general_strategy}
The general approach to incorporate the effect of molecular interactions is identical to the one suggested for solid bodies in previous work~\cite{Argento1997,Sauer2007a} and has been summarized also in our recent contribution~\cite{GrillSSIP}, which is repeated here for convenience.
For a classical conservative system, the total potential energy of the system can be stated taking into account the internal and external energy~$\Pi_\text{int}$ and~$\Pi_\text{ext}$.
The additional contribution from molecular interaction potentials~$\Pi_\text{ia}$ is simply added to the total potential energy as follows.
\begin{equation}\label{eq::total_potential_energy_is_minimized}
 \Pi_\text{TPE}=\Pi_\text{int}-\Pi_\text{ext}+\Pi_\text{ia} \defeq \text{min.}
\end{equation}
Note that the standard parts~$\Pi_\text{int}$ and~$\Pi_\text{ext}$ remain unchanged from the additional contribution.
One noteworthy difference is that internal and external energy are summed over all \textit{individual} bodies in the system whereas the total interaction free energy is summed over all \textit{pairs} of interacting bodies.
In order to shed some light on the basic characteristics of systems with adhesive elastic fibers, the dimensionless parameters of the governing \eqref{eq::total_potential_energy_is_minimized} are identified by means of nondimensionalization and discussed in Appendix~\ref{sec::nondimensionalization}.

According to the \textit{principle of minimum of total potential energy}, the weak form of the equilibrium equations, which serves as basis for a subsequent finite element discretization of the problem, is derived by means of variational calculus.
The very same equation may alternatively be derived by means of the \textit{principle of virtual work}, which also holds for non-conservative systems:
\begin{align}\label{eq::total_virtual_work_is_zero}
 \delta \Pi_\text{int} - \delta \Pi_\text{ext} + \delta \Pi_\text{ia} = 0
\end{align}
Clearly, the evaluation of the interaction potential~$\Pi_{ia}$, or rather its variation~$\delta \Pi_{ia}$, is the crucial step here.
Recall~\eqref{eq::pot_fullvolint} to realize that it generally requires the evaluation of two nested 3D integrals%
\footnote{It is important to mention that, assuming additivity of the involved potentials, systems with more than two bodies can be handled by superposition of all pair-wise two-body interaction potentials.
It is thus sufficient to consider one pair of beams in the following.
The same reasoning applies to more than one type of physical interaction, i.e., potential contribution.}.
The direct approach of incorporating this interaction potential in a computational model using 6D numerical quadrature turns out to be extremely costly and in fact inhibits any application to (biologically) relevant multi-body systems.
See \secref{sec::algorithm_complexity_fullint_vs_SSIP_vs_SBIP} for more details on the algorithmic complexity and the computational cost of this naive, direct approach as well as the novel SBIP approach to be proposed in \secref{sec::method_single_length_specific_integral}.

\subsection{Van der Waals interaction potential between two straight cylinders}\label{sec::basics_vdW_disks_cylinders}
One of the main objectives of this work is that the mutual orientation of two fibers, i.e. the angle~$\alpha$ enclosed by the local tangent vectors of the (potentially curved) fiber centerlines, shall be taken into account in order to improve the accuracy of the reduced interaction law and the overall computational model.
In this section, we review the angle dependency for the simple case of two straight fibers, which will serve as an analytical reference solution in order to verify the specific reduced interaction law to be derived in~\secref{sec::ia_pot_single_length_specific_evaluation_vdW}.
To begin with, recall the analytical solutions for the special cases of parallel and perpendicular cylinders of infinite length (in the regime of small separations).
These expressions agree with the following, more general relationship valid for all mutual angles~$\alpha \in \, ]0,\pi/2]$ stated e.g.~in the textbook~\cite[p.~173]{parsegian2005}:
\begin{equation}\label{eq::pot_ia_vdW_cyl_cyl_skewed_smallseparation}
  \Pi_\text{vdW,cyl-cyl} =  - \frac{A_\text{Ham}}{6} \sqrt{R_1 R_2} \, g_\text{bl}^{-1} / \sin\alpha
      \qquad \text{with} \quad  A_\text{Ham} \defvariable \pi^2 \rho_1 \rho_2 C_\text{vdW}
\end{equation}
Here, $R_1$ and $R_2$ denote the radius of the first and second cylinder, and $g_\text{bl}^{-1}$ denotes their (bilateral) smallest surface-to-surface separation, also known as gap.
For the limiting case of perpendicular cylinders~$\alpha=\pi/2$, this coincides with~$\Pi_{\text{vdW,cyl}\perp\text{cyl,ss}}$ listed in the quick reference table of analytical solutions in our previous article~\cite{GrillSSIP} (alongside other expressions mentioned here).
For the case of parallel cylinders, however, note that the total interaction potential of infinitely long cylinders is infinite and we obtain~$\tilde \pi_{\text{vdW,cyl}\parallel\text{cyl,ss}}$ instead.\\

\noindent\textit{Remark.} Interestingly, the $1 / \sin\alpha$-scaling behavior also holds true for screened electrostatic interaction of two cylinders \cite{brenner1974}, \cite[p.~23]{Langbein1974}.
Indeed, it is shown in \cite[p.~218]{israel2011}, that this relation results from fundamental geometric considerations related to the so-called Derjaguin approximation, and is thus independent of the type of interaction, i.e., the specific form of the point interaction potential law~$\Phi(r)$.\\

In addition to these analytical expressions obtained by means of simple pairwise summation, further theoretical work that relax certain assumptions and consider more advanced aspects like interaction across inhomogeneous or anisotropic media, differences in optical material properties or retardation can be found in the literature.
To give but one example, \cite{Rajter2007} studies cylinders with anisotropic optical properties considering the example of carbon nanotubes.
A review of recent research activities on this topic is given in \cite{Dobson2012}.
This work however focuses on the extension towards curved slender fibers with arbitrary mutual separations/orientations due to their possibly large elastic deformations in 3D, and therefore uses the basic pairwise summation approach as mentioned and motivated in~\secref{sec::two-body_molecular_interaction_SBIP}.

\subsection{Steric repulsion -- mechanical contact}
The prevailing notion of contact between bodies in biophysics is commonly described as excluded volume effect, which means that bodies may approach each other without any influence on each other and only as soon as their surfaces touch, the repulsive contact forces that inhibit any overlap of the bodies' volumes may rise to infinite strength.

Throughout this work, repulsive contact forces will be modeled based on the repulsive part of the LJ potential law (\eqref{eq::pot_ia_LJ_pointpair}), which is an inverse-twelve power law in the separation of the point-like atoms.
A number of alternative force-distance laws can be found in literature, however this approach seems to be most consistent with the modeling of vdW interactions as inverse-six point-pair potential.
In particular, in Sec.~\ref{sec::beam_interaction_formulations_miscellaneous}, the approach of using a repulsive interaction potential will be compared to macroscopic formulations well-established for the mechanical contact interaction of slender fibers respectively beams.

\subsection{Geometrically exact 3D beam theory and corresponding finite element formulations}
We use geometrically exact 3D beam theory to model large elastic deformations of slender fibers in 3D space.
See e.g.~Ref.\cite{Meier2017c} for a recent review of both space-continuous beam theories as well as suitable temporal and spatial discretization schemes.
The interaction approach to be proposed in this article is both independent of the specific beam formulation and the discretization schemes used to describe the mechanics of individual fibers.
In this work the proposed interaction approach is exemplarily applied in combination with geometrically exact beam elements of both shear-deformable (Simo-Reissner) and shear-free (Kirchhoff-Love) type.
The following two subsections briefly summarize the fundamentals required for the remainder of this article.

\subsubsection{Space-continuous beam theory}
\begin{figure}[htpb]%
  \centering
  \def\svgwidth{0.6\textwidth}
  \input{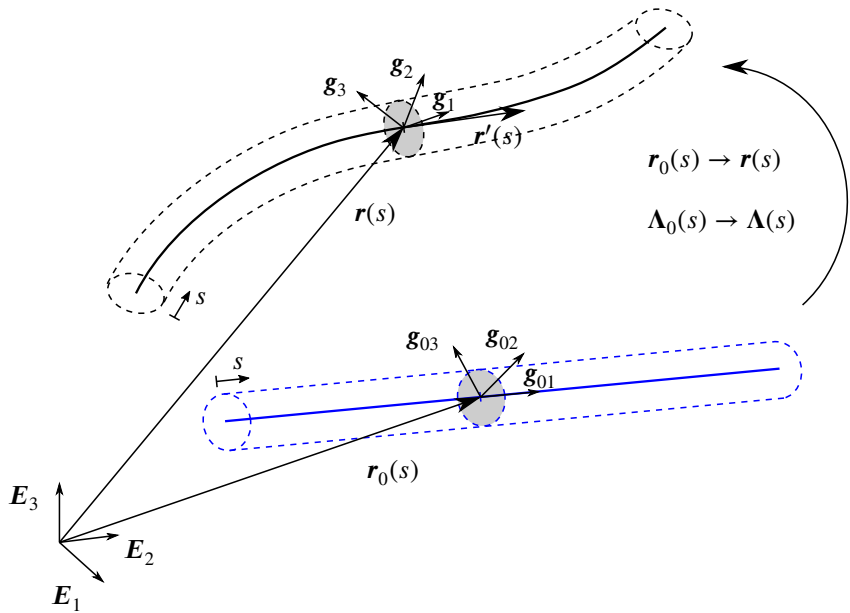}
  \caption{Example configurations and kinematics of the Cosserat continuum formulation of a beam to illustrate the field of centroid positions~$\vr(s,t)$ and material triad field~$\vLambda(s,t)$: Initial, i.e., stress-free (blue) and deformed (black) configuration. Straight configuration in the initial state is chosen exemplarily here without loss of generality.}
  \label{fig::beam_kinematics}
\end{figure}

The configuration of a beam at time~$t$ is uniquely defined by the controid position vector~$\vr(s,t) \in \MR^3$ and the orthonormal frame~$\vLambda(s,t)=[\vg_1,\vg_2,\vg_3] \in SO(3)$ describing the cross-section orientation at each point~$s$ along the 1D Cosserat continuum.
Note that the arc-length parameter~$s$ is hereby defined in the stress-free, initial configuration of the centerline curve~$\vr_0(s) \defvariable \vr(s,t=0)$.
See~\figref{fig::beam_kinematics} for an illustration of these geometrical quantities and the resulting kinematics.
According to this concept of geometry representation, the position~$\vx$ of an arbitrary material point~$P$ of the slender body is obtained from
\begin{equation} \label{eq::position_material_point_Cosserat}
  \vx_\text{P}(s,s_2,s_3,t) = \vr(s,t) + s_2 \, \vg_2(s,t) + s_3 \, \vg_3(s,t).
\end{equation}
Here, the additional convective coordinates~$s_2$ and $s_3$ specify the location of P within the cross-section, i.e., as linear combination of the orthonormal directors~$\vg_2$ and $\vg_3$.
For a minimal parameterization of the triad, e.g.~the three-component rotation pseudo-vector~$\vpsi$ may be used, i.e. $\vLambda(s,t)=\vLambda(\vpsi(s,t))$, such that we end up with six independent degrees of freedom~$(\vr,\vpsi)$ at every centerline location $s$ to define the position of each material point in the body by means of~\eqref{eq::position_material_point_Cosserat}.

Again refer to \figref{fig::beam_kinematics} for a sketch of the kinematics of geometrically exact beam models.
Based on these kinematic quantities, deformation measures as well as constitutive laws can be defined.
Finally, the potential energy of the internal (elastic) forces and moments~$\Pi_\text{int}$ is expressed uniquely by means of the set of six degrees of freedom~$(\vr,\vpsi)$ at each point of the 1D Cosserat continuum.
See e.g.~\cite{jelenic1999,crisfield1999,Meier2017c} for a detailed presentation of these steps.\\

\noindent\textit{Remark on notation.}
Unless otherwise specified, all vector and matrix quantities are expressed in the global Cartesian basis~$\vE_i$. Differing bases as e.g.~the material frame are indicated by a subscript $\left[.\right]_{\vg_i}$.
Quantities evaluated at time~$t\!=\!0$, i.e., the initial stress-free configuration, are indicated by a subscript $0$ as e.g.~in~$\vr_0(s)$.
Differentiation with respect to the arc-length coordinate~$s$ is indicated by a prime, e.g., for the centerline tangent vector~$\vr'(s,t)=\tdiff{\vr(s,t)}{s}$.
Differentiation with respect to time~$t$ is indicated by a dot, e.g., for the centerline velocity vector~$\dot\vr(s,t) = \tdiff{\vr(s,t)}{t}$.
For the sake of brevity, the arguments~$s,t$ will often be omitted in the following.\\

\noindent\textit{Remark on finite 3D rotations.}
To a large extent, the challenges and complexity in the theoretical as well as numerical treatment of the geometrically exact beam theory can be traced back to the presence of large rotations.
In contrast to the much more common vector spaces, the rotation group $SO(3)$ is a nonlinear manifold (with Lie group structure) and lacks essential properties such as additivity and commutativity, which renders standard procedures such as the interpolation or the update of configurations quite intricate.
We thus follow a two-part strategy.
First, we aim to develop and formulate the novel approach in the most general form in \secref{sec::method_single_length_specific_integral}, including also cases such as arbitrary cross-section shapes or inhomogeneous atomic densities, where the involvement of large 3D rotations is inevitable.
In a second step, however, we aim to abstain from the handling of finite 3D rotations wherever possible when proposing specific reduced interaction laws for instance for the case of homogeneous, circular cross-sections considered in \secref{sec::ia_pot_single_length_specific_evaluation_vdW}.
As a result, this will allow to avoid the handling of finite rotations in the interaction potentials and to achieve simpler and more compact numerical formulations whenever possible.

\subsubsection{Spatial discretization based on beam finite elements}\label{sec::discretization_SBIP}
A smooth, i.e., $C^1$-continuous, discrete representation of the centerline curve is inevitable in the context of molecular interaction laws with its typical high gradients in order to ensure robustness of the numerical method also for reasonably coarse discretizations.
This is a well-known general issue and has been discussed in the context of macroscopic beam contact interaction~\cite{Meier2017b} and (surface enrichment of) 2D and 3D solid contact elements based on the LJ interaction potential~\cite{Sauer2011} before.
In the scope of this work, it is addressed by applying a third order Hermite interpolation scheme to discretize the centerline curve~$\vr(s)$.
The corresponding beam finite element formulations of both Simo-Reissner type and Kirchhoff-Love type have been presented in~\cite{Meier2017b} and~\cite{Meier2017c}, respectively, and only the most essential aspects required later in this work will be briefly summarized in the following.

The spatial centerline curve~$\vr$ is approximated by means of the discrete set of the $i$-th node's position vector~$\hat\vdd^i \in \MR^3$ and tangent vector~$\hat\vdt^i \in \MR^3$ ($i=\{1,2\}$) as primary degrees of freedom and the four scalar cubic Hermite polynomials~$H^i_\text{d/t}$ used for interpolation as follows.
\begin{align}\label{eq::centerline_discretization}
  \vr(\xi) \approx \vr_\text{h}(\xi) &= \sum_{i=1}^2 H_\text{d}^i(\xi) \, \hat \vdd^i + \frac{l_\text{ele}}{2} \sum_{i=1}^2 H_\text{t}^i(\xi) \, \hat \vdt^i =: \vdH \, \hat \vdd
\end{align}
Here, $l_\text{ele}$ denotes the initial length of the element.
The newly introduced element-local parameter~$\xi \in [-1;1]$ is biuniquely related to the arc-length parameter~$s \in [s_\text{ele,min}; s_\text{ele,max}]$ describing the very same physical domain of the beam and the scalar factor defining this mapping between both length measures in differential form is called the element Jacobian~$J(\xi)$ with~$\dd s =: J(\xi) \dd \xi$.
Note that on the right hand side of \eqref{eq::centerline_discretization}, all the centerline degrees of freedom of one beam element, i.e., the nodal positions~$\hat\vdd^i$ and tangents~$\hat\vdt^i$ are collected in one vector~$\hat\vdd$ for a more compact notation.
Accordingly, $\vdH$ is the assembled matrix of shape functions, i.e., Hermite polynomials~$H_\text{d}^i$ and $H_\text{t}^i$.
Following a Bubnov-Galerkin scheme, this very same interpolation is applied to the test functions, i.e., the variation of the centerline curve
\begin{align}\label{eq::var_centerline_discretization}
  \delta \vr(\xi) \approx \delta \vr_\text{h}(\xi) &= \sum_{i=1}^2 H_\text{d}^i(\xi) \, \delta \hat \vdd^i + \frac{l_\text{ele}}{2} \sum_{i=1}^2 H_\text{t}^i(\xi) \, \delta \hat \vdt^i =: \vdH \, \delta \hat \vdd.
\end{align}
As mentioned already in the last section, the specific reduced interaction law to be proposed in~\secref{sec::ia_pot_single_length_specific_evaluation_vdW} will be described by the centerline curve (and tangent field) only and avoid the use of the rotation field in favor of a simple and efficient formulation.
Thus, the spatial discretization of the rotation field will not be required and omitted here for the sake of brevity.\\

\noindent\textit{Remark on notation.}
Note that a derivative with respect to the element-local parameter~$\xi$ will be indicated by a short upright prime, e.g.,~$\vr^\shortmid(\xi,t)=\tdiff{\vr(\xi,t)}{\xi}$ in order to differentiate it from the derivative~$\vr'(s,t)$ with respect to~$s$.\\

At the end of this section, we would like to point out that both the general SBIP approach and the reduced interaction law to be proposed in this article are not limited to a specific interpolation scheme and that the presented Hermite interpolation is just one example that we employ throughout this work.

\section{The section-beam interaction potential (SBIP) approach}
\label{sec::method_single_length_specific_integral}

When considering two slender, fiber-like bodies with lengths $l_i$ and cross-sections $A_i$ ($i=1,2$), it is reasonable to split the two-fold volume integral of the interaction potential~\eqref{eq::pot_fullvolint} in integrals across the fiber lengths and cross-sections as follows:
\begin{equation}\label{eq::pot_3variants}
 \Pi_\text{ia} = \iint_{V_1,V_2} \rho_1(\vx_1) \rho_2(\vx_2) \Phi(r) \dd V_2 \dd V_1  =
 \int_{l_1}  \underbrace{\int_{l_2} \, \overbrace{\! \iint_{A_1,A_2} \rho_1(\vx_1) \rho_2(\vx_2) \Phi(r) \dd A_2 \dd A_1}^{=:\tilde{\tilde{\pi}}(\vr_{1-2},\vpsi_{1-2}) \,\, \rightarrow \,\, \text{SSIP}} \dd s_2}_{=:\tilde{\pi}(\vr_{1-2\text{c}},\vpsi_{1-2\text{c}}) \,\, \rightarrow \,\, \text{SBIP}} \dd s_1.
\end{equation}
In our previous work~\cite{GrillSSIP}, the double length-specific interaction potential $\tilde{\tilde{\pi}}(\vr_{1-2},\vpsi_{1-2})$ between two cross-sections characterized by a distance vector $\vr_{1-2}$ and mutual orientation vector $\vpsi_{1-2}$~\cite{Meier2021CosseratPotential} has been approximated analytically by exploiting the short-range nature of the considered class of interaction potentials and the fundamental kinematic assumption~\eqref{eq::position_material_point_Cosserat} characterizing the fiber deformation. Thus, in the proposed computational model only the two integrals across the fiber lengths $l_1$ and $l_2$ had to be solved numerically, a procedure that was denoted as section-section interaction potential (SSIP) approach referring to the physical meaning of $\tilde{\tilde{\pi}}(\vr_{1-2},\vpsi_{1-2})$. The present works aims to follow this path one (essential) step further, by approximating the second of the interacting fibers/beams as a (cylinder-shaped) surrogate body constructed at the position of smallest distance with respect to a given point on the first fiber such that the single length-specific interaction potential ${\tilde{\pi}}(\vr_{1-2c},\vpsi_{1-2c})$ between a cross-section of the first fiber and the entire second fiber can be approximated analytically. This procedure will be denoted as section-beam interaction potential (SBIP) approach, again referring to the physical meaning of ${\tilde{\pi}}(\vr_{1-2c},\vpsi_{1-2c})$, and will only require a 1D integral, i.e. integration across the first fiber's length $l_1$, to be solved numerically. This general approach will be motivated in the following, before a specific expression for the section-beam interaction potential ${\tilde{\pi}}(\vr_{1-2c},\vpsi_{1-2c})$ will be presented in Sec.~\ref{sec::ia_pot_single_length_specific_evaluation_vdW}.

\begin{figure}[htpb]%
  \centering
  \def\svgwidth{0.65\textwidth}
  \input{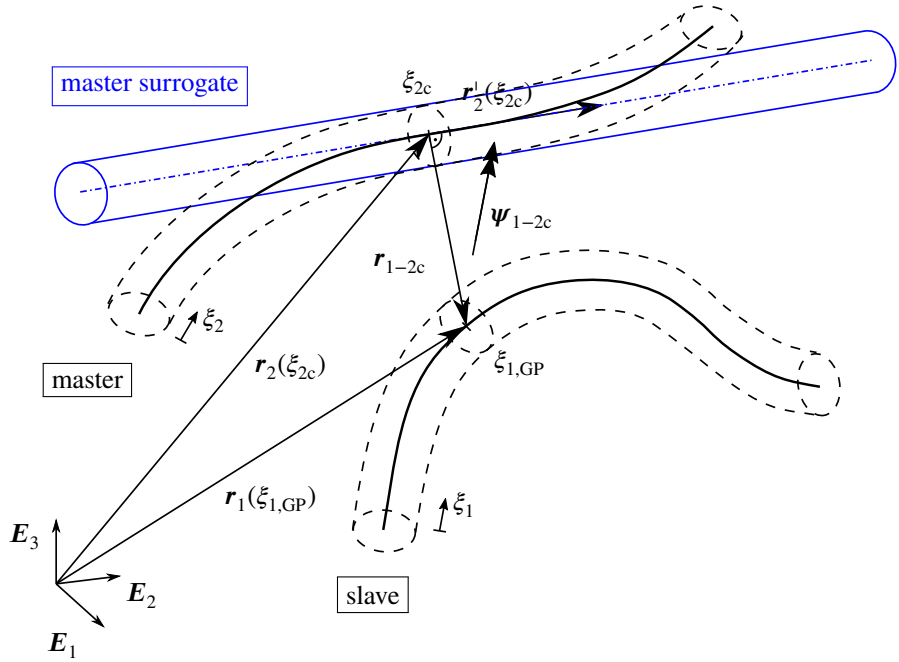}
  \caption{Illustration of the section-beam interaction potential (SBIP) approach.
           The actual, deformed volume of the second interaction partner (``master'') is approximated by a surrogate body (blue) located at the closest point~$\xi_{2\text{c}}$ to a given integration point~$\xi_{1\text{,GP}}$ of the first interaction partner (``slave'').
           Distance vector~$\vr_{1-2\text{c}}$ and relative rotation vector~$\vpsi_{1-2\text{c}}$ uniquely describe the mutual separation and orientation.}
  \label{fig::master_slave_approach_strategy}
\end{figure}

Consider a point-pair interaction potential $\Phi$ with a very steep gradient as for example the inverse power laws with exponent six or twelve from the popular LJ interaction law (\eqref{eq::pot_ia_LJ_pointpair}).
The rapid decay of the potential with increasing distance implies that among all possible point pairs between both bodies only those with the smallest separation contribute significantly to the total interaction potential of both bodies.
When looking at the interaction of two deformable slender bodies such as fibers, this consideration gives rise to an approach where the geometry of the second body is approximated by a surrogate body with simplified geometry located at the point of closest distance from a given point on the first body.
See \figref{fig::master_slave_approach_strategy} for an illustration of the approach using the example of circular cross-sections and therefore a cylinder-shaped surrogate body.
In the region around the closest point, this straight cylinder is expected to be a good approximation for the actual, possibly deformed, beam geometry.
Note, however, that the general SBIP approach to short-ranged beam-beam interactions is not limited to the circular cross-section shape shown in this example.

In accordance with formulations for macroscopic beam contact, the body which is projected onto, i.e., here the one with approximated geometry is referred to as master beam (indicated by subscript m) whereas the first body is called slave beam (indicated by subscript s).
Without loss of generality, the beam with index~$1$ is assumed to be the slave beam whereas index~$2$ is used as a synonym for master.

From a mathematical point of view, the geometrical approximation used in this context is equivalent to a Taylor series expansion of the centerline curve~$\vr_2(\xi_{2\text{c}})$ of the master beam at the closest point~$\xi_{2\text{c}}$ truncated after the second, i.e., linear term:
\begin{align}
  \vr_{2\text{,approx}}(\xi_2) \approx \vr_2(\xi_{2\text{c}}) + \vr_2^{\shortmid}(\xi_{2\text{c}}) \, \xi_2 \, ( + \, \text{H.O.T.} )
\end{align}
Here, the linear term represents the orientation of the surrogate body in the direction of the master beam's tangent vector at the closest point~$\vr_2^{\shortmid}(\xi_{2\text{c}})$.
Recall from the previous section that the short prime denotes a differentiation with respect to the element parameter coordinate, i.e.,~$\vr^{\shortmid}_{i}(\xi_i) = \tdiff{\vr_i(\xi_i)}{{\xi_i}}$.

As stated above, we assume an interaction potential~$\tilde{\pi}(\vr_{1-2\text{c}},\vpsi_{1-2\text{c}})$ for the interaction between all the points within one cross-section of the slave beam and the entire master beam surrogate, i.e., the tangential straight cylinder in the example above. Following~\eqref{eq::pot_3variants}, the total interaction potential is evaluated as an integral along the slave beam centerline curve as follows:
\begin{align}\label{eq::ia_pot_single_integration}
  \Pi_{ia} = \int \limits_0^{l_1} \tilde{\pi}(\vr_{1-2\text{c}},\vpsi_{1-2\text{c}}) \dd s_1.
\end{align}
Generally, such a section-beam interaction potential (SBIP) is a length-specific quantity with dimensions of energy per unit length (of the slave beam).
It is an analytical expression uniquely defined by the mutual configuration, i.e., the distance vector~$\vr_{1-2\text{c}}$ and relative rotation vector~$\vpsi_{1-2\text{c}}$ as illustrated in \figref{fig::master_slave_approach_strategy}.

To evaluate the remaining 1D integral in~\eqref{eq::ia_pot_single_integration}, Gaussian quadrature is applied throughout this work.
The reduction starting from 6D integration (cf.~\eqref{eq::pot_fullvolint}) to 1D integration already indicates the overall gain in efficiency that will be further analyzed and discussed in \secref{sec::algorithm_complexity_fullint_vs_SSIP_vs_SBIP}.
Recall also the 2D integral (two nested 1D integrals) resulting for the general SSIP approach to realize the superior efficiency of this novel SBIP approach that will also be verified in the numerical experiments of \secref{sec::numerical_examples_SBIP}.

In analogy to the previously presented section-section interaction potential (SSIP) approach from Ref.~\cite{GrillSSIP}, the question of how to find an analytical, closed-form expression for the reduced interaction law~$\tilde{\pi}(\vr_{1-2\text{c}},\vpsi_{1-2\text{c}})$ can be considered separately from the generally valid SBIP approach proposed in this section.
Just as for the SSIP laws, such an effective SBIP law~$\tilde \pi$ will depend on the considered type of interaction, the cross-section shape(s) and dimensions, the atom density distributions, and possibly other interaction-specific factors.
An example for how to determine this (single) length-specific potential~$\tilde{\pi}(\vr_{1-2\text{c}},\vpsi_{1-2\text{c}})$ analytically by means of 5D integration starting from the point-point interaction potential law (\eqref{eq::pot_ia_LJ_pointpair}) in case of LJ interaction is presented in \secref{sec::ia_pot_single_length_specific_evaluation_vdW}.
However, other strategies such as postulating the general form of the SBIP law and fitting of the parameters to experimental results e.g.~for the force response in two-fiber systems are considered to be promising alternative ways that could enable a broad variety of future applications of this SBIP approach.\\

\noindent\textit{Discussion of the choice of master and slave side.}
Starting from the problem of two-body interaction that is symmetric with respect to the two interaction partners, the SBIP approach introduces the notion of master and slave, which causes a bias in the formulation and asks for a criterion how to assign these roles.
Both from the mathematical description as a linear Taylor approximation and the illustration in \figref{fig::master_slave_approach_strategy}, it becomes clear that the resulting model error and bias will depend on the magnitude of curvature of the master beam's centerline.
This would give rise to a criterion that chooses the beam with the smaller (maximum or average) curvature as the master beam.
However, such criteria might lead to sudden changes of master and slave over time, which is numerically unfavorable.
Alternatively, one could consider to evaluate the pair of beams in two half passes, where the roles of master and slave switch and only the contributions on the slave side are evaluated in each of the passes (see e.g.~\cite{Sauer2013,Papadopoulos1995}), such that the bias in the formulation is avoided.
However, it can be argued that the model error introduced by the use of the surrogate beam on the master side is negligible, because first, the curvature anyway is assumed to be limited in the underlying beam theory (typically compared to the inverse radius~$1/R$ in case of circular cross-sections) and second, the very short range of the considered interactions naturally limits the impact of the master beam's shape deviation from the surrogate shape, because only the immediate surrounding of the expansion point will contribute noticeably to the total interaction.
Following this assumption that the corresponding model error will be negligible, we apply the simple heuristic that the beam (element) with the smaller (global) identification number (ID) will generally be the slave beam throughout this work and validate this assumption in the numerical example of \secref{sec::num_ex_peeling_pulloff_SBIP}.
The resulting maximal relative difference in the force response on system level turns out to be below $1.5\%$ even for relatively large curvatures, which is considered to be a reasonably small model error.
In addition, this simple criterion based on element IDs ensures a unique decision that does not change in the course of the simulation.\\

\noindent\textit{Remark on self-interactions.}
As already discussed for the SSIP approach, self-interactions, i.e., the interaction of distinct parts of the same beam, can be treated naturally also within the SBIP approach.
Leaving everything else unchanged, the search for and evaluation of (non-neighboring) beam element pairs from one and the same physical beam directly allows to incorporate the effect of self-interaction.
This is considered to be important for long, flexible fibers showing the tendency to large deformations.

\section{Closed-form expression for the disk-cylinder interaction potential}
\label{sec::ia_pot_single_length_specific_evaluation_vdW}
There are different ways to arrive at a closed-form expression for the required SBIP law~$\tilde \pi$.
One of them is the analytical integration of a point-pair potential~$\Phi$ over all point pairs in the section-beam (surrogate) system.
This strategy has been applied in our recent contribution~\cite{Grilldiskcylpot} considering a generic inverse power law~$\Phi_\text{m}(r)=k_\text{m} \, r^{-m}$ with exponent~$m \geq 6$.
Due to the generality, the resulting reduced interaction law~$\tilde \pi$ can be used to model both the adhesive vdW part ($m=6$) and the repulsive part ($m=12$) of the LJ potential.
Moreover, it considers the practically relevant case of circular, undeformable cross-sections and homogeneous densities of the fundamental interacting points in both fibers.
From a geometrical point of view, this leads to the scenario of a disk and a cylinder (cf.~\figref{fig::master_slave_approach_strategy}) with arbitrary mutual configuration, i.e., separation and orientation.
It is of great importance for the accuracy of the SBIP approach that the applied disk-cylinder potential law is accurate for all mutual configurations.

Note that instead of the six degrees of freedom $(\vr_{1-2\text{c}},\vpsi_{1-2\text{c}})$ in the most general scenario~\cite{Meier2021CosseratPotential} considered in the preceding section, this system consisting of a homogeneous circular disk and cylinder (radii $R_1$ and $R_2$) can be described by only three degrees of freedom. Out of different analytical approximations for $(\vr_{1-2\text{c}},\vpsi_{1-2\text{c}})$ as derived in Ref.~\cite{Grilldiskcylpot}, one variant has been identified as particularly appealing and will also be considered throughout the present work. This variant allows for sufficient model accuracy and at the same time for a pleasantly simple and compact closed-form representation of $(\vr_{1-2\text{c}},\vpsi_{1-2\text{c}})$ that only requires two degrees of freedom to describe the fiber interaction: the angle $\alpha:=\arccos{(\vr_1^{\shortmid}(\xi_{1\text{,GP}}) \cdot \vr_2^{\shortmid}(\xi_{2\text{c}})) / (|| \vr_1^{\shortmid}(\xi_{1\text{,GP}}) || \, ||\vr_2^{\shortmid}(\xi_{2\text{c}})|| )}$ enclosed by the tangent vectors $\vr_1^{\shortmid}(\xi_{1\text{,GP}})$ and $\vr_2^{\shortmid}(\xi_{2\text{c}})$ as well as the surface gap function $g_{ul}:=d_{ul}-R_1-R_2$, with $d_{ul}:=|| \vr_1(\xi_{1\text{,GP}} - \vr_2(\xi_{2\text{c}}) || $, between the position vectors $\vr_1(\xi_{1\text{,GP}})$ and $\vr_2(\xi_{2\text{c}})$ defined at a given location $\xi_{1\text{,GP}}$ on the centerline of the slave beam and the associated closest projection point $\xi_{2\text{c}}$ on the centerline of the (surrogate) master beam.

The derivation of this closed-form expression for the disk-cylinder potential law~$\tilde \pi_\text{m,disk-cyl}$ requires a 5D analytical integration of the point-pair potential~$\Phi_\text{m}$, which is a challenging theoretical problem for itself.
Here, we only briefly repeat the problem statement and the final expression found in~\cite{Grilldiskcylpot} for the reader's convenience.\\

\paragraph{Problem statement.}

In the derivation, first an interaction potential $\Pi_\text{m,pt-cyl}$ between a single point on the disk (i.e., the slave beam cross-section) and the cylinder (i.e., the surrogate master beam) is derived before solving a double integral across the disk area.
\begin{align}\label{eq::disk_cyl_pot_m_5Dint}
  \tilde \pi_{\text{m,disk-cyl}} &\defvariable \iint \limits_{A_\text{disk}} \rho_1 \overbrace{\iiint \limits_{V_\text{cyl}} \rho_2 \, \Phi_\text{m}(r) \dd V}^{=: \, \Pi_\text{m,pt-cyl}} \dd A\\
  &\text{with} \quad r=\norm{\vx_1 - \vx_2} \quad \text{and} \quad \vx_1 \in A_\text{disk}, \, \vx_2 \in V_\text{cyl}.  %
\end{align}
Here,~$\vx_1 \in A_\text{disk}$ denotes any point on the disk. On the master side,~$\vx_2 \in V_\text{cyl}$ denotes any point on the surrogate body assumed as an infinitely long auxiliary cylinder oriented along the (normalized) tangent vector~$\vt_2=\vr^\shortmid_2 / \norm{\vr^\shortmid_2}$.

\paragraph{General strategy.}
The general strategy follows the one generally known as point-pairwise summation (see e.g.~\cite{parsegian2005,israel2011} for details and a discussion) and e.g.~applied in~\cite{montgomery2000} for the analytical calculation of vdW forces for certain geometric configurations, e.g., a cylinder and a perpendicular disk.
Since already for such specific scenarios, no exact analytical solution can be found for the integrals, also the proposed derivation made use of the common approach of series expansions in order to find an analytical, closed-form expression for the integral of the leading term(s) of the series.
Due to the rapid decay of the inverse power laws, this is known to yield good approximations for the true solution of the integral.

\paragraph{Approximative solution.}
The final form of the disk-cylinder interaction potential to be used as reduced interaction law in the context of this article reads:
\begin{align}\label{eq::disk-cyl-pot_m}
  \tilde \pi_\text{m,disk-cyl}(g_\text{ul},\alpha) = \hat K_\text{m} \, \rho_1 \, \sqrt{\frac{2 R_1 R_2}{R_1 \, \cos^2 \alpha + R_2}} \, g_\text{ul}^{-m+\frac{9}{2}} \quad \text{with } \hat K_\text{m} \defvariable 4^{-m+\frac{9}{2}} \, K_\text{m}, \quad m \geq 6.
\end{align}
For convenience in later reference, we explicitly state the most common prefactors for the vdW part~$m=6$ and the repulsive part~$m=12$ of the LJ potential as follows:
\begin{align}
  \hat K_6 = \frac{1}{24} \pi^2 \, k_6 \, \rho_2 \quad \text{and} \quad \hat K_{12} = \frac{143}{15 \cdot 2^{14} } \pi^2 \, k_{12} \, \rho_2
\end{align}

\paragraph{Verification.}
An immediate verification of these expressions for the special case~$\alpha=0$ confirms that both~$\tilde \pi_\text{6,disk-cyl}$ and~$\tilde \pi_\text{12,disk-cyl}$ are identical to the independently derived analytical solutions for the interaction potential per unit length~$\tilde \pi_\text{6,cyl$\parallel$cyl}$ and~$\tilde \pi_\text{12,cyl$\parallel$cyl}$ of two infinitely long, parallel cylinders (cf.~Eqs.~(A23) and~(A24) in our previous contribution~\cite{GrillSSIP}).
This is an important finding, as it shows the consistency of the more general expression (\eqref{eq::disk-cyl-pot_m}) valid for all mutual angles~$\alpha$ with previously derived expressions for the important special case~$\alpha=0$.

A comprehensive analysis of the accuracy of \eqref{eq::disk-cyl-pot_m} as well as a comparison to alternative, increasingly complex expressions can be found in Ref.~\cite{Grilldiskcylpot}.
Here, we only briefly summarize the conclusions and finally present the important comparison to the accuracy of the reduced disk-disk interaction potential laws~$\tilde{\tilde{\pi}}$ used together with the SSIP approach in Ref.~\cite{GrillSSIP}.
Most importantly, the pleasantly simple expression from \eqref{eq::disk-cyl-pot_m} shows the correct asymptotic scaling behavior in the decisive regime of small separations and small angles.
It is thus considered the optimal compromise between accuracy and complexity of the expression for the purposes of this work.
In particular, also the theoretically predicted $1/\sin\alpha$ angle dependence (for non-parallel cylinders~$\alpha\neq0$) has been successfully verified.

\begin{figure}[htb]%
  \centering
   \subfigure[Mutual angle $\alpha=0^\circ$]{
    \includegraphics[width=0.45\textwidth]{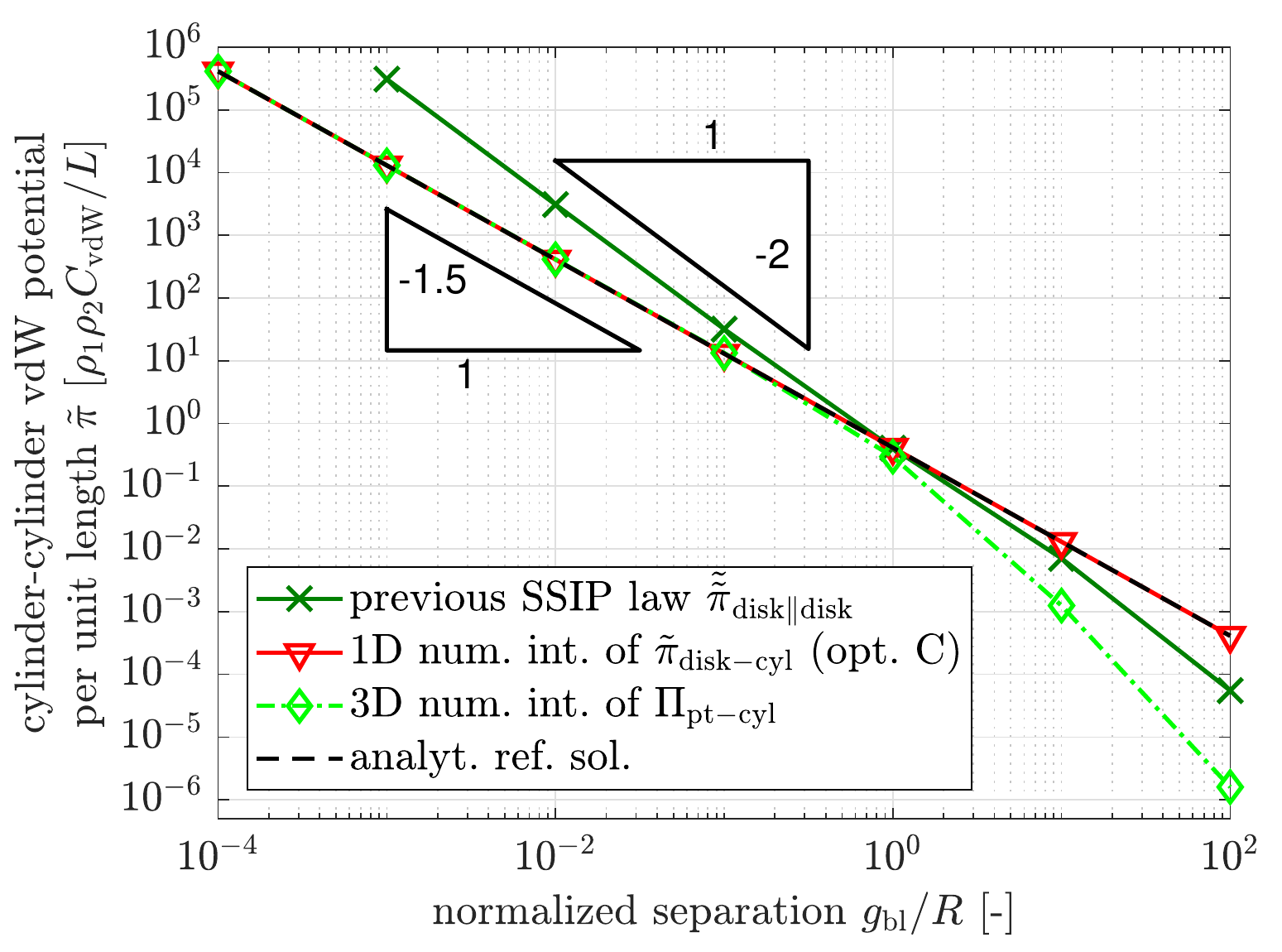}
    \label{fig::cyl-cyl_ia_pot_SBIP_optC_vs_SSIP_over_sep_angle0}
   }
   \subfigure[Mutual angle $\alpha=90^\circ$]{
    \includegraphics[width=0.45\textwidth]{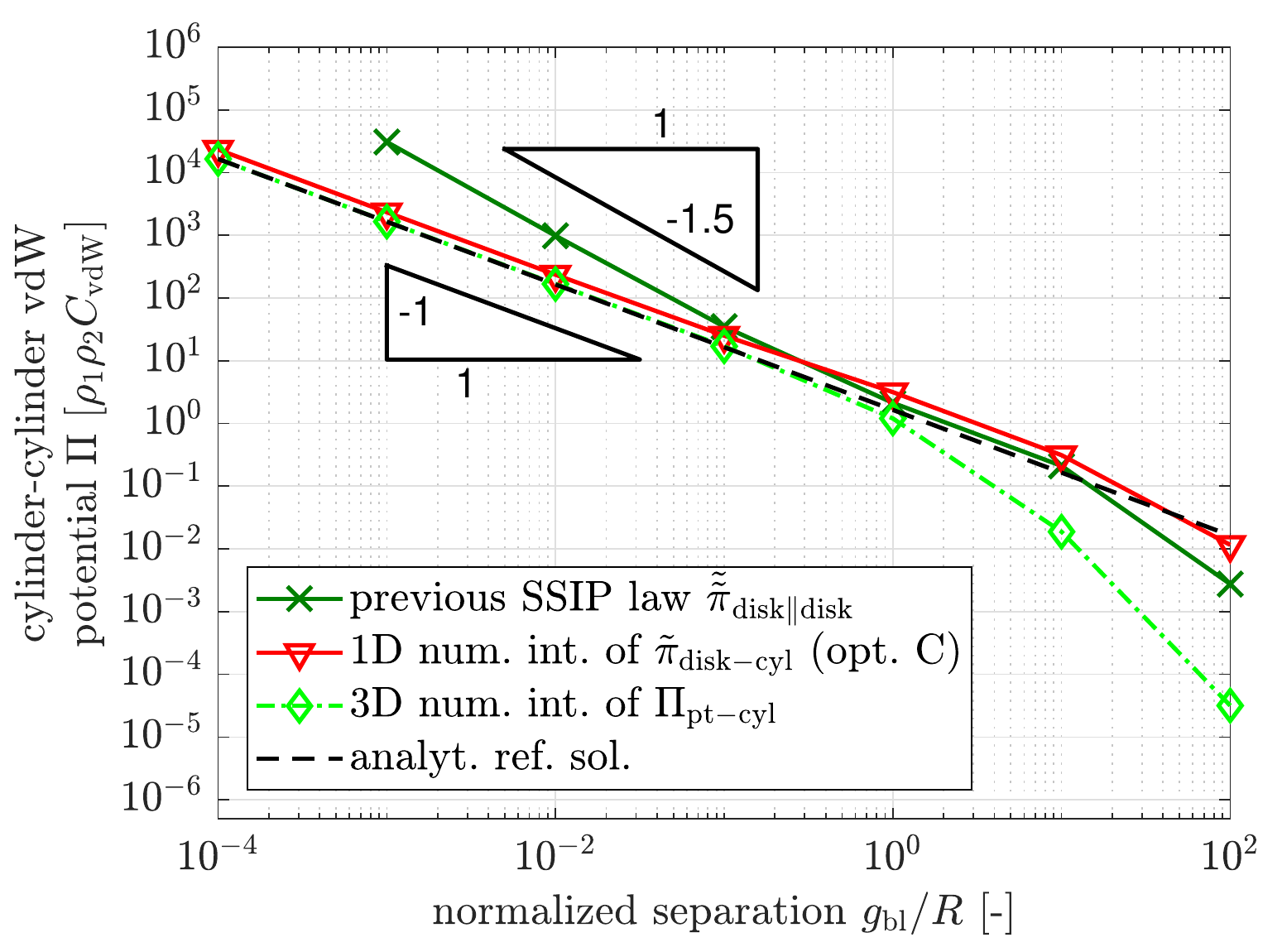}
    \label{fig::cyl-cyl_ia_pot_SBIP_optC_vs_SSIP_over_sep_angle90}
   }
   \caption{Interaction potential of two cylinders as a function of the dimensionless minimal surface separation~$g_\text{bl}/R$ at different mutual angles~$\alpha$.
            Comparison of the previously used SSIP law~$\tilde{\tilde{\pi}}_\text{6,disk$\parallel$disk}$ from Eq.~(18) in Ref.~\cite{GrillSSIP} (used together with the SSIP approach proposed in the same article; dark green line with crosses) with the analytical expression for the disk-cylinder potential~$\tilde \pi_\text{6,disk-cyl}$ \eqref{eq::disk-cyl-pot_m} (used together with the SBIP approach from \secref{sec::method_single_length_specific_integral}; red line with triangles).
            The numerical reference solution obtained via 3D Gaussian quadrature of the point-half space potential~$\Pi_\text{6,pt-hs}$ from Ref.~\cite{Grilldiskcylpot} (green line with diamonds) and the analytical reference solutions summarized in~\secref{sec::basics_vdW_disks_cylinders} (black dashed line) are plotted as reference.}
  \label{fig::cyl-cyl_ia_pot_SBIP_optC_vs_SSIP_over_sep}
\end{figure}

\figref{fig::cyl-cyl_ia_pot_SBIP_optC_vs_SSIP_over_sep} demonstrates that the novel SBIP law~$\tilde \pi_\text{m,disk-cyl}$ from \eqref{eq::disk-cyl-pot_m} in combination with the general SBIP approach from~\secref{sec::method_single_length_specific_integral} is significantly more accurate than the previous method using the SSIP law~$\tilde{\tilde{\pi}}_\text{m,disk$\parallel$disk}$ in combination with the general SSIP approach as proposed in our previous contribution~\cite{GrillSSIP}.
Recall the important result of the SSIP verification therein that the simple and readily available section-section interaction law~$\tilde{\tilde{\pi}}_\text{m,disk$\parallel$disk}$ from Eq.~(18) in Ref.~\cite{GrillSSIP} does in general not yield the correct asymptotic scaling behavior in the limit of small separations, which is decisive in case of short-ranged interactions, and that the orientation of the (disk-shaped) cross-sections would need to be included in the reduced interaction law~$\tilde{\tilde{\pi}}$ to improve this crucial characteristic.
According to Fig.~\ref{fig::cyl-cyl_ia_pot_SBIP_optC_vs_SSIP_over_sep}, the alternative SBIP approach specialized on short-range interactions in combination with the SBIP law~$\tilde \pi_\text{m,disk-cyl}$ derived in~\cite{Grilldiskcylpot} has finally accomplished the goal of reproducing the correct asymptotic scaling behavior as demonstrated for the special cases of two parallel cylinders (expected asymptotic scaling of order 1.5; see Fig.~\ref{fig::cyl-cyl_ia_pot_SBIP_optC_vs_SSIP_over_sep_angle0}) and two perpendicular cylinders (expected asymptotic scaling of order 1; see Fig.~\ref{fig::cyl-cyl_ia_pot_SBIP_optC_vs_SSIP_over_sep_angle90}).

\begin{figure}[htpb]%
  \centering
   \vspace{-5pt}
   \subfigure[]{
    \includegraphics[width=0.45\textwidth]{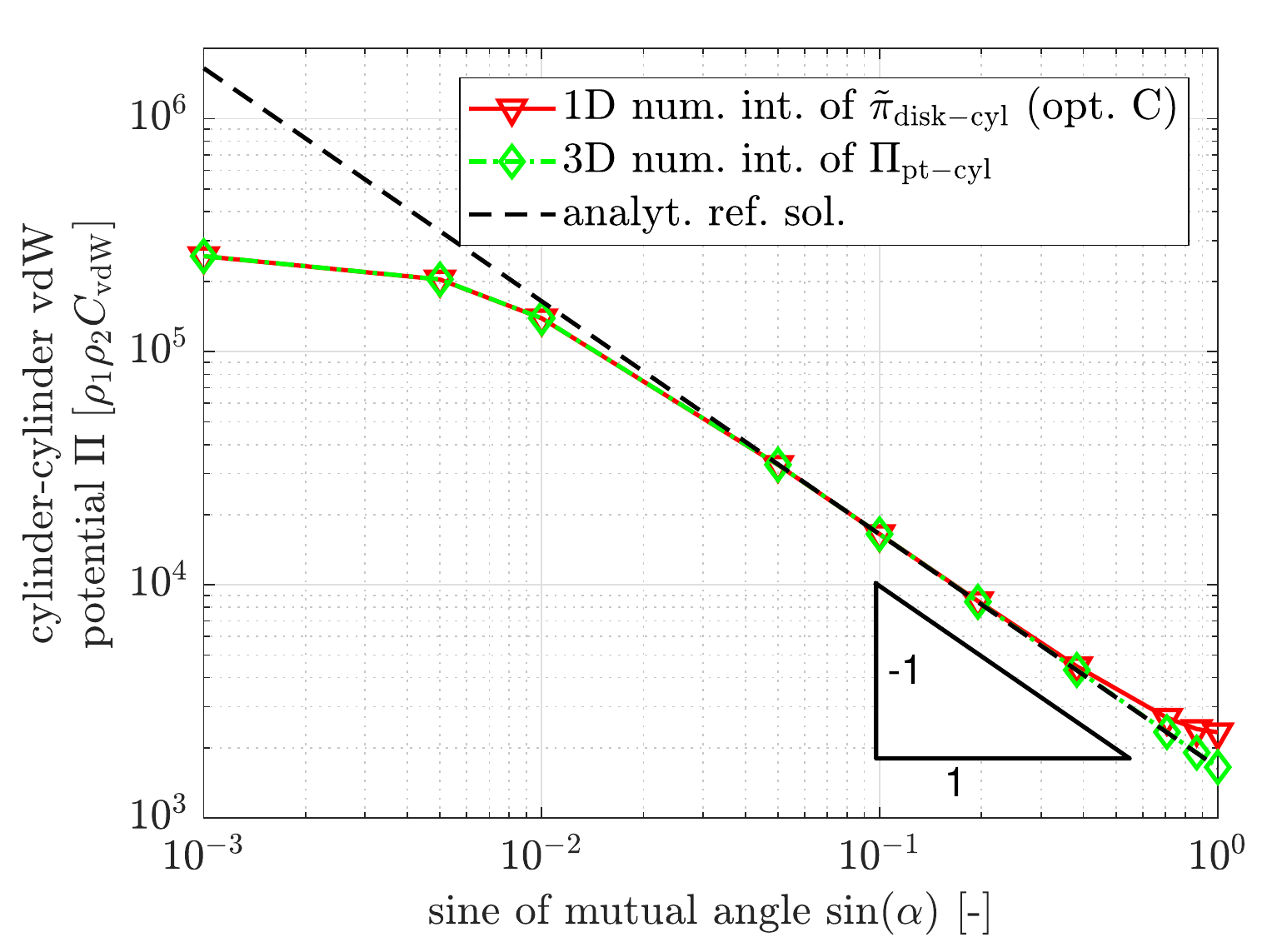}
    \label{fig::cyl-cyl_ia_pot_SBIP_optC_over_angle_sep1e-3}
   }
   \vspace{-5pt}
   \subfigure[]{
    \includegraphics[width=0.45\textwidth]{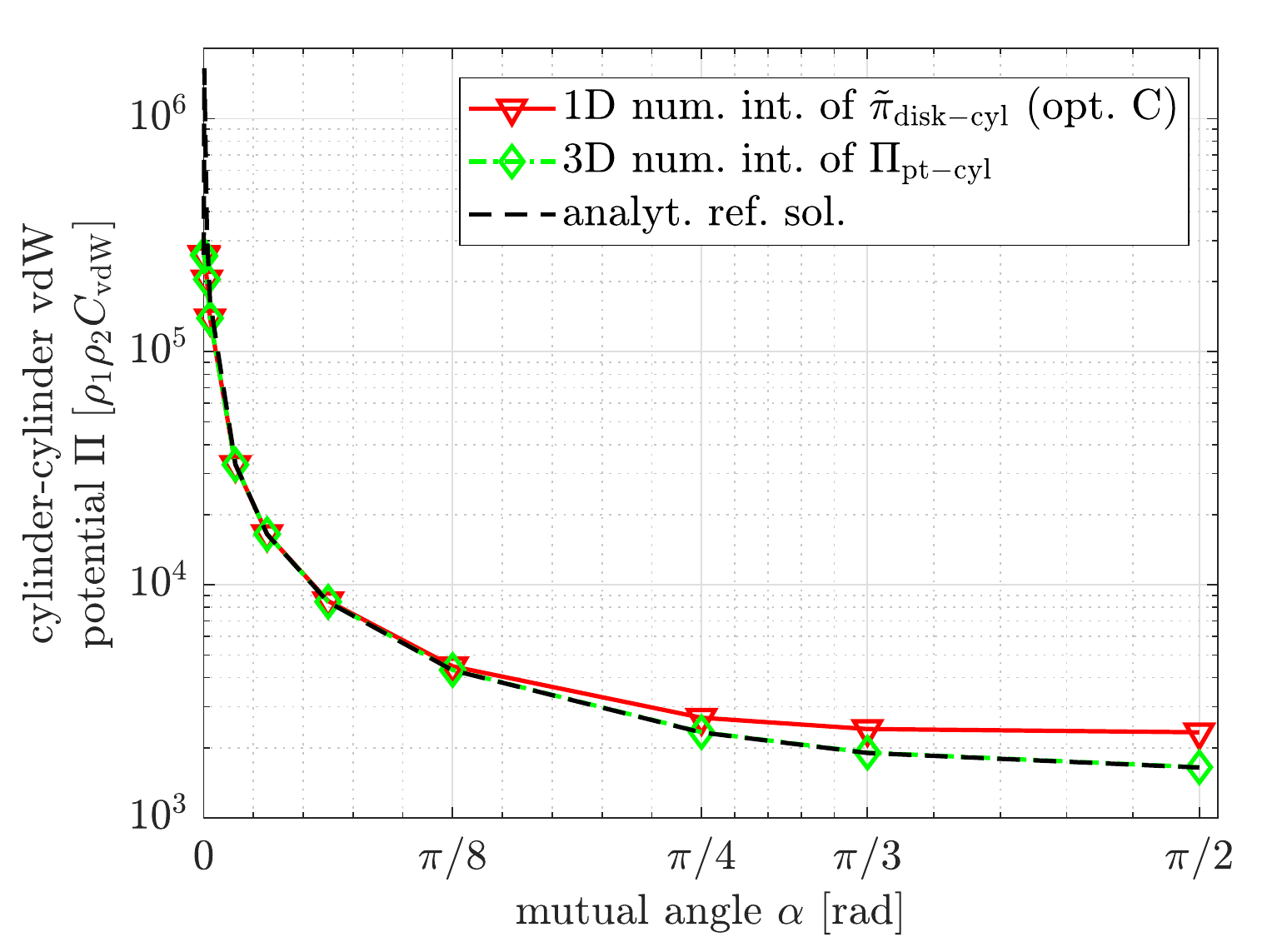}
    \label{fig::cyl-cyl_ia_pot_SBIP_optC_over_angle_semilog_sep1e-3}
   }
   \caption{Interaction potential of two cylinders as a function of the (sine of) the mutual angle at smallest surface separation $g_\text{bl}/R=10^{-3}$. Verification of the correct asymptotic angular dependence of the analytical expression for the disk-cylinder potential~$\tilde \pi_\text{6,disk-cyl}$ from \eqref{eq::disk-cyl-pot_m} (used together with the SBIP approach from \secref{sec::method_single_length_specific_integral}; red line with triangles) by means of a numerical reference solution obtained via 3D Gaussian quadrature of the point-half space potential~$\Pi_\text{6,pt-hs}$ from Ref.~\cite{Grilldiskcylpot} (green line with diamonds) and by means of the analytical reference solution summarized in~\secref{sec::basics_vdW_disks_cylinders} (black dashed line).}
  \label{fig::cyl-cyl_ia_pot_SBIP_optC_over_angle}
\end{figure}

The plots in \figref{fig::cyl-cyl_ia_pot_SBIP_optC_over_angle} complement the analysis above by showing the dimensionless interaction potential as a function of the mutual angle~$\alpha$.
The considered scenario of two cylinders and the three different solutions for the two-cylinder interaction potential are identical to the previous \figref{fig::cyl-cyl_ia_pot_SBIP_optC_vs_SSIP_over_sep}.
Most importantly, the theoretically predicted scaling behavior is confirmed by the numerical reference solution and reproduced by the disk-cylinder potential law~$\tilde \pi_\text{6,disk-cyl}$ (option C) from \eqref{eq::disk-cyl-pot_m}.

\section{Virtual work contribution}
\label{sec::virtual_work_disk-cyl-pot_m}
Recall from~\eqref{eq::total_virtual_work_is_zero} that it is the variation of the two-body interaction energy~$\delta \Pi_\text{ia}$, which needs to be evaluated to incorporate the effects of molecular interactions into both the theoretical and computational framework of nonlinear continuum mechanics.
According to the general SBIP approach proposed in \secref{sec::method_single_length_specific_integral} combined with the generic disk-cylinder interaction potential law~$\tilde \pi_\text{m,disk-cyl}$ from \secref{sec::ia_pot_single_length_specific_evaluation_vdW}, the variation of the two-fiber interaction potential reads
\begin{align}\label{eq::ia_pot_variation_single_integration}
 \delta \Pi_\text{ia} = \int \limits_0^{l_1} \delta \tilde \pi_\text{m,disk-cyl} \dd s_1.
\end{align}
Note that $\delta (\dd s_1)$ vanishes due to the fact that~$\dd s_1 = \norm{ \vr_{01}^\shortmid(\xi_1) } \dd \xi_1$ only depends on the element parameter coordinate~$\xi_1$ and the initial (``0'') configuration of the slave beam, but not on the current configuration, i.e., current values of the primary degrees of freedom.
For the sake of brevity, the subscripts ``m'' and ``disk-cyl'' as well as the function arguments~$(g_\text{ul},\alpha)$ of~$\tilde \pi$ will be omitted throughout this section.
The variation of~$\tilde \pi$ consists of the summands, each defined by two factors
\begin{align}\label{eq::virtual_work_SBIP_disk-cyl-pot}
 \delta \Pi_\text{ia} = \int \limits_0^{l_1}\diff{\tilde{\pi}}{{g_\text{ul}}} \, \delta g_\text{ul} + \diff{\tilde{\pi}}{{(\cos\alpha)}} \, \delta (\cos\alpha) \dd s_1,
\end{align}
which will be determined subsequently in the next steps. First, the derivatives of~$\tilde \pi$ can be expressed in recursive manner:
\begin{align}\label{eq::disk-cyl-pot_m_firstderiv_gap_ul}
  \diff{\tilde{\pi}}{{g_\text{ul}}} &= \frac{(-m+\frac{9}{2})}{g_\text{ul}} \, \tilde \pi\\
  \diff{\tilde{\pi}}{{(\cos\alpha)}} &= - \frac{R_1 \cos\alpha}{R_1 \cos^2\alpha + R_2} \, \tilde \pi\label{eq::disk-cyl-pot_m_firstderiv_cos_alpha}
\end{align}
Note that the remaining two factors~$\delta g_\text{ul}$ and~$\delta (\cos\alpha)$ are known from the macroscopic line contact formulation proposed in~\cite{meier2016} and its combination with a point contact formulation presented in a unified ABC beam contact formulation~\cite{Meier2017a}, respectively.
Both are reproduced here for the sake of completeness and a unified notation:
\begin{align}\label{eq::var_gap_ul}
  \delta g_\text{ul} = \delta d_\text{ul} = \delta \vd_\text{ul}^T \vn_\text{ul} =
                      \left( \delta \vr_1^T(\xi_1) - \delta \vr_2^T(\xi_\text{2c}) \right) \vn_\text{ul}
\end{align}
\begin{align}\label{eq::var_cosalpha}
  \delta (\cos\alpha) & = \left( \delta \vr_1^{\shortmid T}(\xi_1) \, \vv_{\alpha1}  + \delta \left( \vr_2^{\shortmid T}(\xi_\text{2c}) \right) \vv_{\alpha2} \right) \, \sgn(\vt_1^T \vt_2)
\end{align}
In the previous equations, we have introduced the unit tangent vectors~$\vt_i \defvariable \vr^\shortmid_i / \norm{\vr^\shortmid_i}$, $i=1,2$, the unilateral unit normal vector $\vn_\text{ul} \defvariable \vd_\text{ul} / d_\text{ul}$ with unilateral distance vector $\vd_\text{ul} \defvariable \vr_1(\xi_1) - \vr_2(\xi_\text{2c})$ as well as the auxiliary vectors
\begin{align}
  \vv_{\alpha1} \defvariable \frac{1}{ \norm{ \vr_{1}^{\shortmid} } }\left( \vI_{3 \times 3} - \vt_1 \otimes \vt_1^{T} \right) \, \vt_2 \quad \text{and}
  \quad \vv_{\alpha2} \defvariable \frac{1}{ \norm{ \vr_{2}^{\shortmid} } }\left( \vI_{3 \times 3} - \vt_2 \otimes \vt_2^T \right) \, \vt_1.
\end{align}
Note the difference between the notations~$\delta \left( \vr_2^\shortmid(\xi_{2\text{c}}) \right)$ and~$\delta \vr_2^\shortmid(\xi_{2\text{c}})$ (see~\eqref{eq::var_total_tangent_master}), which originates from the fact that~$\xi_{2\text{c}}$ is the result of a (closest) point-to-curve projection, i.e., it depends on the primary variables of our problem. Thus, $\delta \left( \vr_2^\shortmid(\xi_{2\text{c}}) \right)$ represents a total variation, and $\delta \vr_2^\shortmid(\xi_{2\text{c}})$ a partial variation at fixed $\xi_{2\text{c}}$. In contrast to~$\delta g_\text{ul}$ in~\eqref{eq::var_gap_ul}%
\footnote{
In the final step of~\eqref{eq::var_gap_ul}, the orthogonality condition~$\vr_2^{\shortmid T}(\xi_\text{2c}) \, \vn_\text{ul} \equiv 0$ has been exploited and the additional contribution from the variation of the (closest-point) arc-length coordinate on the master side~$\delta \xi_\text{2c}$ vanishes.
}%
, the additional contribution from~$\delta \xi_\text{2c}$ must actually be computed and included to ensure a variationally consistent formulation in~\eqref{eq::var_cosalpha}.
Also for the later reference, all the expressions required in this respect are given here as
\begin{align}\label{eq::var_total_pos_master}
  \delta ( \vr_2(\xi_{2\text{c}}) ) &= \delta \vr_2 (\xi_{2\text{c}}) + \vr_2^\shortmid (\xi_{2\text{c}}) \, \delta \xi_\text{2c}\\
  \delta ( \vr_2^\shortmid(\xi_{2\text{c}}) ) &= \delta \vr_2^\shortmid (\xi_{2\text{c}}) + \vr_2^{\shortmid\shortmid} (\xi_{2\text{c}}) \, \delta \xi_\text{2c} \label{eq::var_total_tangent_master}
\end{align}
using the following expression for the variation of the slave beam parameter coordinate
\begin{align}
  \delta \xi_\text{2c} = \frac{1}{p_{2,\xi_2}} \Big( -\delta \vr_1^T \vr_2^\shortmid + \delta \vr_2^T \vr_2^\shortmid - \delta \vr_2^{\shortmid T} \underbrace{ (\vr_1 - \vr_2 ) }_{= \, \vd_\text{ul}} \Big),
\end{align}
where the derivative of the scalar orthogonality condition~$p_2 \defvariable \vr_2^{\shortmid T} \vd_\text{ul} \defeq 0$ with respect to~$\xi_2$ reads
\begin{align}\label{eq::orthogonalitycondition_master_deriv}
  p_{2,\xi_2} = \vr_2^{\shortmid\shortmid T} \vd_\text{ul} - \vr_2^{\shortmid T} \vr_2^\shortmid.
\end{align}
At this point, we have gathered all the required pieces that allow us to evaluate the virtual work contribution from molecular interactions~$\delta \Pi_\text{ia}$ according to~\eqref{eq::ia_pot_variation_single_integration}.
As discussed along with the general SBIP approach in \secref{sec::method_single_length_specific_integral}, the 1D integral along the slave beam is evaluated numerically, e.g., by means of Gaussian quadrature.
Note that the correctness of the presented and implemented expressions of this section has been verified to be consistent with the corresponding interaction energy~$\Pi_\text{ia}$ (see~\eqref{eq::ia_pot_single_integration} with (\ref{eq::disk-cyl-pot_m})) by means of an automatic differentiation tool~\cite{Trilinos2012}.

In a next step, the contribution~$\delta \Pi_\text{ia}$ of the interaction potential to the weak form~\eqref{eq::total_virtual_work_is_zero} needs to be discretized in space. The discrete counterpart~$\delta \Pi_\text{ia,h}$ of the space-continuous form~$\delta \Pi_\text{ia}$ is obtained via substitution of the centerline interpolation scheme from Eqs.~(\ref{eq::centerline_discretization}) and (\ref{eq::var_centerline_discretization}) into~\eqref{eq::virtual_work_SBIP_disk-cyl-pot}.
This allows to identify the discrete residual vectors~$\vdr_\text{ia,i}$ of the interacting beam elements~$i=\{1,2\}$ that finally result from the molecular interactions. For quasi-static problems, this step is sufficient to transfer our mechanical problem into a discrete set of nonlinear algebraic equations that need to be solved numerically for the discrete (nodal) primary variables~$\hat \vdd$. If tangent-based solution schemes, e.g. Newton-Raphson, shall be applied for this purpose, the required linearization of these residual vectors~$\Delta[ \delta \Pi_\text{ia} ]$ with respect to the vector of primary degrees of freedom~$\hat \vdd$ is provided in \appref{sec::linearization_SBIP}.

\noindent\textit{Discussion of the special treatment required for master beam endpoints.}\\
Recall from \secref{sec::method_single_length_specific_integral} and \ref{sec::ia_pot_single_length_specific_evaluation_vdW} that the cylinder used as the surrogate for the master beam has been assumed to have an infinite length.
Due to the very short range of the interactions considered here, this is an excellent approximation in almost all cases.
In the special case that the result of the closest-point projection is a master beam endpoint, however, this approximation overestimates the true contribution to the interaction potential approximately by a factor of two.
Again, given the short range of interactions considered here, the resulting model error can be interpreted as if the master beam was slightly longer%
\footnote{
by the length of the cut-off radius longer, to be more precise
}
than it actually should be.

Due to this short range of interactions and the rarity of this event involving the endpoints of slender fibers among all those cases involving the points between the endpoints, the influence of this model error on the total two-body interaction potential is expected to be negligible in almost all applications.
Nevertheless, we can think of the worst case scenario, where two straight, parallel, adhesive fibers of finite length (with equilibrium inter-axis separation) slide along each other in axial direction and the only effective force would be the one at the fiber endpoints, where the influence of the second beam on an exemplarily considered cross-section of the first beam rapidly decreases, because the second beam ends.
Whereas the unmodified SBIP approach using~$\tilde \pi_\text{m,disk-cyl}$ would yield zero force, it could be augmented by a special treatment of master beam endpoints that subtracts half of the interaction potential contribution at any integration point where the result of the closest-point projection is a master beam endpoint.
This procedure probably needs to be smooth such that the transition from the full disk-cylinder interaction potential contribution to half that value at the master beam endpoint needs to be smeared out over a small, yet finite length of the beam.
Due to the expected negligible effect in almost all applications, this augmentation of the SBIP approach is left for future work, but this discussion as well as the described worst-case scenario should prove useful when implementing, calibrating and verifying this model enhancement.

\section{Beam Interaction Formulations from a Meta-Level Perspective}
\label{sec::beam_interaction_formulations_miscellaneous}
This section aims to take a step back and look at beam interaction formulations from a meta-level perspective in order to get an overview of the previously existing approaches and the new one proposed in this article.

\subsection{Classification and comparison of approaches for beam-beam interactions}
\label{sec::beam_interaction_formulations_classification_comparison}
\begin{figure}[p]%
  \rotatebox{90}{
    \begin{minipage}[c][\textwidth][c]{\textheight}
      \def\svgwidth{0.2\textwidth}
      \begin{center}
        \begin{tabular}{|c|c|c|c|}\hline
          point-point & section-section (SSIP) & section-beam surrogate (SBIP) & beam surrogate-beam surrogate\\\hline
          &&&\\
          &&&\\
          \input{beam_beam_interaction_classification_point-point.pdf_tex} & \input{beam_beam_interaction_classification_section-section.pdf_tex} & 
\begingroup%
  \makeatletter%
  \providecommand\color[2][]{%
    \errmessage{(Inkscape) Color is used for the text in Inkscape, but the package 'color.sty' is not loaded}%
    \renewcommand\color[2][]{}%
  }%
  \providecommand\transparent[1]{%
    \errmessage{(Inkscape) Transparency is used (non-zero) for the text in Inkscape, but the package 'transparent.sty' is not loaded}%
    \renewcommand\transparent[1]{}%
  }%
  \providecommand\rotatebox[2]{#2}%
  \newcommand*\fsize{\dimexpr\f@size pt\relax}%
  \newcommand*\lineheight[1]{\fontsize{\fsize}{#1\fsize}\selectfont}%
  \ifx\svgwidth\undefined%
    \setlength{\unitlength}{142.12288933bp}%
    \ifx\svgscale\undefined%
      \relax%
    \else%
      \setlength{\unitlength}{\unitlength * \real{\svgscale}}%
    \fi%
  \else%
    \setlength{\unitlength}{\svgwidth}%
  \fi%
  \global\let\svgwidth\undefined%
  \global\let\svgscale\undefined%
  \makeatother%
  \begin{picture}(1,0.74463001)%
    \lineheight{1}%
    \setlength\tabcolsep{0pt}%
    \put(0,0){\includegraphics[width=\unitlength,page=1]{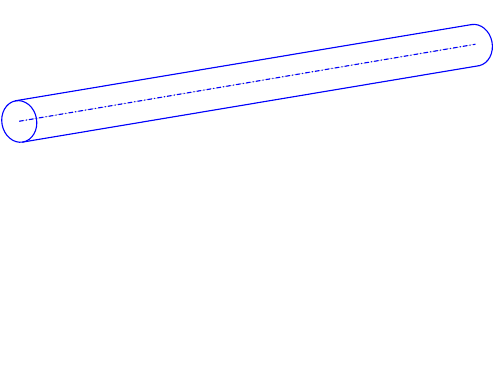}}%
    \put(0.00061085,0.19193422){\color[rgb]{0,0,0}\makebox(0,0)[lt]{\lineheight{0}\smash{\begin{tabular}[t]{l}\fbox{master}\end{tabular}}}}%
    \put(0.50088127,0.0912232){\color[rgb]{0,0,0}\makebox(0,0)[lt]{\lineheight{0}\smash{\begin{tabular}[t]{l}\fbox{slave}\end{tabular}}}}%
    \put(0,0.69330995){\color[rgb]{0,0,1}\makebox(0,0)[lt]{\lineheight{0}\smash{\begin{tabular}[t]{l}\fbox{beam surrogate}\end{tabular}}}}%
    \put(0,0){\includegraphics[width=\unitlength,page=2]{beam_beam_interaction_classification_section-surrogate.pdf}}%
  \end{picture}%
\endgroup%
 & 
\begingroup%
  \makeatletter%
  \providecommand\color[2][]{%
    \errmessage{(Inkscape) Color is used for the text in Inkscape, but the package 'color.sty' is not loaded}%
    \renewcommand\color[2][]{}%
  }%
  \providecommand\transparent[1]{%
    \errmessage{(Inkscape) Transparency is used (non-zero) for the text in Inkscape, but the package 'transparent.sty' is not loaded}%
    \renewcommand\transparent[1]{}%
  }%
  \providecommand\rotatebox[2]{#2}%
  \newcommand*\fsize{\dimexpr\f@size pt\relax}%
  \newcommand*\lineheight[1]{\fontsize{\fsize}{#1\fsize}\selectfont}%
  \ifx\svgwidth\undefined%
    \setlength{\unitlength}{142.63947748bp}%
    \ifx\svgscale\undefined%
      \relax%
    \else%
      \setlength{\unitlength}{\unitlength * \real{\svgscale}}%
    \fi%
  \else%
    \setlength{\unitlength}{\svgwidth}%
  \fi%
  \global\let\svgwidth\undefined%
  \global\let\svgscale\undefined%
  \makeatother%
  \begin{picture}(1,0.83978557)%
    \lineheight{1}%
    \setlength\tabcolsep{0pt}%
    \put(0.00144698,0.78865137){\color[rgb]{0,0,1}\makebox(0,0)[lt]{\lineheight{0}\smash{\begin{tabular}[t]{l}\fbox{beam surrogate}\end{tabular}}}}%
    \put(0.0860657,0.01399947){\color[rgb]{0,0,1}\makebox(0,0)[lt]{\lineheight{0}\smash{\begin{tabular}[t]{l}\fbox{beam surrogate}\end{tabular}}}}%
    \put(0,0){\includegraphics[width=\unitlength,page=1]{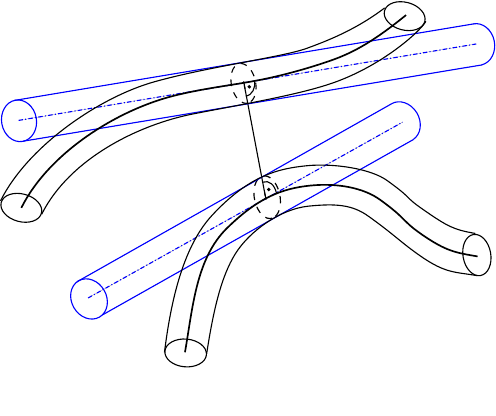}}%
  \end{picture}%
\endgroup%
\\
          &&&\\\hline
          3D $\times$ 3D = 6D & 1D $\times$ 1D = 2D & 1D $\times$ 0D = 1D & 0D $\times$ 0D = 0D\\\hline
          &&&\\
          $\Pi_\text{ia} = \sum_{i\in \mathcal{B}_1} \sum_{j\in \mathcal{B}_2} \Phi(r_{ij})$&&&\\
          $\Pi_\text{ia} = \int \limits_{V_1} \int \limits_{V_2} \rho_1 \rho_2 \Phi(r) \dd V_2 \dd V_1$ & $\Pi_\text{ia} = \int \limits_0^{l_1} \int \limits_0^{l_2} \tilde{\tilde{\pi}}(\vr_{1-2},\vpsi_{1-2}) \dd s_2 \dd s_1$ & $\Pi_\text{ia} = \int \limits_0^{l_1} \tilde{\pi}(\vr_{1-2\text{c}},\vpsi_{1-2\text{c}}) \dd s_1$ & $\Pi_\text{ia} = \Pi_\text{ia}(\vr_{1\text{c}-2\text{c}},\vpsi_{1\text{c}-2\text{c}})$\\
          $r = \norm{\vx_1-\vx_2}$&&&\\
          &&&\\\hline
          atomistic resolution~\cite{Sauer2007a} & long- and short-range interactions~\cite{GrillSSIP} & short-range interactions (\secref{sec::method_single_length_specific_integral}) & - \\\hline
          - & - & macroscopic line contact\cite{meier2016} & macroscopic point contact\cite{wriggers1997}\\\hline
      \end{tabular}
      \end{center}
      \caption{Classification of formulations for beam-beam interactions.}
      \label{fig::beam_beam_interaction_classification_overview}
    \end{minipage}
  }
\end{figure}

A classification and comparison of beam-beam interaction formulations is provided in \figref{fig::beam_beam_interaction_classification_overview}.
The leftmost column depicts the approach of point-pairwise summation -- or corresponding nested volume integration -- of the fundamental interaction potentials for pairs of molecules or charges.
The second and third column show the formulations based on section-section interaction potentials (SSIP)~\cite{GrillSSIP}, and based on section-beam interaction potentials (SBIP) proposed in \secref{sec::method_single_length_specific_integral}, respectively.
Finally, the rightmost column depicts a possible further dimensional reduction of the problem to the interaction of two beam surrogates, which would allow to evaluate the two-body interaction potential by a single function evaluation.
\footnote{In the context of molecular interactions discussed in this article, the required total interaction potential could e.g. be described by the analytical cylinder-cylinder potential stated in \eqref{eq::pot_ia_vdW_cyl_cyl_skewed_smallseparation}.}
From left to right, the resolution of details decreases and likewise the algorithmic complexity determined by the dimensionality of the underlying problem decreases.
This overview of four distinct, logical categories of beam-beam interaction formulations therefore illustrates the tradeoff between resolution of details ranging from atomistic view to trivial meso/macroscopic bodies on the one hand and the dimensionality and thus main driver for the computational cost on the other hand.
The ultimate goal for the derivation and choice of (a class of) formulations however is to outsmart this natural conflict of objectives by a consistent dimensional reduction of the fully resolved problem (on the left) to the minimal possible description that is yet able to reproduce the most important, characteristic features.
Based on the detailed theoretical considerations in Ref.~\cite{GrillSSIP} and \secref{sec::method_single_length_specific_integral}, this optimal choice is given as the SSIP approach for long-range and the SBIP approach for short-range molecular interactions of beams.

This new overall picture of beam interaction formulations also allows to classify previous approaches to macroscopic contact of beams and interpret them in the context of molecular interactions, which are also the origin of the macroscopic contact forces and resulting non-penetrability of objects that we observe in everyday life.
Interestingly, the very first numerical formulation of (macroscopic) beam contact is based on the concept of determining the one bilateral closest-point pair between both beams and evaluating an heuristic penalty force law as a function of the closest point-pair separation in order to preclude any (noticeable) penetrations~\cite{wriggers1997}.
Given this new overall picture of beam interaction formulations, such an approach can be interpreted as the consistent, most extreme dimensional reduction of the problem motivated by the short range of interactions and the resulting possibility to evaluate the total interaction potential for a pair of surrogate bodies approximating the shape of the actual beams.
However, this new perspective likewise reveals the well-known limitations of this kind of approach with respect to describing arbitrary mutual configurations as the illustrative examples of two parallel straight beams or one straight beam and a surrounding helical beam demonstrate (see e.g.~the discussion in Ref.~\cite{meier2016}).
The non-uniqueness of the bilateral closest-point pair in such situations is a confirmation of the oversimplification of the general beam-beam interaction problem by such an approach.
Nevertheless, this category of surrogate-surrogate interaction formulations is the most efficient theoretically possible class of formulations and due to its validity for a certain range of mutual orientations, this efficiency can be exploited in combined approaches such as the ABC formulation~\cite{Meier2017a} that handle the problematic mutual configurations differently.
This recognition of the superior efficiency of the existing, combined macroscopic contact formulation asks for a more detailed discussion of the applicability of such an heuristic approach to preclude penetration also in micro- and nanoscale problem settings, which will thus be given in the following section.

\subsection{Brief comparison of micro- and macroscopic approaches to beam contact}
\label{sec::micro_vs_macroscopic_approaches_beam_contact}
Modeling the steric repulsion that prevents a penetration of distinct fibers has a long history in the field of computational contact mechanics and has first been addressed in Ref.~\cite{wriggers1997}.
The paradigm of these macroscopic continuum models is that the smallest surface separation or gap must be equal to or greater than zero which is formulated as an inequality constraint.
With the development of the novel SSIP and SBIP approaches to molecular interactions of fibers, an alternative modeling strategy has arisen, which is motivated by the rather microscale perspective of LJ interactions between all material points in the slender continua.
This asks for a brief assessment and comparison of the modeling approaches.

On the one hand, penalty-based models for beam contact are well-established formulations with optimized efficiency as well as robustness.
On the other hand, the SSIP and SBIP approaches are based on first principles in form of the LJ law and are thus expected to better resolve the actual contact force distributions, especially in the case of nano- to micro-scale applications.
This becomes obvious if we recall the purely heuristic nature of the penalty force law and the resulting (small) negative gap values, i.e., tolerated penetrations.
It will most likely depend on the specific application whether the associated model error has a significant or rather negligible influence on the results.
In order to answer this question in the context of the authors' recent work e.g.~on biopolymer filament mechanics, it seems most important to look at the adhesive force laws to be applied in combination with the models for steric repulsion.
The SSIP law modeling long-ranged electrostatic attraction~\cite[Eq.~35]{GrillSSIP} is an inverse power law in the inter-axis separation~$d$ rather than the smallest surface separation~$g=d-R_1-R_2$ and thus expected not to be very sensitive to small changes in the gap values in case of contacting fibers~$g\approx0$.
On the contrary, the SSIP law~\cite[Eq.~31]{GrillSSIP} as well as the SBIP law~\eqref{eq::disk-cyl-pot_m} modeling the short-range vdW adhesion are inverse power laws in the gap itself and therefore highly sensitive with respect to the gap~$g$.
Indeed, the thorough validation of both adhesion models using the example of two straight slender fibers in Ref.~\cite{GrillPeelingPulloff} as well as an unsuccessful attempt to use penalty beam contact in combination with vdW adhesion to model the peeling process of two slender fibers%
\footnote{The resulting peeling force values showed a noticeable unphysical dependence on both the type of the penalty force law and the value of the penalty parameter~$\varepsilon$.
}
confirm these considerations.
Moreover, refer to~\secref{sec::regularization_SBIP} and Ref.~\cite{GrillSSIP} for a detailed discussion of the importance to correctly resolve the regime of small gap values by means of a suitable regularization strategy to remedy the inherent singularity of the (reduced) vdW interaction laws for zero separation~$g \to 0$.

To conclude, the choice of a proper computational model for steric repulsion between contacting fibers is closely linked to the type of adhesion and will most likely also depend on the specific application.
For the reasons outlined above, the authors decided to apply the penalty-based line contact formulation together with the rather long-ranged SSIP law for electrostatic attraction, whereas the SSIP/SBIP approach based on the repulsive part of the LJ law will be applied in combination with the short-ranged SSIP/SBIP law for vdW adhesion, respectively.
Nevertheless, a more detailed analysis of the similarities and differences of existing, macroscopic beam contact formulations and the novel approaches based on molecular steric repulsion is considered an interesting aspect of future work in this field.

\section{Regularization and selected further aspects}\label{sec::discretization_algorithmic_aspects}
This section discusses the numerical regularization scheme as well as further (algorithmic) aspects that are of special importance for the application of the novel SBIP approach and the proposed interaction law~$\tilde\pi$.

\subsection{Regularization of the reduced disk-cylinder interaction law in the limit of zero separation}\label{sec::regularization_SBIP}
Due to the inherent singularity of molecular interaction potentials in the limit of zero separation, a numerical regularization is required in order to solve the governing, nonlinear system of equations resulting from \eqref{eq::total_virtual_work_is_zero} in a robust manner.
Generally, such a numerical regularization is a standard approach in (beam) contact formulations (see e.g.~\cite{durville2012,Meier2017a}) and in the specific context of the LJ potential considered here, it has first been applied in~\cite{Sauer2011}.
In analogy to the regularization of the section-section interaction potential law in our previous contribution~\cite{GrillSSIP}, a quadratic/linear extrapolation of the \textit{section-beam} interaction potential/force law will be applied here in the range of very small gap values~$g_\text{ul} < g_\text{ul,reg}$ below a certain regularization parameter~$g_\text{ul,reg} \in \MR^+$.
All the additionally required expressions to compute the regularized section-beam interaction potential law~${\tilde{\pi}}_\text{reg}$ are provided in \appref{sec::regularization_SBIP_expressions}.

If this regularization parameter is chosen sufficiently small, which means smaller than any gap value occurring in any converged configuration of any time/load step throughout the entire simulation, such a regularization scheme will not influence the results at all and can thus be considered a mere auxiliary means to enable and improve the iterative process of solving the nonlinear system of equations.
Altogether, the necessity of a suitable regularization scheme due to its significantly positive effect on robustness and efficiency will be confirmed also by the numerical examples to be presented in \secref{sec::numerical_examples_SBIP}.

\subsection{Objectivity and conservation properties}
Recall from~\secref{sec::virtual_work_disk-cyl-pot_m} that the final contributions to the discrete element residual vectors~$\vdr_\text{ia}$ resulting from the general SBIP approach in combination with the reduced interaction law from~\secref{sec::ia_pot_single_length_specific_evaluation_vdW} have the same abstract form as in the case of macroscopic beam contact formulations~\cite{Meier2017a}.
Most importantly, they are functions of the unilateral gap~$g_\text{ul}$ and the mutual angle of the tangent vectors~$\alpha$.
Due to this fact, the proofs presented in~\cite[Appendix B]{Meier2017a} likewise hold in this case and it is thus straightforward to conclude that also the SBIP approach in combination with the here proposed reduced interaction law fulfills objectivity, global conservation of linear and angular momentum as well as global conservation of energy.

\subsection{Algorithmic complexity}\label{sec::algorithm_complexity_fullint_vs_SSIP_vs_SBIP}
The following discussion focuses on the part of evaluating the total interaction potential~$\delta \Pi_\text{ia}$ (and likewise its linearization~$\Delta[\delta \Pi_\text{ia}]$) as this is the one determined by the computational approach to molecular interactions of fibers.
Depending on many other factors, this part may or may not be the dominating one in the entire algorithmic framework required for a nonlinear finite element solver in structural dynamics.
Based on the experiences with the novel SBIP approach, the previously proposed SSIP approach, and the attempt of directly evaluating~$\delta \Pi_\text{ia}$ via 6D numerical integration (see \eqref{eq::pot_fullvolint}), it can however be stated that in a best case scenario the computational cost of evaluating this part is still a considerable one and in the worst case scenario -- using direct 6D numerical integration -- it becomes so costly that it is actually unfeasible for any system of practical relevance.
This initial general assessment shows both the urgent need for an efficient approach and also the high leverage of any potential improvement in this respect, which has actually been the main motivation for the development of the novel SBIP approach.

To narrow down the broad topic of algorithm efficiency, this analysis can be restricted to the evaluation of one element pair, because the number of interacting element pairs can be considered a fixed number for now.
This number mainly depends on the spatial distribution of the fibers and the range of interaction, i.e., the cut-off radius, which means that it will be limited due to both the short range of interactions considered in this article and the non-penetrability constraint restricting the closest packing of fibers.
Note that the associated important question of an efficient search for element pairs and the selection of element pairs to be finally evaluated will be discussed in the following Secs.~\ref{sec::search_partitioning_SBIP} and \ref{sec::filter_criterion_ele_pair_proximity}.

At this point, recall from the analysis in our previous contribution~\cite{GrillSSIP} that the algorithmic complexity of the evaluation of one element pair in case of direct 6D numerical integration of \eqref{eq::pot_fullvolint} will be~$\bigO ( n_\text{GP,tot,ele-length}^2 \cdot n_\text{GP,tot,transverse}^4 )$.
Here, $n_\text{GP,tot,ele-length}$ and $n_\text{GP,tot,transverse}$ denote the number of integration points in axial and transverse direction, respectively.
The SSIP approach already reduces this complexity significantly to~$\bigO ( n_\text{GP,tot,ele-length}^2 )$ due to the replacement of the 4D numerical integration over the cross-section areas by an effective section-section interaction potential law.
By the novel SBIP approach, this is finally reduced even further and for the remaining 1D integral along the slave beam (cf.~\eqref{eq::ia_pot_single_integration}), we obtain
\begin{equation}
  \bigO \left( n_\text{GP,tot,ele-length} \right)
\end{equation}
complexity.
Bearing in mind the typically large number of integration points required to integrate the (short-ranged) molecular interaction laws with its high gradients with sufficient accuracy, this linear complexity makes a significant difference as compared to the quadratic complexity of the SSIP approach.
Based on the experience of the numerical examples studied throughout this work, typical values are given as~$n_\text{GP,tot,ele-length} = 10 \ldots 100$. This is thus the factor we can expect to save from the reduced dimensionality of numerical integration when comparing the proposed SBIP approach to the previously derived SSIP approach, which itself offers potential savings by a factor of $10^4 \ldots 10^8$ as compared to direct 6D numerical integration~\cite{GrillSSIP}.
Of course, this comes at the cost of the closest point-to-curve projection required in case of the novel SBIP approach.
This projection consists of solving the scalar nonlinear orthogonality condition (cf.~\eqref{eq::orthogonalitycondition_master_deriv}) e.g.~by means of Newton's method, which however turns out to be rather insignificant as compared to evaluating the terms of the integrand.
The net saving will thus be slightly smaller than the number of (axial) integration points per element, but still significant.

In addition to that, there is another positive effect to be considered.
Due to the additional analytical integration step in the derivation of the reduced SBIP law from~\secref{sec::ia_pot_single_length_specific_evaluation_vdW} as compared to the SSIP laws, the (inverse) exponent of the effective interaction law and thus integrand is reduced by one: cf.~$-m+9/2$ in \eqref{eq::disk-cyl-pot_m} vs.~$-m+7/2$ in Eq.~(A12) of Ref.~\cite{GrillSSIP}, for~$m \geq 6$.
In turn, this makes the integrand smoother and less integration points are required to achieve the same accuracy of the numerical integration.
Especially for the short-ranged interactions e.g.~from the LJ interaction with~$m=6$ and $m=12$, this makes a significant difference in the decisive regime of small separations and contributes to the superior efficiency of the SBIP approach as compared to the SSIP approach or even the direct 6D numerical integration.

In this respect, it seems noteworthy to point out the clear separation of the general SSIP/SBIP approach and the applied reduced interaction law.
Generally, the complexity of the specific expression does not necessarily depend on whether it is a SSIP or SBIP law.
However, some conclusions like the one just made for the resulting exponent of the power law -- if consistently derived from an inverse-power point pair potential law -- are possible.
Likewise, assuming homogeneous, circular cross-sections we can state that the mutual configuration of the disk-disk system has four degrees of freedom whereas the disk-cylinder system can be described by three degrees of freedom as discussed in Ref.~\cite{Grilldiskcylpot}.
However, this does not allow to estimate the complexity of the specific expressions even in the hypothetical case of exact analytical interaction laws.
Given the concrete examples of expressions for short-range interactions presented in~\secref{sec::ia_pot_single_length_specific_evaluation_vdW} and our previous contribution~\cite{GrillSSIP}, respectively, it is important to underline that they are based on different simplifying assumptions and thus naturally have a different accuracy~\cite{Grilldiskcylpot}.
The fact that the SSIP law expressions are simpler than their SBIP law counterparts must thus be seen in the light of this tradeoff between simplicity and accuracy.
Nevertheless, when comparing the performance in the numerical example to be presented in~\secref{sec::num_ex_peeling_pulloff_SBIP}, one will notice this effect of less operations being required to evaluate the simpler yet less accurate specific SSIP law as compared to the SBIP law.%
\footnote{
Note that actually the evaluation of the discrete residual vector and, predominantly, the tangent stiffness matrix should be considered when comparing simplicity of expressions and number of required operations.
For the sake of clarity, however, this argument is made on the level of reduced interaction laws knowing well that the judgment holds true also for the resulting residual vector and stiffness matrix.
In fact, the differences in simplicity increase due to the two differentiation steps.
}
This contrary effect diminishes the observable net speed-up resulting from the superior efficiency of the general SBIP approach over the general SSIP approach described above.

To conclude this important assessment of the algorithm's efficiency, it can be stated that the general SBIP approach is significantly more efficient than the SSIP approach (which in turn is still significantly more efficient than a direct 6D numerical integration).
This holds even despite the superior accuracy of the applied SBIP law~(\eqref{eq::disk-cyl-pot_m}), which is therefore also slightly more complex as compared to the simple SSIP law from Eq.~(A12) of our previous contribution~\cite{GrillSSIP}.
In the numerical example of peeling two adhesive fibers to be presented in~\secref{sec::num_ex_peeling_pulloff_SBIP}, the combination of the novel SBIP approach and the specific SBIP law turns out to be approximately a factor of 4 faster than its SSIP counterpart.

\subsection{Search for interacting pairs and partitioning for parallel computing}\label{sec::search_partitioning_SBIP}
The search for interacting beam element pairs follows an efficient standard algorithm commonly referred to as bucket search (see e.g.~Ref.~\cite{Wriggers2006} for details), which has already been used in combination with the SSIP approach~\cite{GrillSSIP}.
Due to the very short range of the interactions such as vdW and repulsive steric forces considered in this article, the requirements for the search algorithm are very similar to those from macroscopic (beam) contact formulations.
The resulting small cut-off radius is beneficial with respect to both minimizing the number of interaction pairs to be evaluated and an effective partitioning of the problem to parallelize the evaluation on multiple processors without excessive cost for communication between the processors.
Hereby, the partitioning of the problem is based on the spatial arrangement of the beam elements and uses the same subdivision of the computational domain into buckets already obtained from the search algorithm.
A repartitioning and thus redistribution of the interaction pairs to the processors is done only if the spatial distribution of the beam elements has changed so much that -- considering the cut-off radius -- there is a chance that new interaction pairs need to be identified and evaluated.
Generally, the computational cost of these steps of search and partitioning turned out to be insignificant as compared to the evaluation of the interaction pairs.
Therefore, the parallelization of the pair evaluation on multiple processors indeed reduces the overall computation time significantly.

\subsection{A criterion to sort out element pairs separated further than the cut-off radius before the actual evaluation}\label{sec::filter_criterion_ele_pair_proximity}
The following is applied as an additional step after the search for spatially proximate and thus potentially interacting element pairs.
Motivated by the critical influence of the pair evaluation on the overall computational cost, this additional filtering step aims to sort out element pairs that are identified by the rather rough and conservative bucket search algorithm, but are further separated than the cut-off radius and will thus not contribute to the total interaction energy.
The key idea is thus to skip the entire evaluation of the element pair, i.e., the loop over the integration points on the slave side, which otherwise would only after the closest point-to-curve projection check the cut-off radius and skip the evaluation of terms for this specific integration point.

To achieve high net savings, the applied criterion must be cheap to evaluate and is thus only based on the nodal positions and an estimate of the actual, deformed centerline geometry of the elements by means of so-called spherical bounding boxes.
A very similar filter criterion has been applied in the context of macroscopic beam contact~\cite{Meier2017a}, where further details can be found.
The only difference lies in the distance threshold value that is used.
Here, we skip the pair evaluation if the minimal distance between the spherical bounding boxes is more than twice the cut-off radius, which should be on the safe side to not miss any contributions also in the case of strongly deformed elements.
Still, the resulting decrease in the overall runtime observed for the numerical examples from~\secref{sec::numerical_examples_SBIP} was up to 30\%, which is quite remarkable and underlines the effectiveness of this additional filtering step.
As an additional benefit, it has been observed that the number of non-unique or ill-posed closest point-to-curve projections significantly decreased and in fact vanished for the numerical examples considered in the context of this work.

\section{Numerical examples}\label{sec::numerical_examples_SBIP}
All presented formulations and algorithms have been implemented in C++ and integrated into the existing computational framework of the in-house research code BACI~\cite{BACI2020}.
Note that the implementation of the SBIP approach as well as the disk-cylinder potential has been verified by means of a second, independent implementation in MATLAB~\cite{MATLAB2017b}.
Furthermore, the correct implementation of the discrete element residual vectors as well as tangent stiffness matrices has been verified by means of an automatic differentiation tool provided via the package Sacado, which is part of the Trilinos project~\cite{Trilinos2012}.

At this point, it is also important to emphasize that the novel SBIP approach seamlessly integrates into an existing nonlinear finite element solver for solid and structural mechanics.
It does neither depend on any specific beam (finite element) formulation nor time discretization scheme, which underlines the versatility of this novel approach.
In the following numerical examples, it has been used in combination with geometrically exact, Hermitian Simo-Reissner beam elements~\cite{Meier2017b} and both in statics as well as an (implicit) time integration framework.
The resulting nonlinear system of equations is highly challenging to solve, mainly due to the competition of the strongly nonlinear, deformation-dependent adhesive and repulsive forces, which also leads to physical instabilities such as snapping free and snapping into contact.
We used a Newton-Raphson algorithm with an additional step size control, which is described in more detail in Appendix C of our previous contribution~\cite{GrillSSIP}.

\subsection{Peeling and pull-off behavior of two adhesive elastic fibers}\label{sec::num_ex_peeling_pulloff_SBIP}
\begin{figure}[htpb]%
  \centering
  \subfigure[Problem setup: undeformed configuration.]{
    \def\svgwidth{0.13\textwidth}
\begingroup%
  \makeatletter%
  \providecommand\color[2][]{%
    \errmessage{(Inkscape) Color is used for the text in Inkscape, but the package 'color.sty' is not loaded}%
    \renewcommand\color[2][]{}%
  }%
  \providecommand\transparent[1]{%
    \errmessage{(Inkscape) Transparency is used (non-zero) for the text in Inkscape, but the package 'transparent.sty' is not loaded}%
    \renewcommand\transparent[1]{}%
  }%
  \providecommand\rotatebox[2]{#2}%
  \newcommand*\fsize{\dimexpr\f@size pt\relax}%
  \newcommand*\lineheight[1]{\fontsize{\fsize}{#1\fsize}\selectfont}%
  \ifx\svgwidth\undefined%
    \setlength{\unitlength}{90bp}%
    \ifx\svgscale\undefined%
      \relax%
    \else%
      \setlength{\unitlength}{\unitlength * \real{\svgscale}}%
    \fi%
  \else%
    \setlength{\unitlength}{\svgwidth}%
  \fi%
  \global\let\svgwidth\undefined%
  \global\let\svgscale\undefined%
  \makeatother%
  \begin{picture}(1,2.5)%
    \lineheight{1}%
    \setlength\tabcolsep{0pt}%
    \put(0,0){\includegraphics[width=\unitlength,page=1]{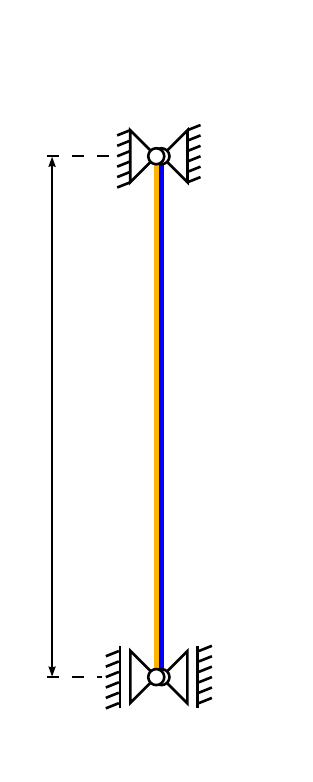}}%
    \put(0.72220523,0.16035767){\color[rgb]{0,0,0}\makebox(0,0)[lt]{\lineheight{0}\smash{\begin{tabular}[t]{l}$u_x, F_x^{br}$\end{tabular}}}}%
    \put(0.19677776,1.13176587){\color[rgb]{0,0,0}\makebox(0,0)[lt]{\lineheight{0}\smash{\begin{tabular}[t]{l}$l$\end{tabular}}}}%
    \put(0.71669108,2.17380371){\color[rgb]{0,0,0}\makebox(0,0)[lt]{\lineheight{0}\smash{\begin{tabular}[t]{l}$u_x, F_x^{tr}$\end{tabular}}}}%
    \put(0,0){\includegraphics[width=\unitlength,page=2]{num_ex_vdW_twoparallelbeams_pulloff_from_contact_problem_setup.pdf}}%
    \put(0.54009603,2.26407878){\color[rgb]{0,0,0}\makebox(0,0)[lt]{\lineheight{0}\smash{\begin{tabular}[t]{l}$y$\end{tabular}}}}%
    \put(0.81509603,2.03602702){\color[rgb]{0,0,0}\makebox(0,0)[lt]{\lineheight{0}\smash{\begin{tabular}[t]{l}$x$\end{tabular}}}}%
    \put(0,0){\includegraphics[width=\unitlength,page=3]{num_ex_vdW_twoparallelbeams_pulloff_from_contact_problem_setup.pdf}}%
  \end{picture}%
\endgroup%

    \label{fig::num_ex_vdW_twoparallelbeams_problem_setup}
  }
  \hspace{0.5cm}
  \subfigure[Quasi-static force-displacement curve. Force values to be interpreted as multiple of a reference point load that causes a deflection of~$l/4$ if applied at the fiber midpoint.]{
   \includegraphics[width=0.45\textwidth]{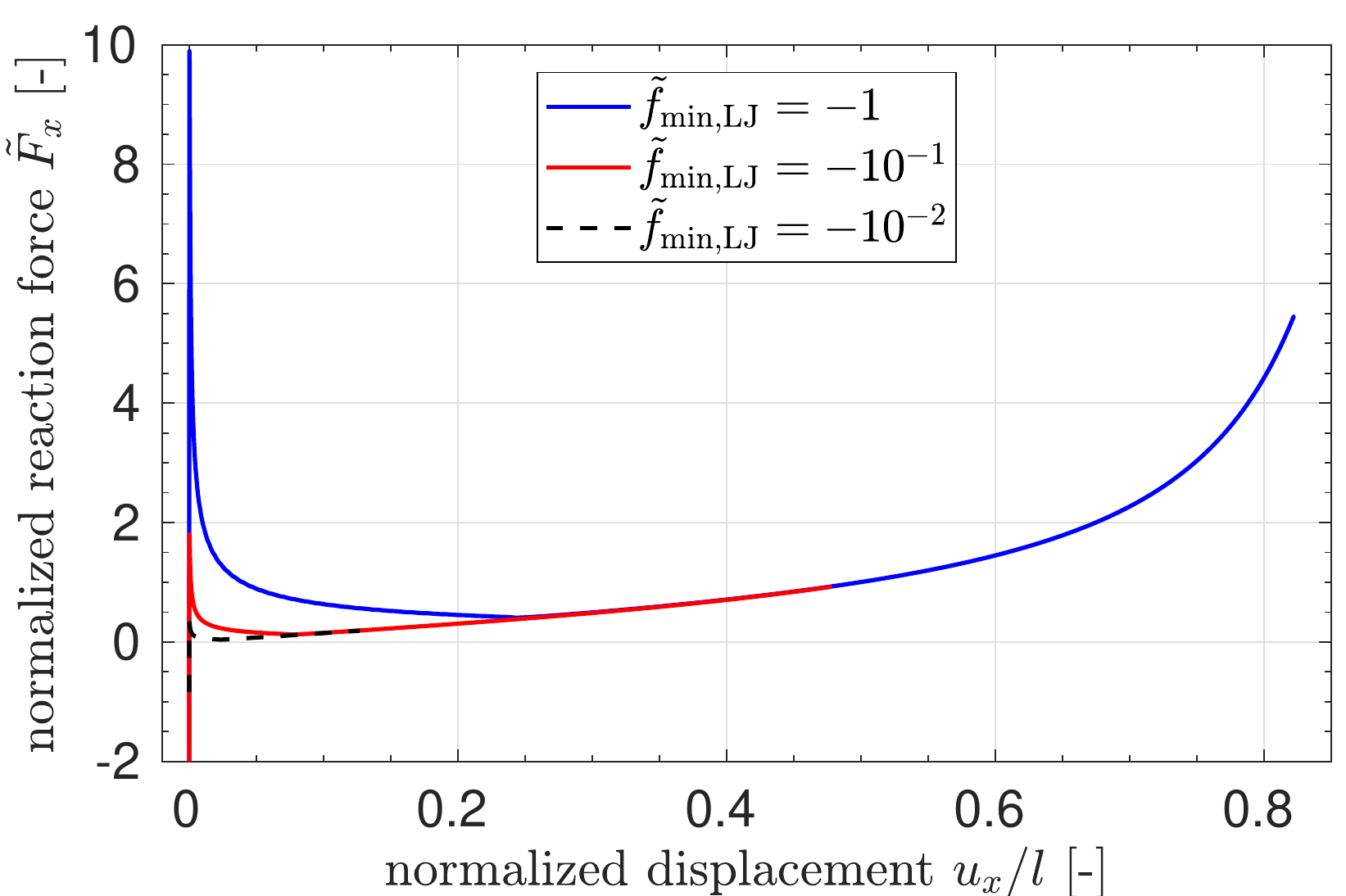}
   \label{fig::num_ex_vdW_twoparallelbeams_pulloff_force_over_displacement}
  }
  \hfill
  \subfigure[Detail view for small displacement values.]{
   \includegraphics[width=0.3\textwidth]{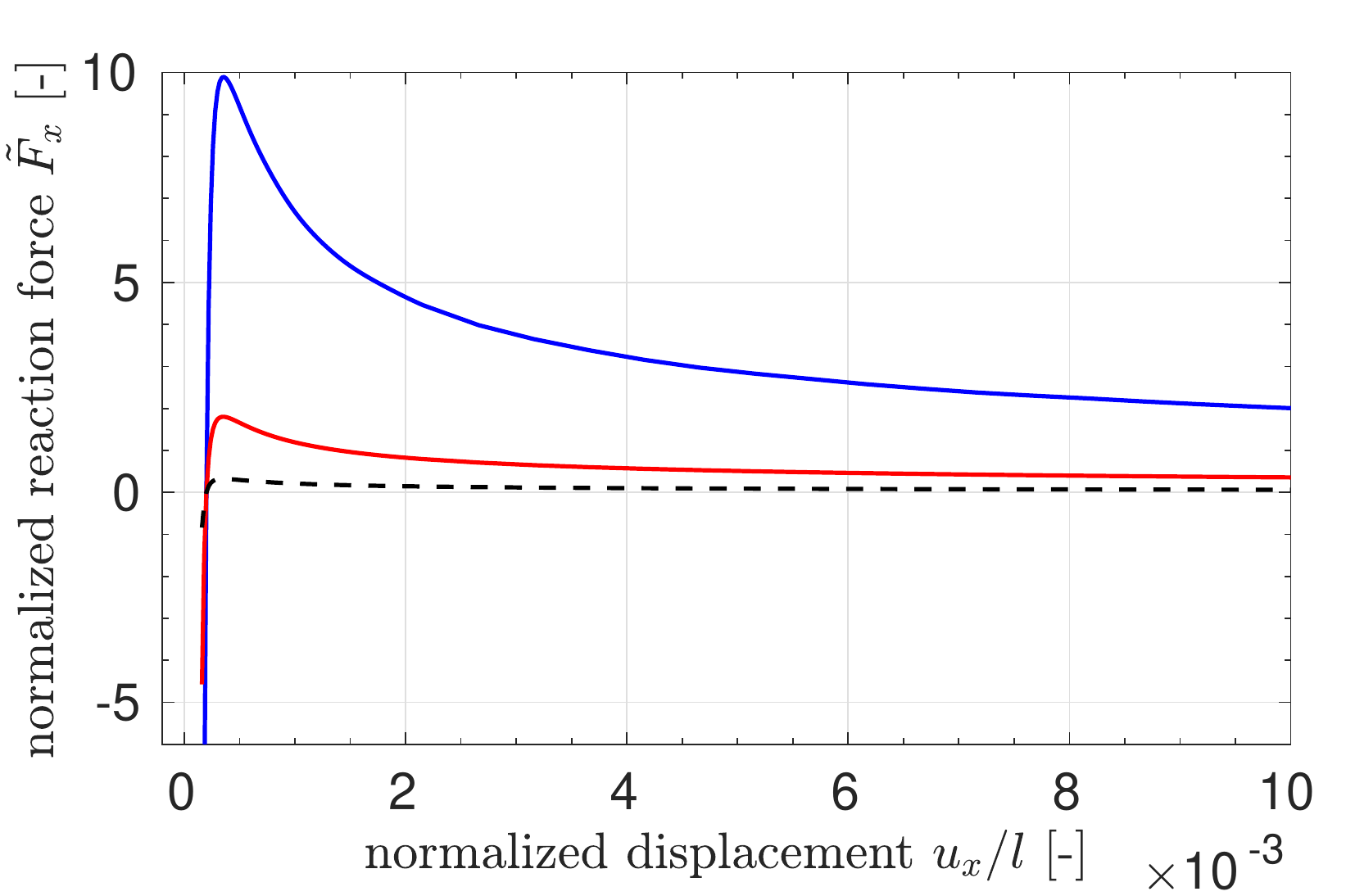}
   \label{fig::num_ex_vdW_twoparallelbeams_pulloff_force_over_displacement_zoom}
  }
  \caption{Numerical peeling experiment with two adhesive elastic fibers interacting via the LJ potential.}
  \label{fig::num_ex_vdW_twoparallelbeams_peeling}
\end{figure}

This numerical example has first been studied in the authors' previous contribution~\cite{GrillPeelingPulloff} where the SSIP approach~\cite{GrillSSIP} has been used to model vdW adhesion and steric repulsive forces.
\figref{fig::num_ex_vdW_twoparallelbeams_problem_setup} shows the problem setup consisting of two parallel, straight fibers interacting via a LJ potential.
The idea is to study the entire process of separating these adhesive fibers starting from contact along the entire length up to the point where they would suddenly snap free.
Therefore, a displacement~$u_x$ in $x$-direction is prescribed at both ends of the right fiber and the sum of the reaction forces~$F_x \defvariable F_x^\text{tr}+F_x^\text{br}$ in this direction is measured in order to obtain the characteristic, quasi-static force-displacement curve.

Most importantly, this simple setup with two initially straight beams allows to verify the accuracy of the SBIP approach by means of analytical reference solutions.
The theoretical work for the scenario of infinitely long and parallel rigid cylinders presented in Ref.~\cite[Appendix A.2.2]{GrillSSIP} is able to predict both the equilibrium spacing~$g_\text{LJ,eq,cyl$\parallel$cyl}$ and the maximal magnitude of adhesive forces per unit length~$\tilde{f}_\text{min,LJ,cyl$\parallel$cyl}$, i.e. the effective local pull-off force per unit length.
It will be shown that these local quantities are excellently reproduced by the SBIP approach whereas the previously used simple SSIP law fails to do so (without additional calibration).
While the results from both approaches agree very well in a \textit{qualitative} manner, only the SBIP approach is thus able to yield the \textit{quantitatively} correct pull-off force and other important global quantities characterizing this numerical example.

Moreover, it will be shown that the bias introduced by the choice of master and slave is negligible even for the considerably large fiber deformations in this example.
First, this is confirmed by the symmetries of the example, which are excellently preserved in the numerical solutions.
And second, using a modified setup with one rigid, straight beam and one deformable beam, we can quantify the error introduced by approximating the master beam geometry as a straight cylinder, because one of the two possible choices for master and slave will be the exact solution of the problem.
For the other choice, we obtain a maximum relative error of $1.4\%$ in the global pulling force even for the rather extreme scenario of maximal adhesive force and thus strongly bent fibers to be considered here.
Altogether, this is a very important confirmation that the fundamental modeling approach of approximating one of the fibers as cylinder when calculating the two-fiber interaction potential leads to a very high accuracy in the case of very short-ranged interactions considered here.

\subsubsection{Setup and parameters}
At this point, only the differences in the setup compared to the original numerical peeling experiment mentioned above will be presented.
Refer to \cite[Sec.~4]{GrillPeelingPulloff} for a complete presentation of the setup, numerical methods and parameter values.
Most importantly, the LJ interaction between the fibers is modeled by means of the novel SBIP approach from \secref{sec::method_single_length_specific_integral} in combination with the proposed disk-cylinder interaction potential law~$\tilde \pi_\text{m,disk-cyl}$ from \eqref{eq::disk-cyl-pot_m}, which is used for both the attractive~$m=6$ as well as the repulsive~$m=12$ part of the LJ interaction.
The two prefactors~$k_6$ and $k_{12}$ specifying the LJ point-pair potential law will be varied to study their influence on the system response.
Instead of using these prefactors, however, it seems more meaningful and intuitive in this context to use an equivalent set of parameters, which describe the equilibrium spacing~$g_\text{LJ,eq,cyl$\parallel$cyl}$ and the maximal magnitude of adhesive forces per unit length~$\tilde{f}_\text{min,LJ,cyl$\parallel$cyl}$, i.e. the effective local pull-off force per unit length, of straight, parallel fibers.
According to the theoretical work for the scenario of infinitely long and parallel rigid cylinders presented in~\cite[Appendix A.2.2]{GrillSSIP}, these two alternative sets of two LJ parameters are bi-uniquely related%
\footnote{%
taking into account the atom densities~$\rho_i$ and radii~$R_i$ of the fibers $i=1,2$
}
as follows:
\begin{align}
  g_\text{LJ,eq,cyl$\parallel$cyl} &\approx 1.1434 \, \left( - k_{12}/k_6 \right)^{\frac{1}{6}}\label{eq::gap_LJ_eq_parallel_cyl_asfunctionof_prefactors}\\
  \tilde{f}_\text{min,LJ,cyl$\parallel$cyl} &\approx 0.7927 \, \rho_1 \rho_2 \sqrt{\frac{2 R_1 R_2}{R_1+R_2}} \, k_6 \, \left( - k_6/k_{12} \right)^{\frac{5}{12}}\label{eq::force_LJ_min_parallel_cyl_asfunctionof_prefactors}
\end{align}
Again refer to~\cite[Appendix A.2.2]{GrillSSIP} for the rather lengthy exact expressions of the scalar prefactors that are given here in their approximate floating point representation.
For the sake of brevity, the subscript ``cyl$\parallel$cyl'' will be omitted in the remainder of this section, however, keep in mind that these quantities describe the academic case of infinitely long and parallel, rigid cylinders.
Since the meaning of these two parameters is much more intuitive in the context of fiber adhesion, we can now argue that the equilibrium surface spacing~$g_\text{LJ,eq}$ is a fundamental property of the type of physical interaction (and maybe also the material combination of both fibers and surrounding media) and keep it fixed for now with an exemplarily chosen value of~$g_\text{LJ,eq}=10^{-3}=0.05 \cdot R = 2\times10^{-4} \cdot l$, which corresponds to five percent of the fiber radius.
The minimal LJ force per unit length~$\tilde{f}_\text{min,LJ}$ on the other hand is a viable measure for the strength of adhesion and will be varied to study this important influence on the system behavior.
Unless otherwise stated, two integration segments with ten Gauss points each is used for numerical integration of the LJ contributions in each of the 64 beam elements per fiber.
It has been verified by means of refinement studies that the influence of the spatial discretization error and numerical integration error is negligible for all results to be presented in the following.
The regularization strategy proposed in~\secref{sec::regularization_SBIP} has been applied with a regularization separation~$g_\text{ul,reg} = 8 \times 10^{-4} = 0.04 \cdot R$, which is smaller than the equilibrium spacing~$g_\text{LJ,eq}$ given above and thus -- as has been argued in~\secref{sec::regularization_SBIP} -- led to identical results as compared to the non-regularized interaction law, yet ensures robustness and a significant reduction of the number of required nonlinear iterations by almost a factor of five.
In order to further reduce the computational cost, the very short range of the LJ interaction has been exploited by applying a cut-off radius of~$r_c=0.1=5R$, which again had no influence on the results.
As mentioned above, all other parameters including the geometrical and material properties of the fibers remain unchanged as compared to \cite[Sec.~4]{GrillPeelingPulloff}.

\subsubsection{Qualitative quasi-static system behavior}
\begin{figure}[htpb]%
  \centering
  \subfigure[$u_x/l=0$]{
    \includegraphics[width=0.25\textwidth]{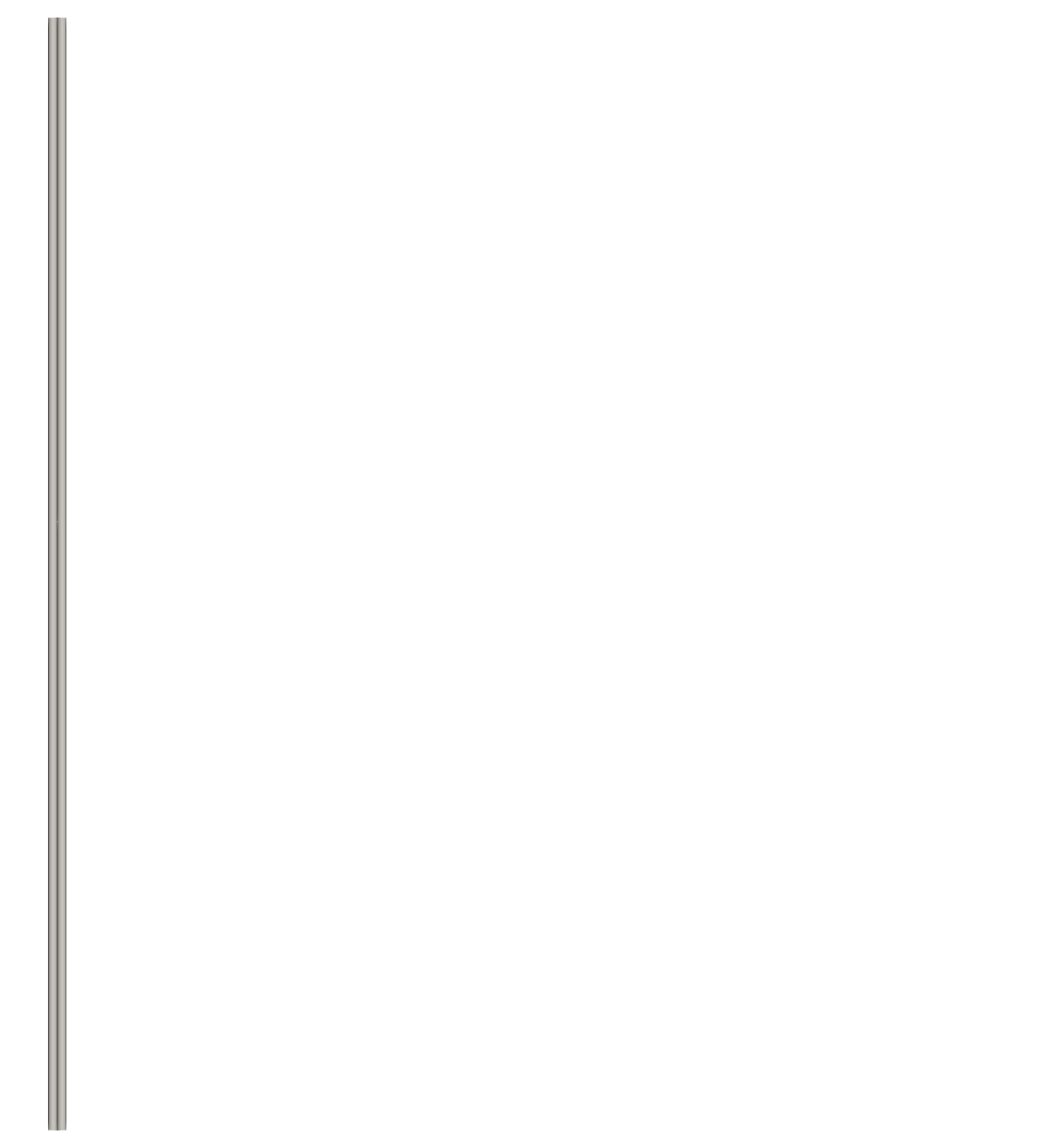}
    \label{fig::num_ex_vdW_twoparallelbeams_snapshot_reldisp0}
    \hspace{-2.8cm}
  }
  \subfigure[$u_x/l=10^{-2}$]{
    \includegraphics[width=0.25\textwidth]{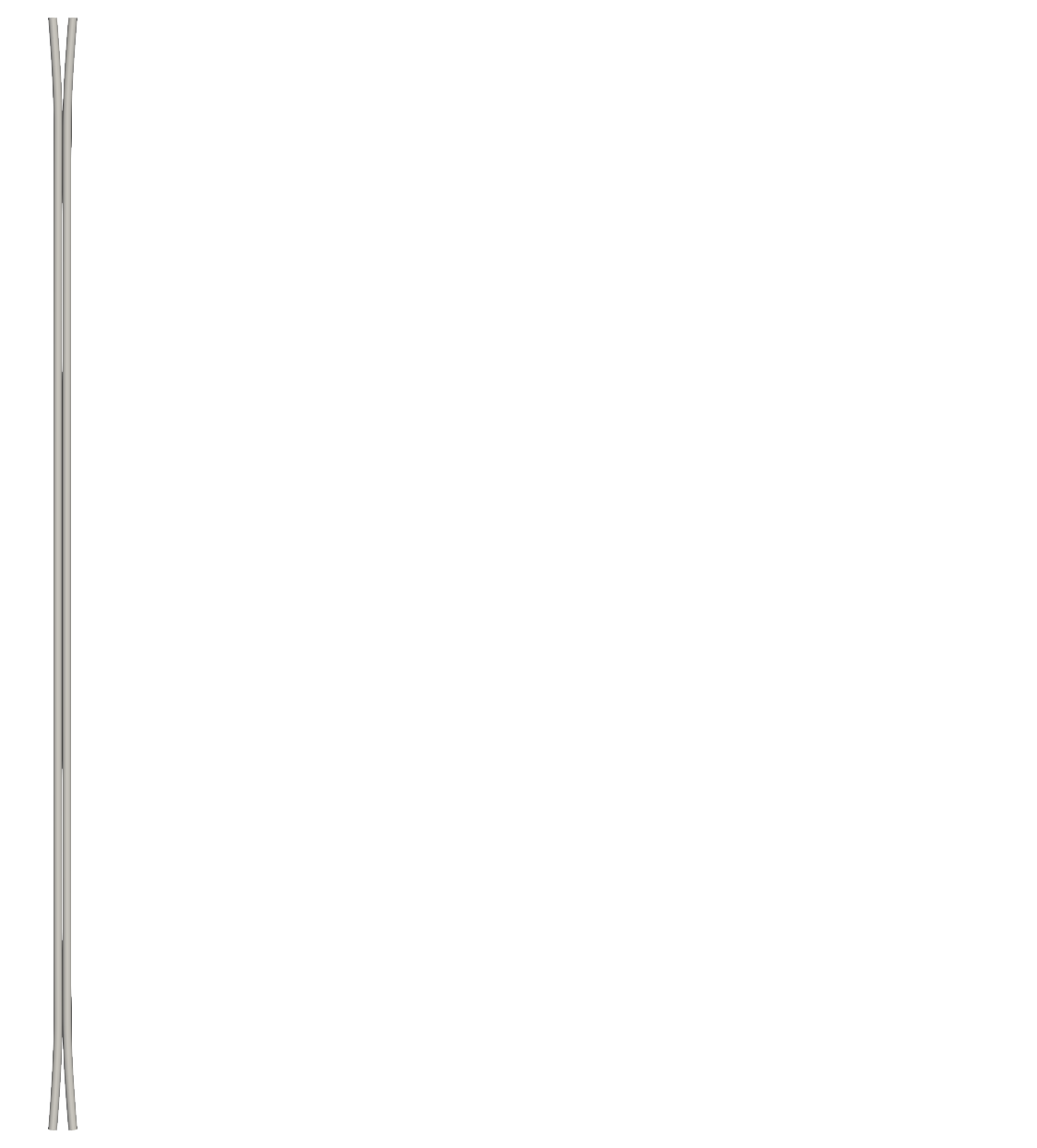}
    \label{fig::num_ex_vdW_twoparallelbeams_snapshot_reldisp0_01}
  \hspace{-2.7cm}
  }
  \subfigure[$u_x/l=0.1$]{
    \includegraphics[width=0.25\textwidth]{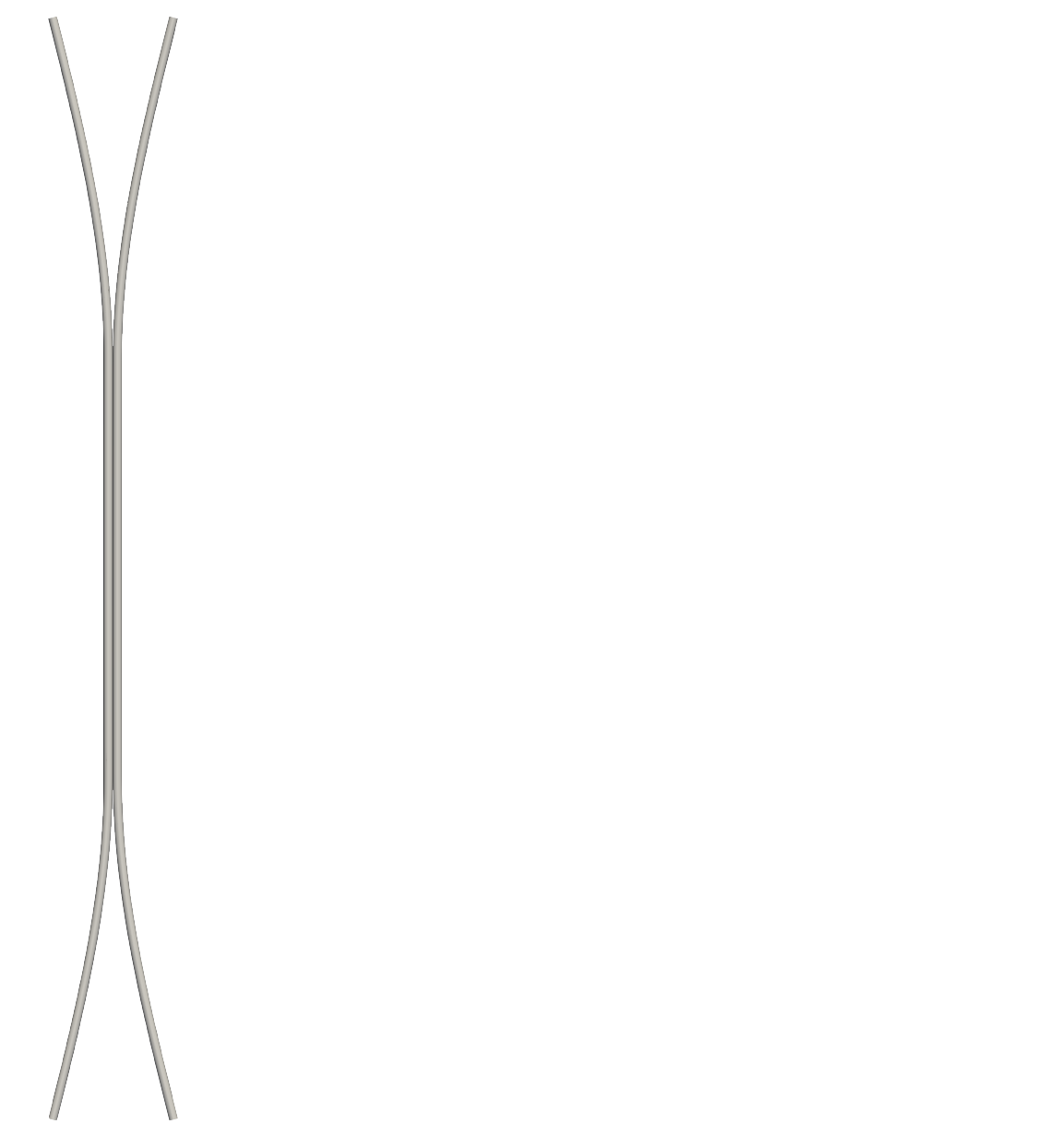}
    \label{fig::num_ex_vdW_twoparallelbeams_snapshot_reldisp0_1}
  \hspace{-2.5cm}
  }
  \subfigure[$u_x/l=0.2$]{
    \includegraphics[width=0.25\textwidth]{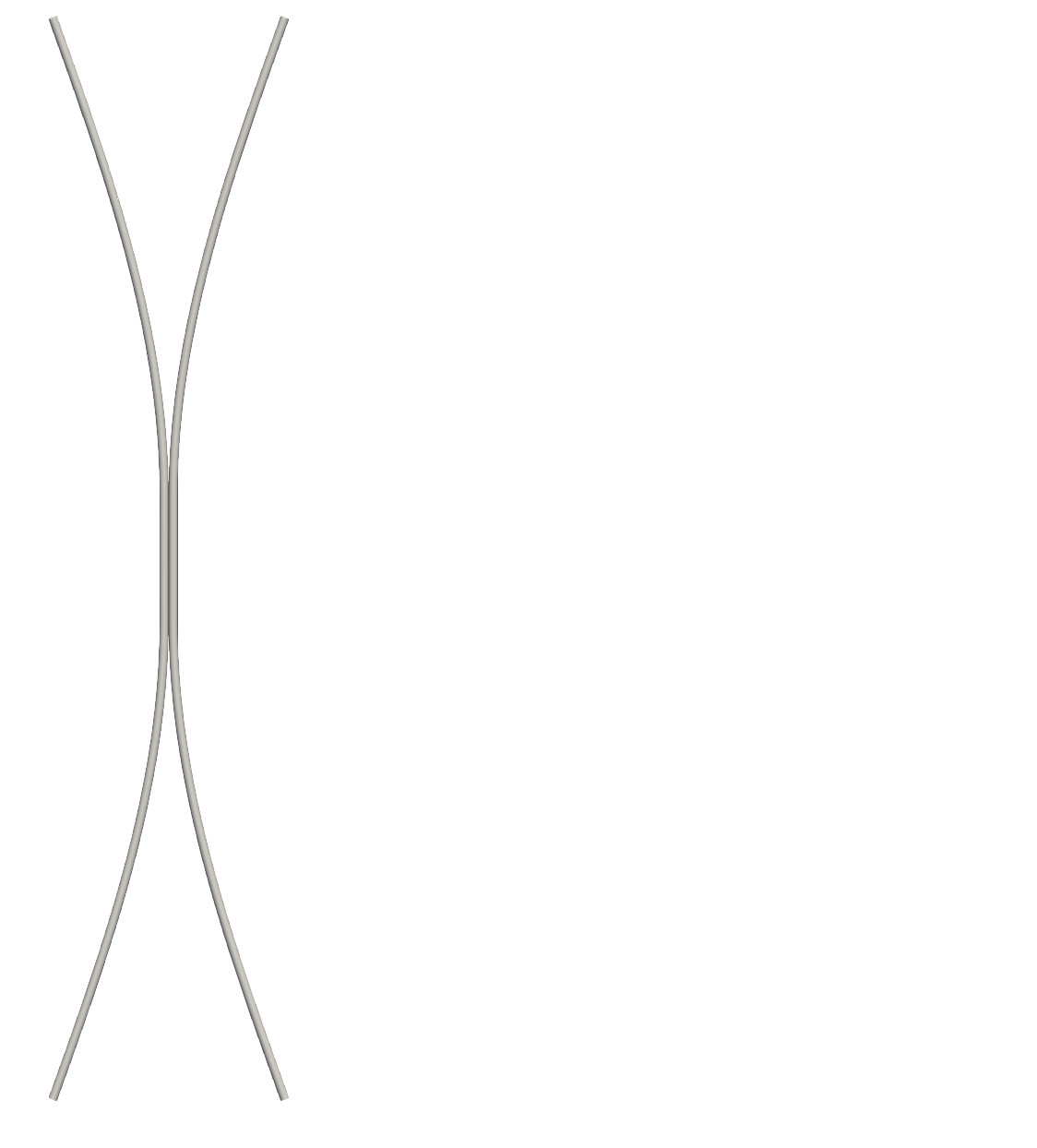}
    \label{fig::num_ex_vdW_twoparallelbeams_snapshot_reldisp0_2}
  \hspace{-2.3cm}
  }
  \subfigure[$u_x/l=0.5$]{
    \includegraphics[width=0.25\textwidth]{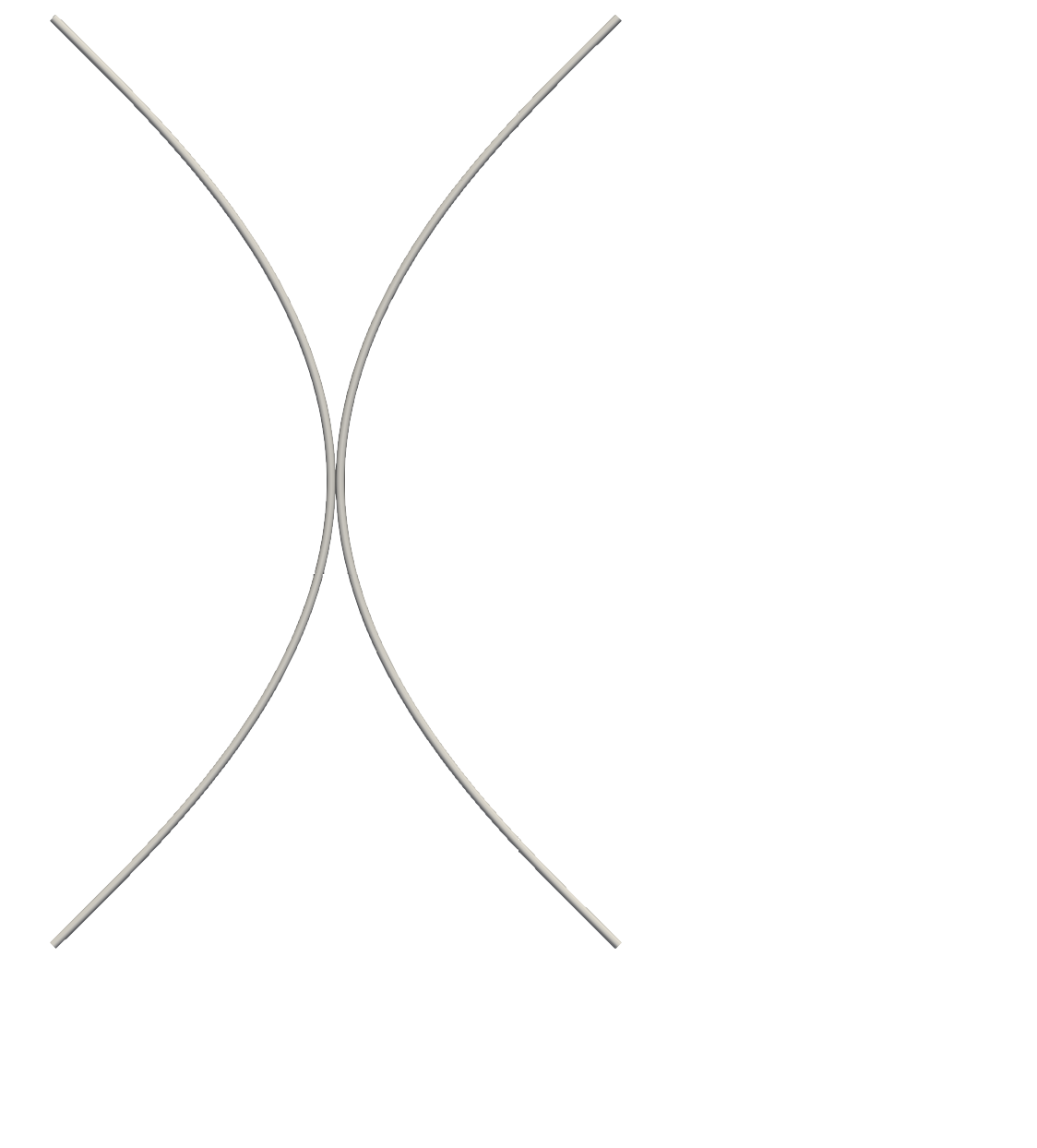}
    \label{fig::num_ex_vdW_twoparallelbeams_snapshot_reldisp0_5}
  \hspace{-1cm}
  }
  \subfigure[$u_x/l\approx0.8218$: ultimately before snapping free]{
    \includegraphics[width=0.25\textwidth]{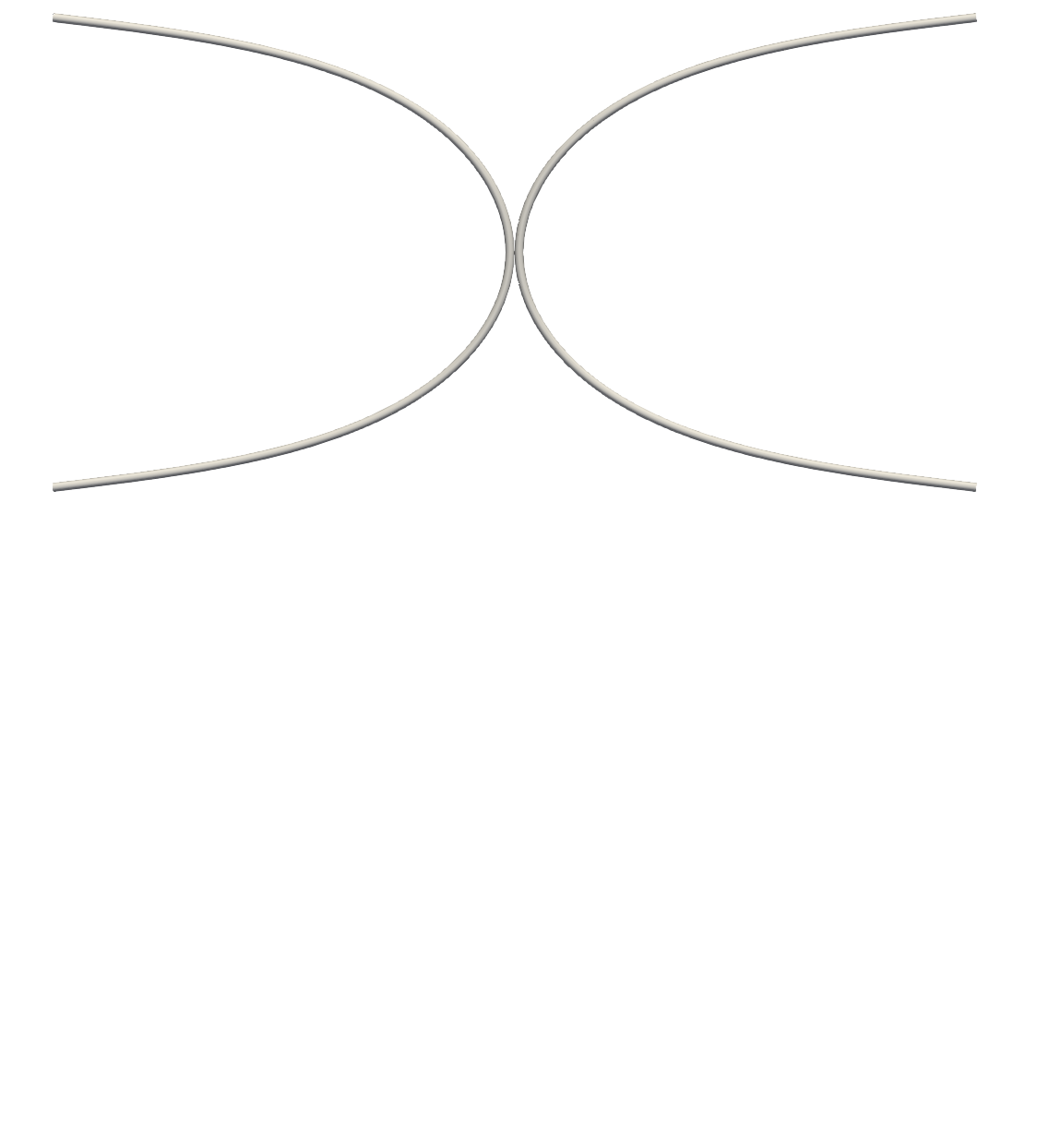}
    \label{fig::num_ex_vdW_twoparallelbeams_snapshot_reldisp0_8218_finalstate}
  }
  \caption{A selection of characteristic equilibrium configurations obtained for the case~$\tilde{f}_\text{min,LJ}=-1$.}
  \label{fig::num_ex_vdW_twoparallelbeams_peeling_snapshots}
\end{figure}
Let us first look at the case of~$\tilde{f}_\text{min,LJ}=-1$, which is the strongest attractive strength considered in this numerical experiment.
The resulting equilibrium configurations for exemplarily chosen displacement values~$u_x$ are shown in~\figref{fig::num_ex_vdW_twoparallelbeams_peeling_snapshots}.
Note that the configurations are symmetric with respect to both a vertical and a horizontal middle axis, which also implies for the individual reaction force components on the left/right (``l''/``r'') side and top/bottom (``t''/``b'') that~$F_x^\text{tl} = F_x^\text{bl} = - F_x^\text{tr} = - F_x^\text{br}$ and $F_y \equiv 0$, as expected from the symmetric setup of the experiment and already observed in~\cite{GrillPeelingPulloff}.
It seems worth mentioning that this symmetry is not broken by the choice of master and slave beam as required by the SBIP approach -- more on this important aspect in \secref{sec::num_ex_influence_master_slave}.
The corresponding force-displacement curve is shown in \figref{fig::num_ex_vdW_twoparallelbeams_pulloff_force_over_displacement} (blue line) and a magnified view of the small displacement value range is provided in \figref{fig::num_ex_vdW_twoparallelbeams_pulloff_force_over_displacement_zoom}.
Overall, the shape of the curve is very similar to the one obtained using the previous SSIP approach in~\cite{GrillPeelingPulloff} and also the three characteristic phases of \textit{initiation of fiber deformation}, \textit{peeling}, and \textit{pull-off} as obtained and analyzed for the case of electrostatic attraction in~\cite{GrillPeelingPulloff} are recognizable here.
Moreover, the sharp force maximum for a very small displacement value in the initiation phase is confirmed by this study using the novel SBIP approach and more accurate disk-cylinder interaction law~$\tilde \pi_\text{m,disk-cyl}$.
It can thus be concluded that the known limitation regarding the accuracy of the previously applied, simple SSIP law~$\tilde{\tilde{\pi}}$ does not affect the qualitative analysis and conclusions drawn in Ref.~\cite{GrillPeelingPulloff}.%
\footnote{%
This had already been verified by means of a few simulation runs using a prototype of the novel SBIP approach proposed in the present work.
}
A detailed comparison, including a quantitative analysis of the differences in the global system response for the identical set of parameter values will follow in a dedicated section below.
Following up on the qualitative results, note that the very short range of the LJ interactions leads to the fact that fibers interact almost exclusively if or where they are in contact and that there is no perceptible far range effect as observed and measured in form of a second branch (``separated fibers'') in the force-displacement plot e.g.~for the case of electrostatic attraction in~\cite{GrillPeelingPulloff}.
This observation is also in agreement with the common notion that short-ranged interactions such as vdW adhesion only have an influence on the state of being/remaining in contact, and not the process of coming into contact from an initially separated state.

\begin{figure}[htpb]%
  \centering
  \subfigure[$u_x/l=0.1$]{
    \includegraphics[width=0.5\textwidth]{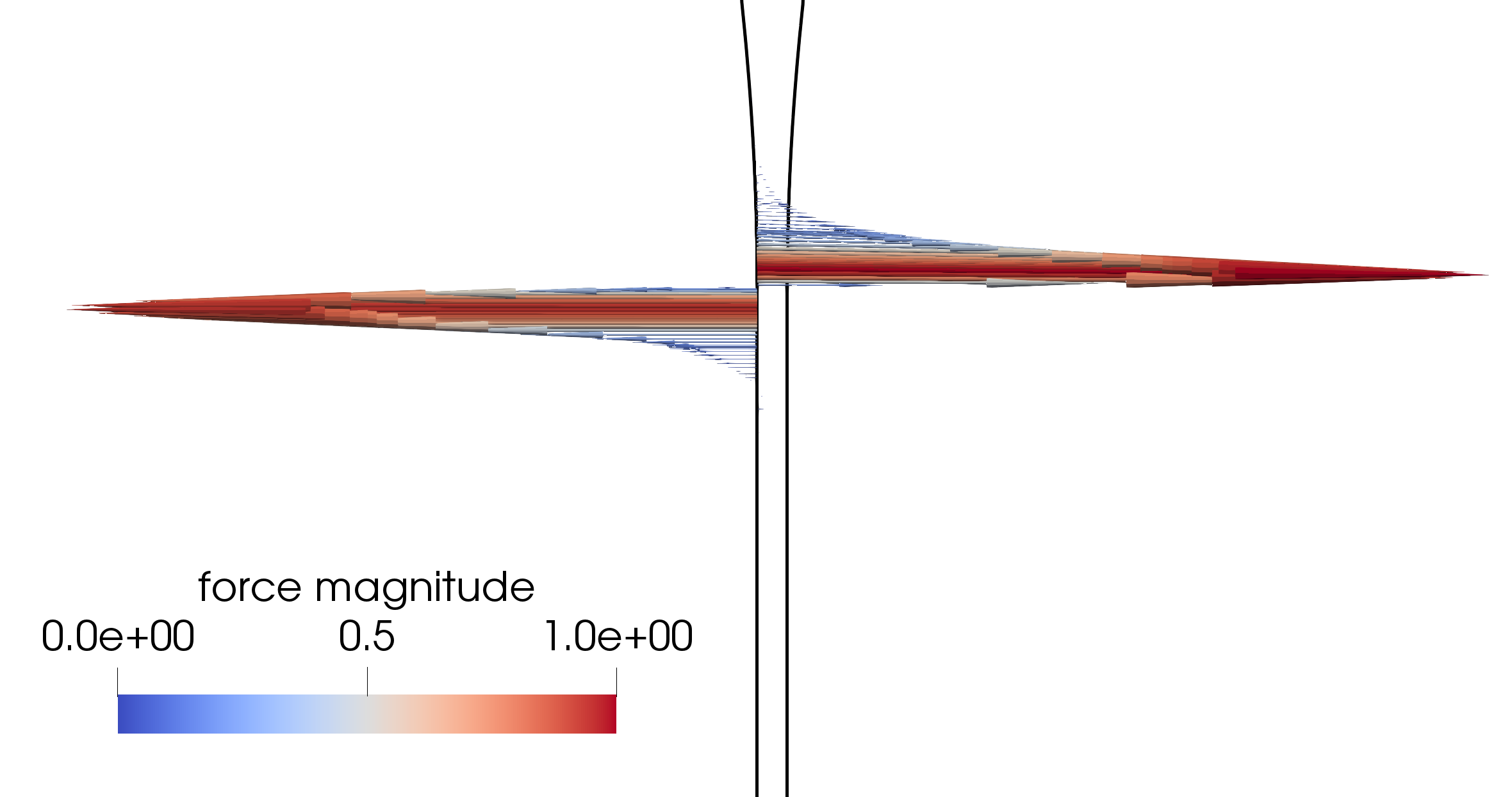}
    \label{fig::num_ex_vdW_twoparallelbeams_snapshot_forcedistribution_reldisp0_1}
  }
  \subfigure[$u_x/l\approx0.8218$: ultimately before snapping free]{
    \includegraphics[width=0.42\textwidth]{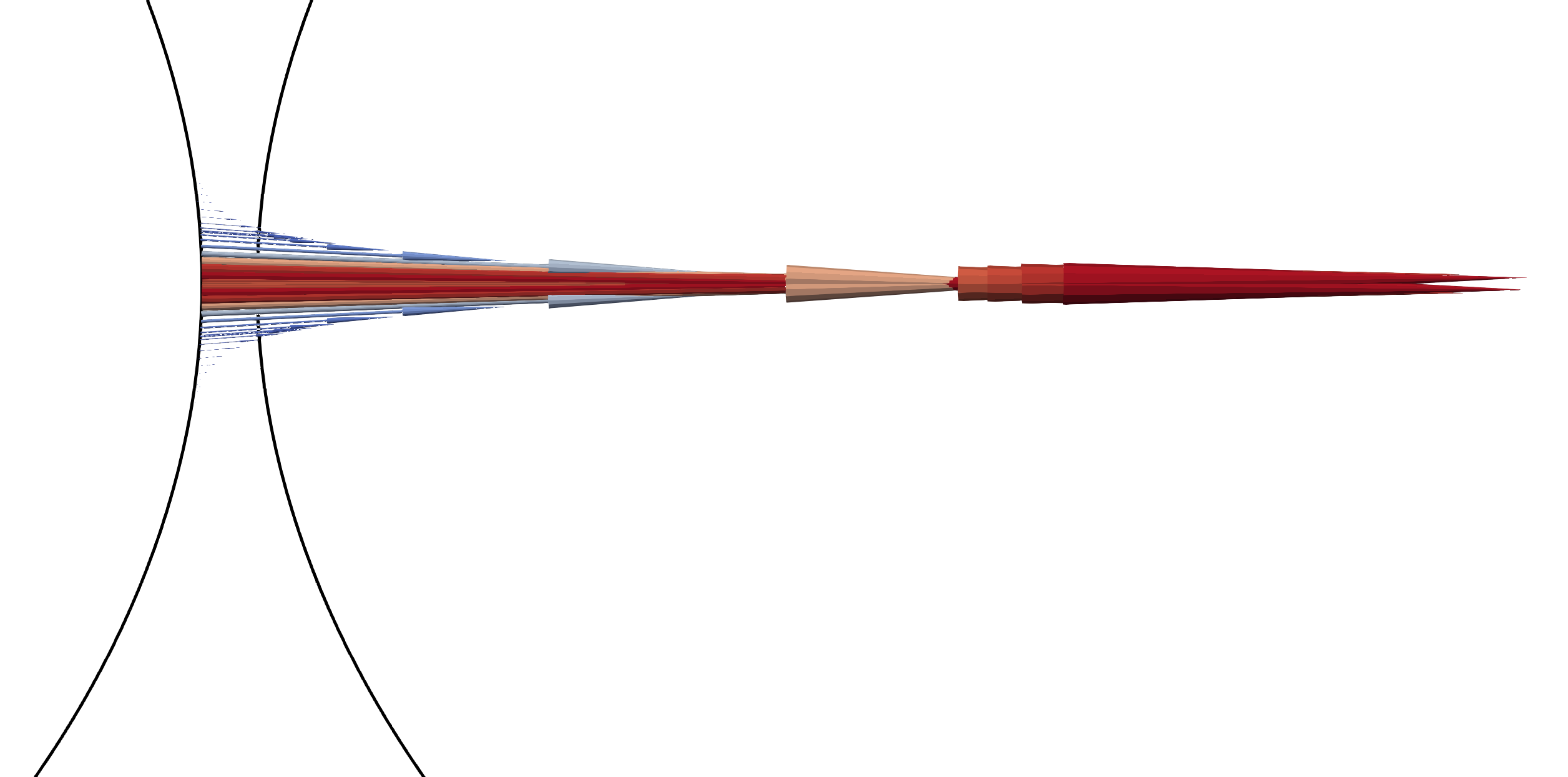}
    \label{fig::num_ex_vdW_twoparallelbeams_snapshot_forcedistribution_finalstate}
  }
  \caption{Detail study of the resulting line force distributions for two displacement values obtained for the case~$\tilde{f}_\text{min,LJ}=-1$. For clarity, the fibers are depicted as their centerlines and forces are shown for the left fiber only. Note that both pictures show details of the entire system and are not to scale.}
  \label{fig::num_ex_vdW_twoparallelbeams_snapshot_forcedistribution}
\end{figure}

A quantitative analysis of the results will be given in the following section discussing the influence of the strength of adhesion, however, two brief aspects are stated here as an immediate verification of the results.
First, the specified parameter value for the equilibrium gap of infinitely long and parallel, rigid cylinders~$g_\text{LJ,eq}=10^{-3}$ is indeed recovered as a simulation result in the middle parts of the fibers wherever they are approximately parallel.
And second, as can be seen from the visualization of the resulting LJ force distributions in~\figref{fig::num_ex_vdW_twoparallelbeams_snapshot_forcedistribution}, the maximum magnitude of the observed interaction forces per unit length agrees very well with the specified parameter value~$\tilde{f}_\text{min,LJ}=-1$.

The interesting shape of the interaction force distribution shown in \figref{fig::num_ex_vdW_twoparallelbeams_snapshot_forcedistribution_reldisp0_1} reveals the competing nature of the adhesive and repulsive forces.
At the point, where the smallest surface separation of the fibers transitions from below to beyond the equilibrium distance, there is a sharp transition from strong repulsive to strong adhesive forces.
Due to the bending resistance of the fibers, there must be a repulsive region when separating the two fibers.
Only in the middle part, the fibers maintain the net force-free equilibrium surface separation~$g_\text{LJ,eq}$.
\figref{fig::num_ex_vdW_twoparallelbeams_snapshot_forcedistribution_finalstate} finally shows a state where the fibers have been pulled further apart than~$g_\text{LJ,eq}$ even at theirs closest points such that solely adhesive forces prevail.

\subsubsection{Influence of the strength of adhesion}
\begin{figure}[htpb]%
  \centering
  \subfigure[]{
    \includegraphics[width=0.25\textwidth]{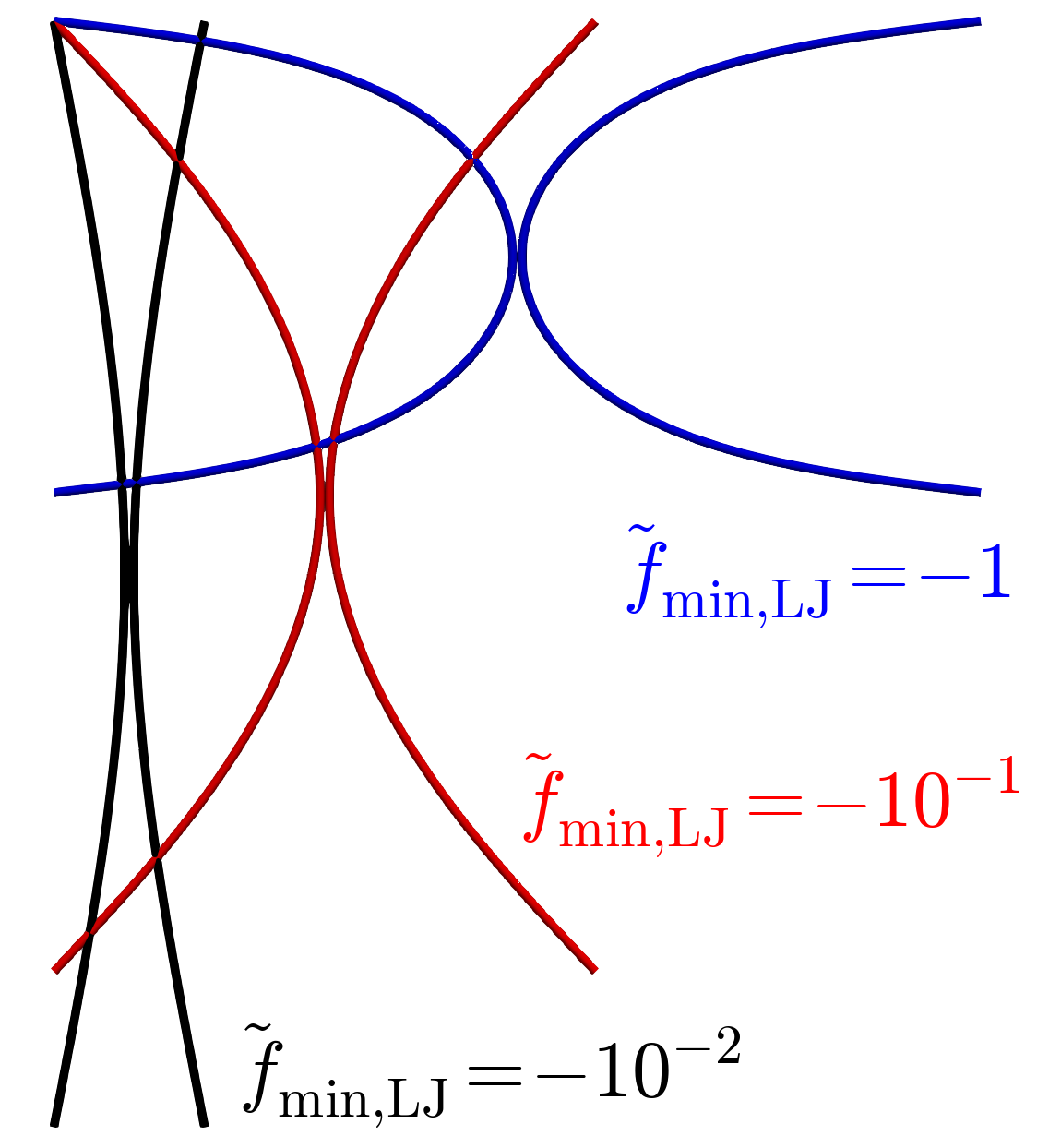}
    \label{fig::num_ex_vdW_twoparallelbeams_snapshot_comparison_finalstates}
  }
  \caption{Comparison of the final configurations before snapping free for three values of the min.~LJ force per unit length~$\tilde{f}_\text{min,LJ}$.}
  \label{fig::num_ex_vdW_twoparallelbeams_peeling_snapshots_comparison_finalstates}
\end{figure}

A more general, theoretical study of parameter influences in systems of adhesive fibers by means of nondimensionalization of the governing partial differential equations is provided in~\appref{sec::nondimensionalization}.
Here, it is complemented by a numerical study considering the influence of the adhesive strength for this fundamental two-fiber peeling and pull-off example.
Thus, keeping all other parameters unchanged, the minimal LJ force per unit length (i.e. the maximal adhesive force magnitude) of infinitely long and parallel cylinders~$\tilde{f}_\text{min,LJ}$ is varied over two orders of magnitude such that both limits of ``strong adhesion/low fiber stiffness'' and ``weak adhesion/high fiber stiffness'' can be observed.
The resulting force-displacement curves for~$\tilde{f}_\text{min,LJ}=\{-1,-0.1,-0.01\}$ (blue solid line, red solid line, black dashed line) are shown in \figref{fig::num_ex_vdW_twoparallelbeams_pulloff_force_over_displacement} and \ref{fig::num_ex_vdW_twoparallelbeams_pulloff_force_over_displacement_zoom} and the corresponding final equilibrium configurations ultimately before snapping free are displayed in \figref{fig::num_ex_vdW_twoparallelbeams_peeling_snapshots_comparison_finalstates}.
As expected, an increase of the adhesive strength generally leads to increased reaction force values and higher displacement values before the fibers would finally snap free.
Specifically, we obtain normalized force peak values of~$\tilde F_{x\text{,max}} \approx 0.32$ at~$u_x/l \approx 3.4 \times 10^{-4}$ for $\tilde{f}_\text{min,LJ}=-0.01$, $\tilde F_{x\text{,max}} \approx 1.8$ at~$u_x/l \approx 3.6 \times 10^{-4}$ for $\tilde{f}_\text{min,LJ}=-0.1$, and $\tilde F_{x\text{,max}} \approx 9.9$ at~$u_x/l \approx 3.6 \times 10^{-4}$ for $\tilde{f}_\text{min,LJ}=-1$.
The peak force values thus increase by a factor of approximately 5.6 and 31 if the minimal LJ force per unit length is increased by a factor 10 and 100, respectively.
However, the location of the force peak is not influenced by the varying adhesive strength, which meets our expectations from keeping the equilibrium surface separation~$g_\text{LJ,eq}$ fixed.
Finally, the maximum normalized displacement values~$u_{x\text{,max}}/l$ before the fibers would snap free are observed to be approximately 0.13, 0.48, and 0.82 for $\tilde{f}_\text{min,LJ}=\{-0.01,-0.1,-1\}$, respectively.
Note however that these maximum displacement values are the result of a quasi-static analysis and that the exact point of snapping free will depend on the dynamics of the system.
A final interesting observation based on the force-displacement curves in \figref{fig::num_ex_vdW_twoparallelbeams_pulloff_force_over_displacement} is that both pairs of curves for $\tilde{f}_\text{min,LJ}=\{-0.01,-0.1\}$ and $\tilde{f}_\text{min,LJ}=\{-0.1,-1\}$ approximately correspond to each other in a certain, limited displacement range~$u_x/l \approx \, [ 0.08, 0.13 ]$ and $u_x/l \approx \, [ 0.25, 0.48 ]$, respectively, before the system with weaker adhesive strength would snap free.

\subsubsection{Influence of the choice of master and slave}\label{sec::num_ex_influence_master_slave}
Following up on the discussion of how to assign the roles of master and slave in \secref{sec::method_single_length_specific_integral}, this numerical example is used to study the influence of this choice and thus bias in the SBIP approach.
For this sake, all three cases shown in \figref{fig::num_ex_vdW_twoparallelbeams_pulloff_force_over_displacement} have been repeated with switched roles of master and slave and the resulting differences in the reaction force values along the entire force-displacement curve have been evaluated.
The maximal relative differences along the entire curve are approximately $\{ 4\times 10^{-6}, 5\times 10^{-6}, 3\times10^{-5}\}$ for $\tilde{f}_\text{min,LJ}=\{-0.01,-0.1,-1\}$, respectively, and thus are of the same order of or even smaller than the spatial discretization error that has been verified by doubling the number of elements.
This is a very important confirmation that the introduction of master and slave and the corresponding approximation of the master beam's geometry as cylinder in the calculation of the two-fiber interaction potential does not break the inherent symmetries of the problem setup and is thus a reasonable modeling assumption.

To quantify the model error associated with replacing the master beam by a surrogate cylinder, the setup of the example is now slightly modified by constraining the left fiber's deformation to zero and thus preserving the initial straight cylinder shape.
Using the left fiber as the master beam will thus eliminate the corresponding model error and serves as a reference solution for the subsequent simulation run with switched roles of master and slave.
This results in maximal relative differences in the pulling force values of approximately $\{0.13\%, 0.43\%, 1.42\%\}$ for $\tilde{f}_\text{min,LJ}=\{-0.01,-0.1,-1\}$, once again measured along the entire force-displacement curve.
Given the considerable magnitude of curvature for the highest adhesion force value, which is even slightly more than in the original example visualized in \figref{fig::num_ex_vdW_twoparallelbeams_peeling_snapshots_comparison_finalstates}, this is considered to be a reasonably small error level, which will presumably be negligible in most practical applications of the novel SBIP approach.

\subsubsection{Comparison of SSIP and SBIP approach and corresponding reduced interaction laws}
\begin{figure}[htpb]%
  \centering
  \subfigure[Quasi-static force-displacement curve. Force values to be interpreted as multiple of a reference point load that causes a deflection of~$l/4$ if applied at the fiber midpoint.]{
   \includegraphics[width=0.45\textwidth]{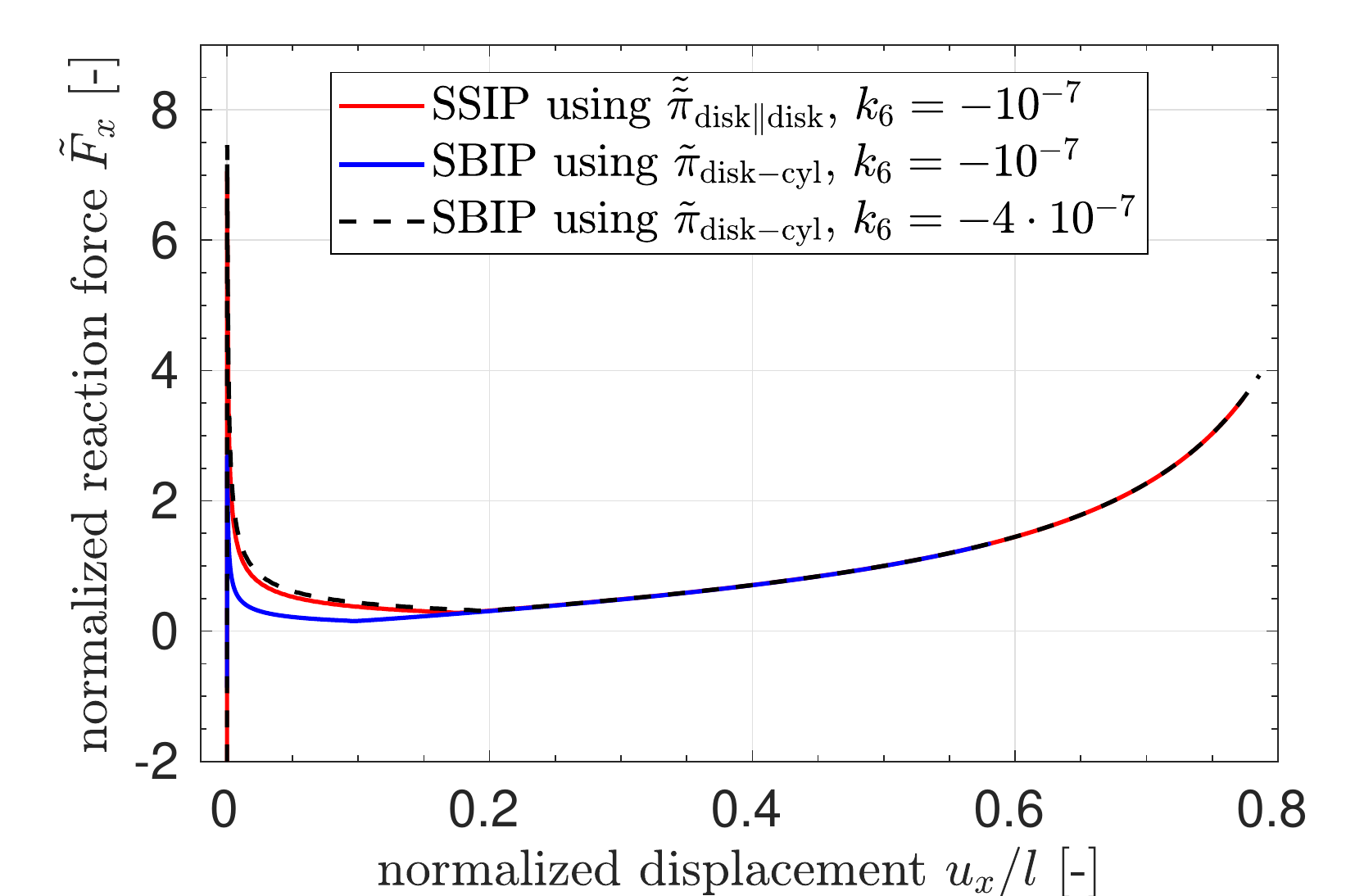}
   \label{fig::num_ex_vdW_twoparallelbeams_pulloff_SSIP_vs_SBIP_force_over_displacement}
  }
  \hspace{0.2cm}
  \subfigure[Detail view for small separations.]{
   \includegraphics[width=0.3\textwidth]{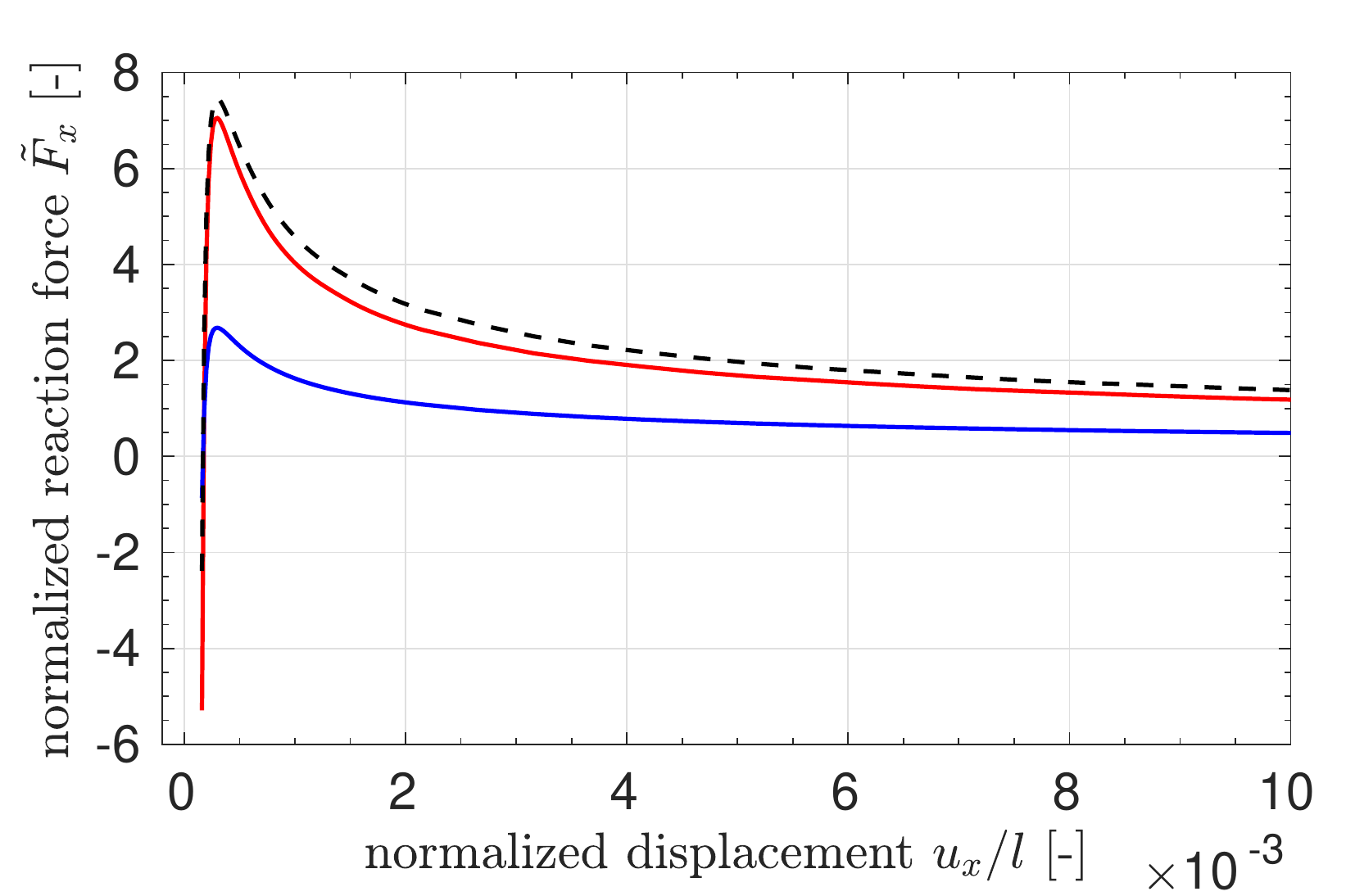}
   \label{fig::num_ex_vdW_twoparallelbeams_pulloff_SSIP_vs_SBIP_force_over_displacement_zoom}
  }
  \caption{Comparison of the results obtained in~\cite{GrillPeelingPulloff} via the previous SSIP law~$\tilde{\tilde{\pi}}_\text{m,disk$\parallel$disk}$ from \cite{GrillSSIP} with the results for the proposed SBIP law~$\tilde \pi_\text{m,disk-cyl}$ from~\eqref{eq::disk-cyl-pot_m} and corresponding approach from~\secref{sec::method_single_length_specific_integral}.}
  \label{fig::num_ex_vdW_twoparallelbeams_peeling_SSIP_vs_SBIP}
\end{figure}

At this point, recall from the prior analysis of the model accuracy using the example of two cylinders in \secref{sec::ia_pot_single_length_specific_evaluation_vdW} that the previously used simple SSIP law~$\tilde{\tilde{\pi}}_\text{m,disk$\parallel$disk}$ from Ref.~\cite{GrillSSIP} overestimates both the total interaction potential and force in the decisive regime of small separations and that the novel SBIP law~$\tilde \pi_\text{m,disk-cyl}$ from~\eqref{eq::disk-cyl-pot_m} on the contrary ensures the correct scaling behavior and is thus significantly more accurate.
The influence of this model error on the global system behavior shall be briefly investigated by means of the numerical peeling and pull-off experiment considered in this section.
\figref{fig::num_ex_vdW_twoparallelbeams_peeling_SSIP_vs_SBIP} shows the corresponding force-displacement curves for the original set of LJ parameters~$k_6=-10^{-7}$, $k_{12}=5\times10^{-25}$ from \cite{GrillPeelingPulloff}, using either the previous SSIP law~$\tilde \pi_\text{m,disk-cyl}$ (red solid line) or the proposed SBIP law~$\tilde \pi_\text{m,disk-cyl}$ from~\eqref{eq::disk-cyl-pot_m} and corresponding approach from~\secref{sec::method_single_length_specific_integral} (blue solid line).
The results confirm the assessment summarized above because the peak force value~$\tilde F_{x\text{,max}}$ is overestimated by a factor of approximately 2.6 and the maximum displacement value before snapping free~$u_{x\text{,max}}/l$ is overestimated by a factor of approximately 1.3 if using the simple SSIP law from the authors' previous contribution \cite{GrillSSIP}.
On the other hand, the location of the force peak at $u_x/l \approx 3.0 \times 10^{-4}$, which is determined by the ratio of adhesive and repulsive contributions, is identical for both approaches.
Most importantly, however, is that the qualitative shape of the force-displacement curve is very similar for both approaches and indeed almost the identical force-displacement curve can be obtained from the consistent SBIP law~$\tilde \pi_\text{m,disk-cyl}$ if a different parameter value set $k_6=-4\times 10^{-7}$, $k_{12}=2\times10^{-24}$ is used (black dashed line)\footnote{These calibrated parameter values have been determined by trial and error.}.
This confirms the argument given in \cite{GrillSSIP} that the simple SSIP law~$\tilde{\tilde{\pi}}_\text{m,disk$\parallel$disk}$ can be calibrated to compensate for the overestimation of the interaction potential in the small yet decisive range of separations around the equilibrium separation of the LJ potential.
In this manner, the simple SSIP law~$\tilde{\tilde{\pi}}_\text{m,disk$\parallel$disk}$ yields results, which -- in good approximation -- agree with the consistent SBIP law on the system level, e.g. the reaction force-displacement curve studied in this numerical example.

After the accuracy, let us now briefly compare the efficiency of both approaches, which has been discussed on the theoretical level of algorithmic complexity in~\secref{sec::algorithm_complexity_fullint_vs_SSIP_vs_SBIP}, by looking at the computational cost for one simulation run of the numerical example considered here.
Keeping in mind the general limitations of such a simple performance comparison and especially the unoptimized implementation of both approaches, we found that the SBIP approach is approximately a factor of 3.8 faster than the SSIP approach%
\footnote{%
Average computation time per nonlinear iteration for each of the approximately $1.7\times10^4$ iterations of the full run of either ``SSIP using $\tilde{\tilde{\pi}}_{\mathrm{disk}\parallel\mathrm{disk}}$, $k_\mathrm{6}=-10^{-7}$'' or ``SBIP using $\tilde \pi_\mathrm{disk-cyl}$, $k_\mathrm{6}=-4 \cdot 10^{-7}$'' from \figref{fig::num_ex_vdW_twoparallelbeams_peeling_SSIP_vs_SBIP}. Apart from these two values identical set of parameters, identical code framework, build, system environment, number of processors including parallel distribution and hardware.
}.
The advantage of the SBIP approach that less integration points will be required for the same level of numerical integration error due to the smaller exponent of the reduced interaction law as argued in~\secref{sec::algorithm_complexity_fullint_vs_SSIP_vs_SBIP} has not been exploited in this comparison to isolate the net effect of the decreased algorithmic complexity of the approaches on the one hand and increased complexity of the reduced interaction law on the other hand.
It is thus confirmed by this numerical experiment that the novel SBIP approach in combination with the proposed disk-cylinder interaction law~$\tilde \pi_\text{m,disk-cyl}$ is significantly more efficient than the previous SSIP approach for the case of short-ranged interactions.

\subsection{Adhesive nanofiber-grafted surfaces}\label{sec::num_ex_adhesive_surfaces}
This set of numerical examples deals with bioinspired adhesive surfaces, which have recently gained a lot of attention (see e.g.~\cite{Brodoceanu2016} for a review).
This topic is motivated by the fascinating skills of geckos \cite{Autumn2000}, spiders \cite{Kesel2003}, mussels \cite{Lee2007}, and other animals to stick to surfaces and defy significant detachment forces such as for instance their body weight when sitting on the ceiling or steeply inclined surfaces.
The origin of this extraordinary adhesion is known to lie on the molecular scale and vdW forces between hierarchical arrays of nano-hairs and the surface have been shown to play a major role \cite{Autumn2002,Gao2005}.

In this context, the following computational study suggests and analyzes a possible design for artificial adhesive surfaces based on grafted helical nanofibers.
Hereby, nature's ubiquitous pattern of helical fibers is used to achieve the desired large ratio between strong adhesion under load on the one hand and easy removal of surfaces on the other hand.
This numerical example has been studied in full detail in \cite{GrillDiss} and only a selection of results will be presented in the following.
These results aim to demonstrate the capability of the novel SBIP approach to handle arbitrary mutual configurations of fibers in 3D by means of an initially curved spatial geometry as well as large, non-trivial deformations of fibers.
Also the efficiency of the novel approach will be showcased by applying it to a large-scale system regarding both the number of fibers and time steps.

\subsubsection{Setup and parameters}\label{sec::num_ex_adhesive_surfaces_setup_params}
\begin{figure}[htpb]%
  \centering
  \begin{minipage}{0.5\textwidth}
    \subfigure[Perspective view with translucent top surface.]{
      \includegraphics[width=\textwidth]{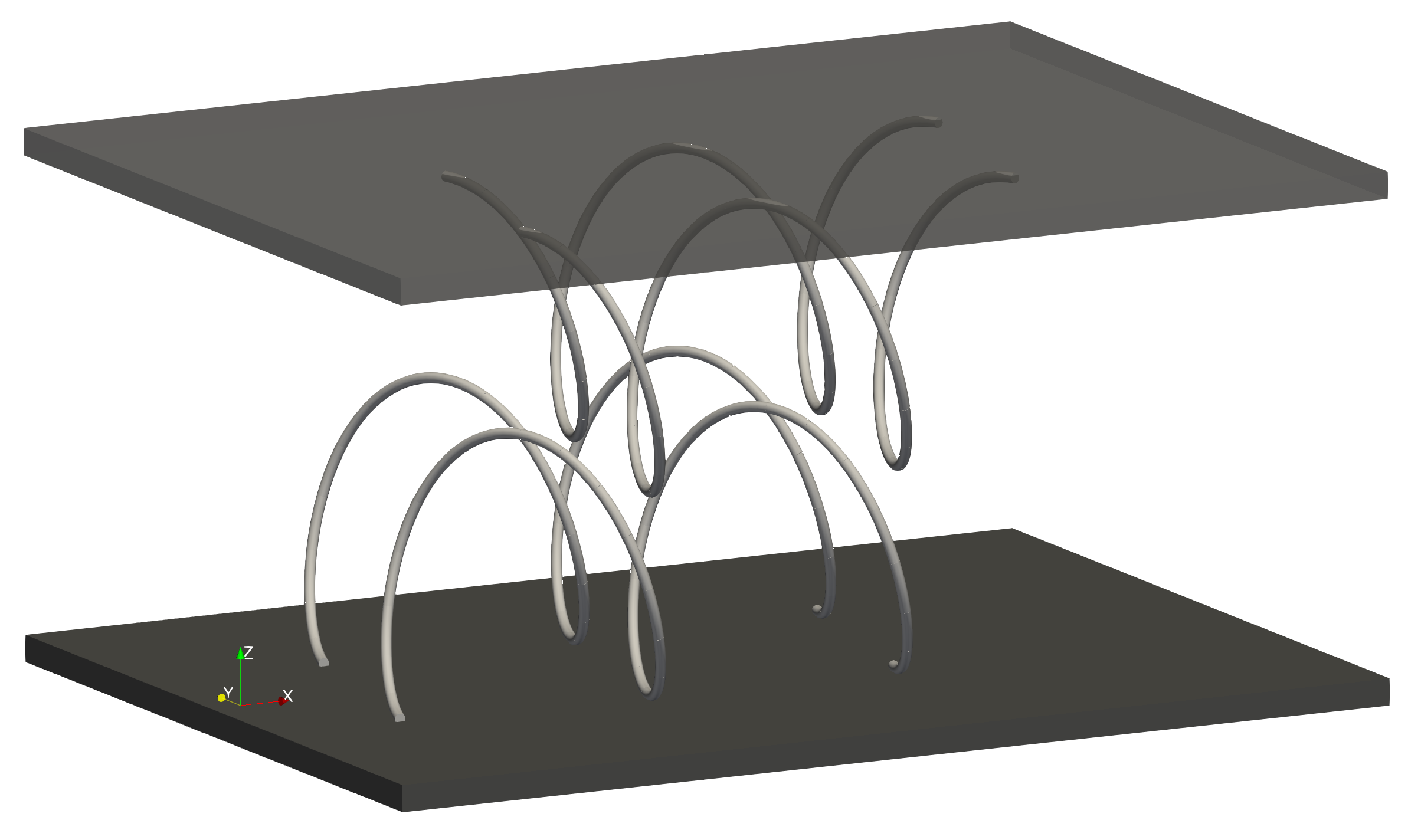}
      \label{fig::num_ex_adhesive_surfaces_initial_config_perspective}
    }
  \end{minipage}
  \begin{minipage}{0.25\textwidth}
  \centering
    \subfigure[View along $x$-axis.]{
      \includegraphics[width=0.75\textwidth]{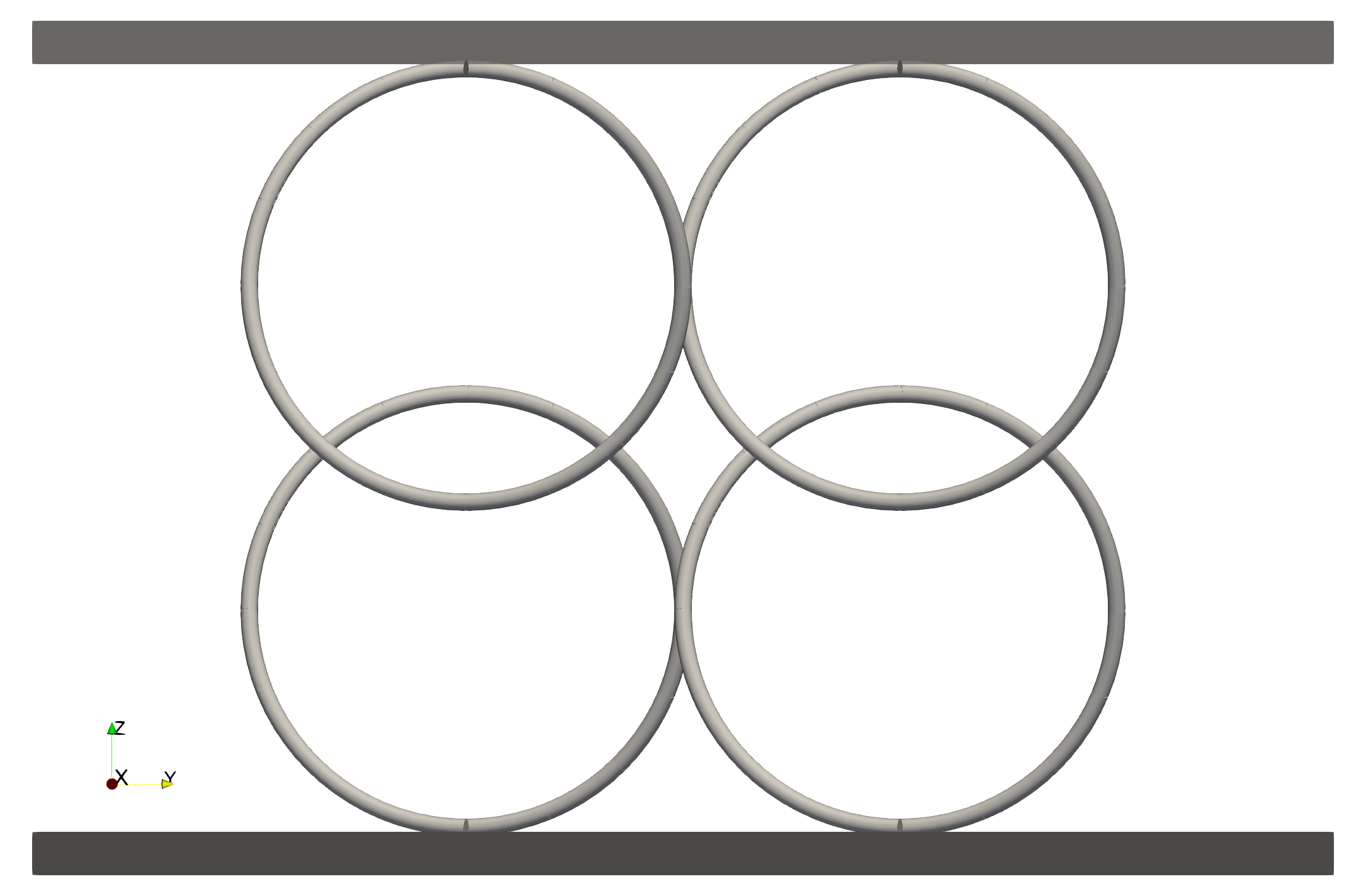}
      \label{fig::num_ex_adhesive_surfaces_initial_config_Xview}
    }
    \subfigure[View along $y$-axis.]{
      \includegraphics[width=\textwidth]{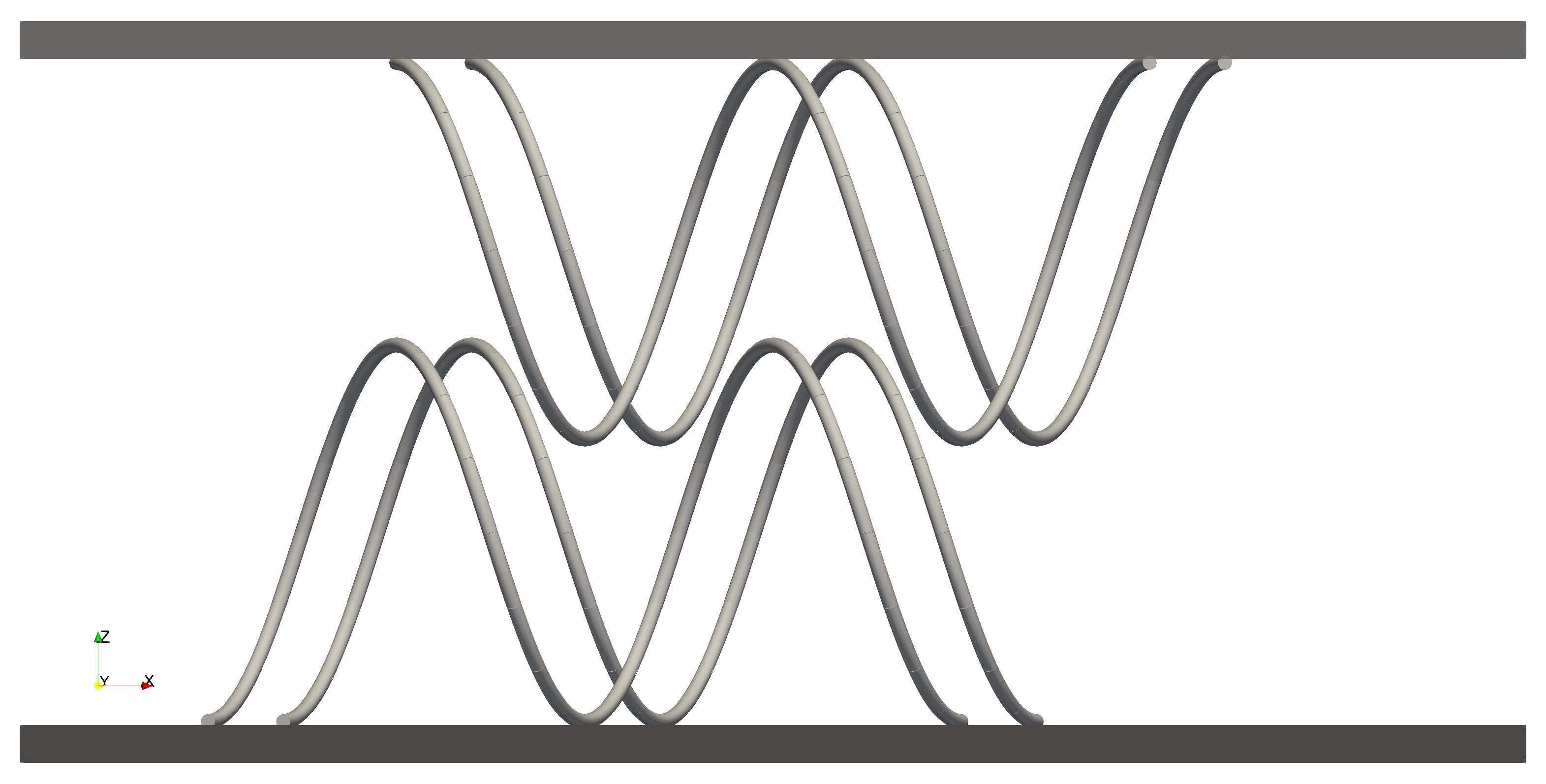}
      \label{fig::num_ex_adhesive_surfaces_initial_config_Yview}
    }
  \end{minipage}
  \caption{Initial, undeformed configuration of the numerical experiments studying adhesive nanofiber-grafted surfaces.}
  \label{fig::num_ex_adhesive_surfaces_setup}
\end{figure}

\figref{fig::num_ex_adhesive_surfaces_setup} shows the setup of this numerical example from different view angles.
As mentioned above, the system is mainly composed of helical fibers with a helix diameter of~$D_h=2R_h=1$ and a pitch, i.e., height of one complete turn, of~$P_h=1$, respectively.
The grafting onto surfaces is modeled by respective Dirichlet boundary conditions applied to the fibers and the dark colored surfaces shown in \figref{fig::num_ex_adhesive_surfaces_setup} are only used for visualization and not considered in the simulation.
In this first minimal setup, 2x2x2 fiber loops are considered, which refers to 2 surfaces with 2 helices each with 2 complete turns, i.e., loops each.
Two different scenarios will be studied.
In the normal loading, also referred to as "Pull" scenario, the surfaces will be moved together until the surfaces of the fibers touch and subsequently pulled apart without any rotation.
In the second, the ``Twist \& Pull'' scenario, the surfaces will be twisted by $75^\circ$ before they are being pulled apart.
The full specification of this numerical example including the time curves used for the prescribed boundary conditions can be found in Ref.~\cite{GrillDiss}.

\subsubsection{Results and discussion}
\begin{figure}[htpb]%
  \centering
  \vspace{-2.3cm}
  \subfigure[$t=2$, $u_z=-u_{z,\text{max}}$: minimum separation of surfaces]{
    \includegraphics[width=0.4\textwidth]{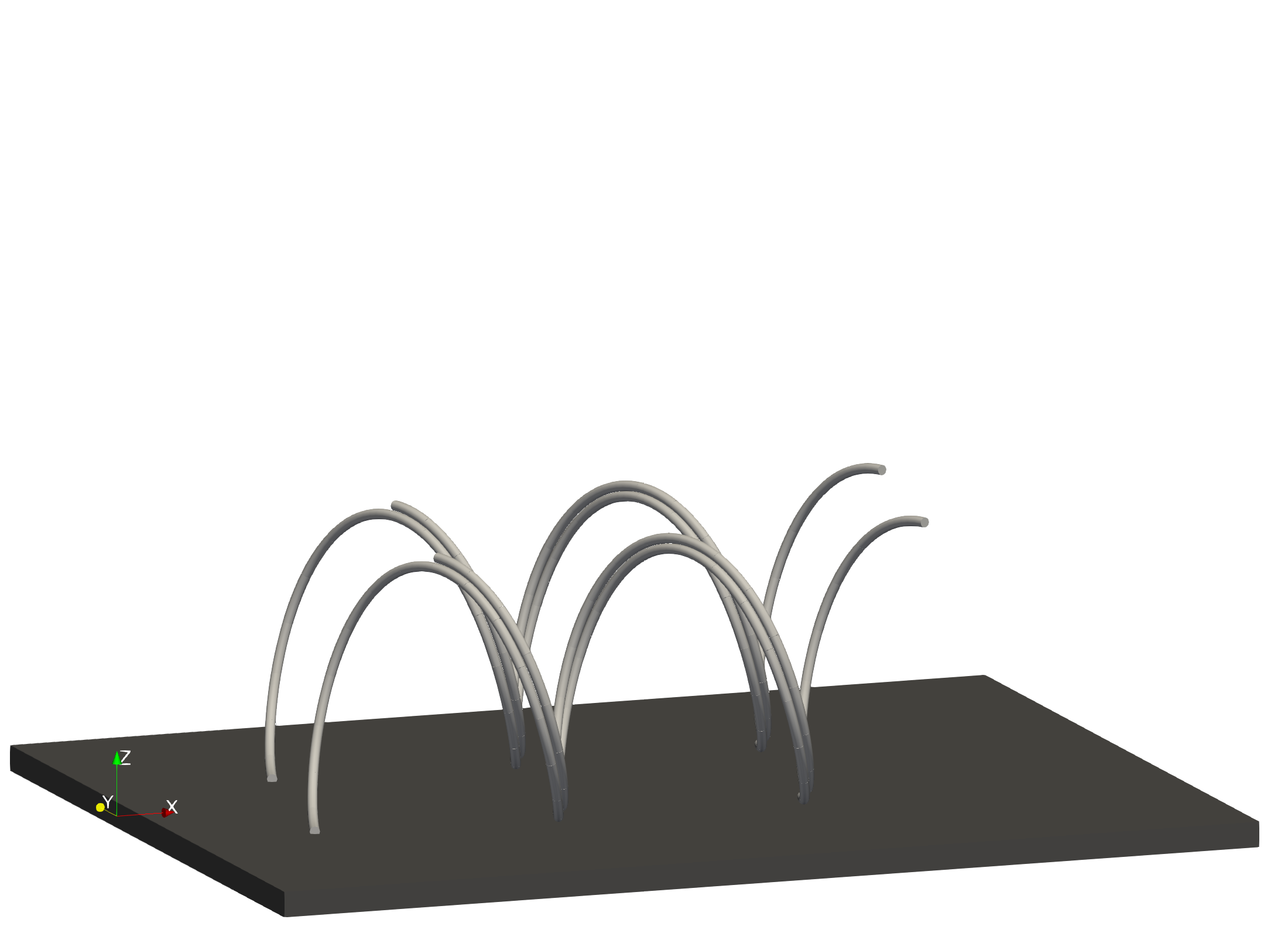}
    \label{fig::num_ex_adhesive_surfaces_simulation_snapshots_Pull_time2_step03531}
  }
  \vspace{-0.5cm}
  \subfigure[$t=2.5$, $u_z=-0.5 \, u_{z,\text{max}}$: early stage of pulling surfaces apart]{
    \includegraphics[width=0.4\textwidth]{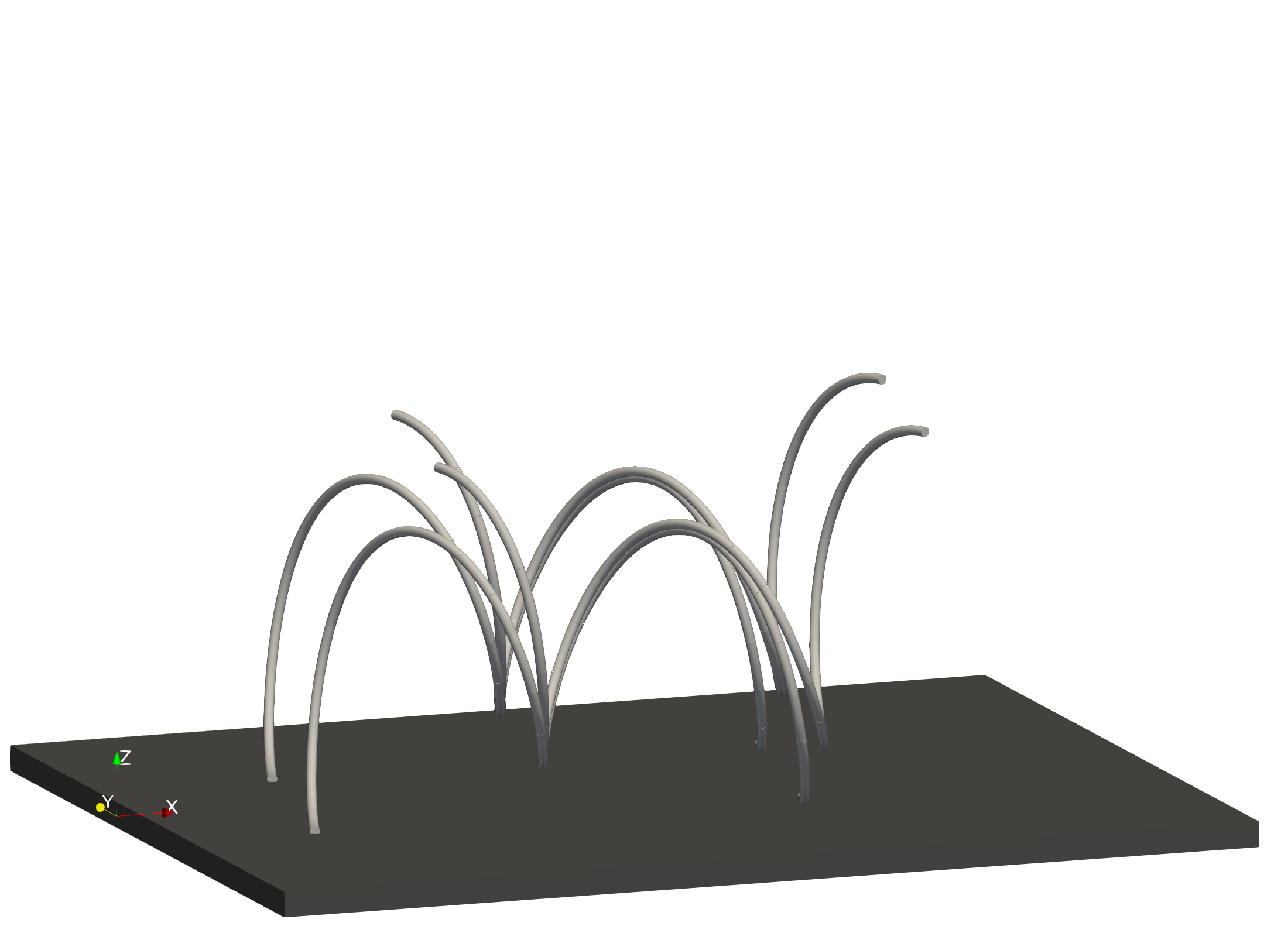}
    \label{fig::num_ex_adhesive_surfaces_simulation_snapshots_Pull_time2_5_step0946}
  }
  \subfigure[$t=3.5$, $u_z=0.5 \, u_{z,\text{max}}$: intermediate stage of pulling surfaces apart]{
    \includegraphics[width=0.4\textwidth]{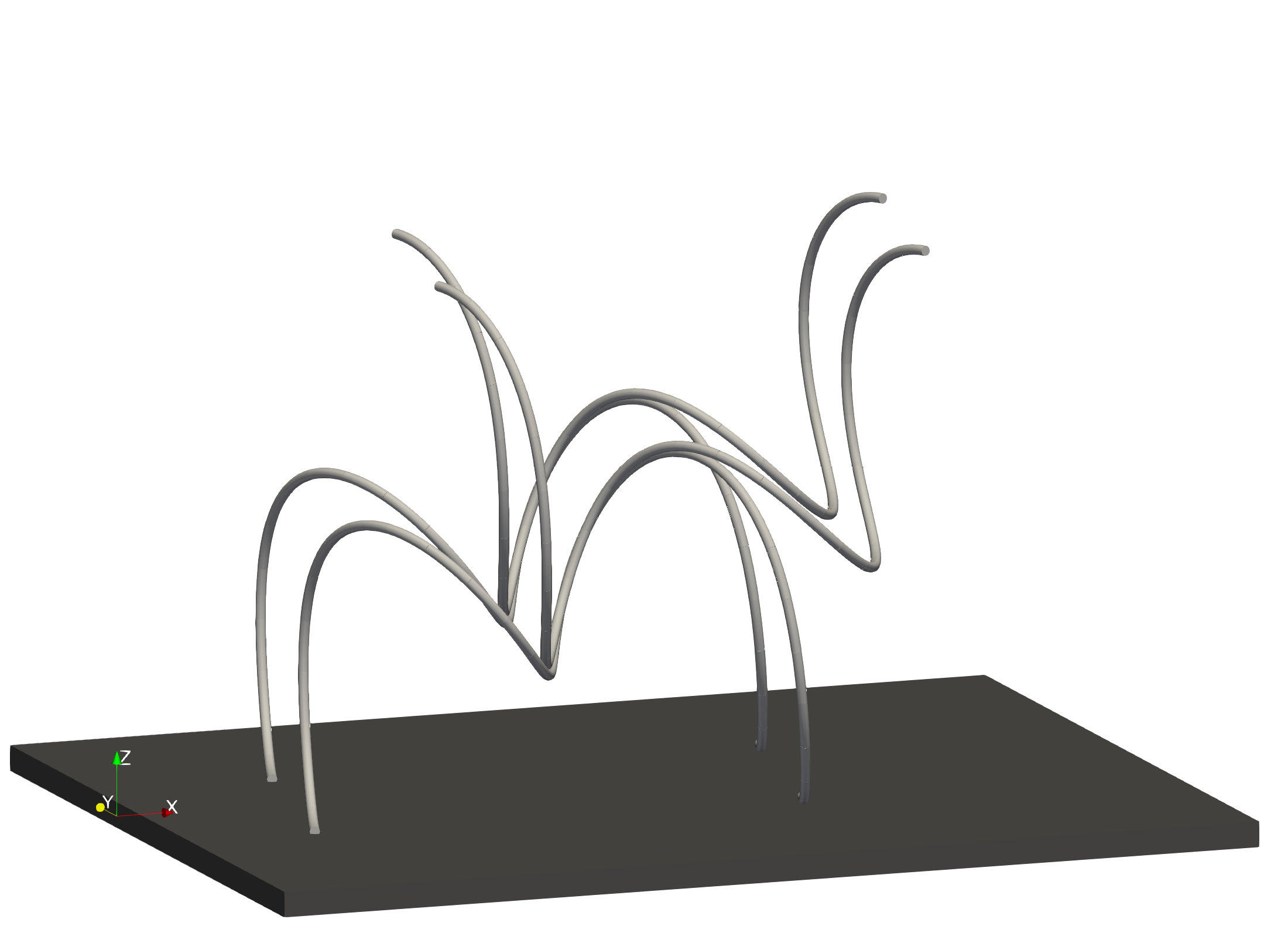}
    \label{fig::num_ex_adhesive_surfaces_simulation_snapshots_Pull_time3_5_step1052}
  }
  \subfigure[$t=4.09$, $u_z=1.09 \, u_{z,\text{max}}$: detachment of the last fiber points]{
    \includegraphics[width=0.4\textwidth]{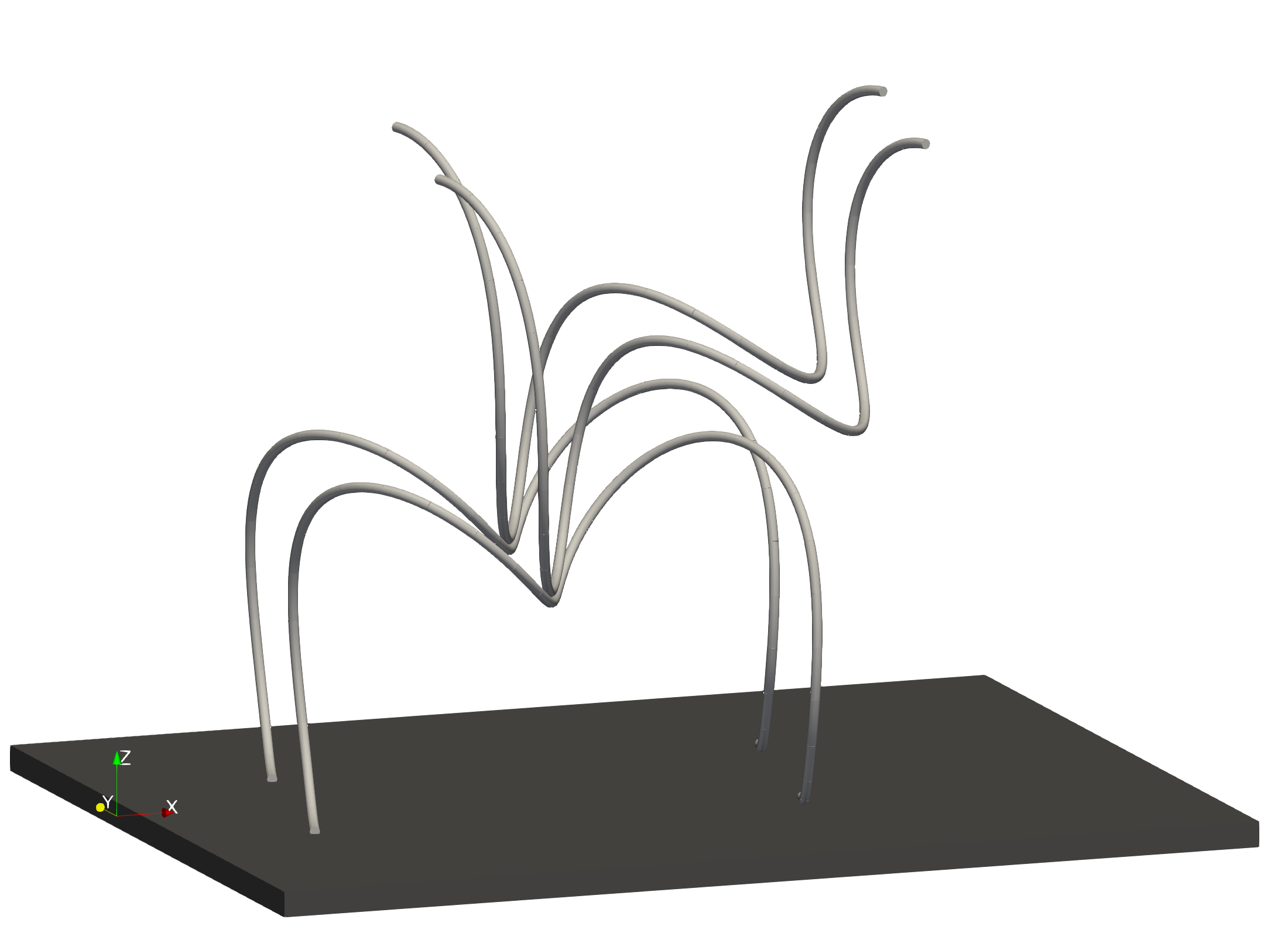}
    \label{fig::num_ex_adhesive_surfaces_simulation_snapshots_Pull_time4_09_step1136}
  }
  \caption{Sequence of simulation snapshots of the ``Pull'' scenario. Top surface is hidden for better visibility of the fibers.}
  \label{fig::num_ex_adhesive_surfaces_simulation_snapshots_2x2x2_Pull}
\end{figure}

\figref{fig::num_ex_adhesive_surfaces_simulation_snapshots_2x2x2_Pull} shows selected simulation snapshots for the normal loading referred to as ``Pull'' scenario.
The configuration in \figref{fig::num_ex_adhesive_surfaces_simulation_snapshots_Pull_time2_step03531} corresponds to the end of the approach (and equilibration) phase at~$t=2$, where the helical fibers of both surfaces are perfectly aligned and their surfaces touch.
In the subsequent pull-off phase shown in \figref{fig::num_ex_adhesive_surfaces_simulation_snapshots_Pull_time2_5_step0946} - \ref{fig::num_ex_adhesive_surfaces_simulation_snapshots_Pull_time4_09_step1136}, the fibers are continuously peeled and the contact length decreases while the fibers are strongly deformed.
Only after approximately~$t=4.09$, the last contact points of the fibers detach and the separation process is completed.

\begin{figure}[htpb]%
  \centering
  \vspace{-2.3cm}
  \subfigure[$t=2.5$, $\psi_z=18.75^\circ$: early stage of twisting surfaces]{
    \includegraphics[width=0.4\textwidth]{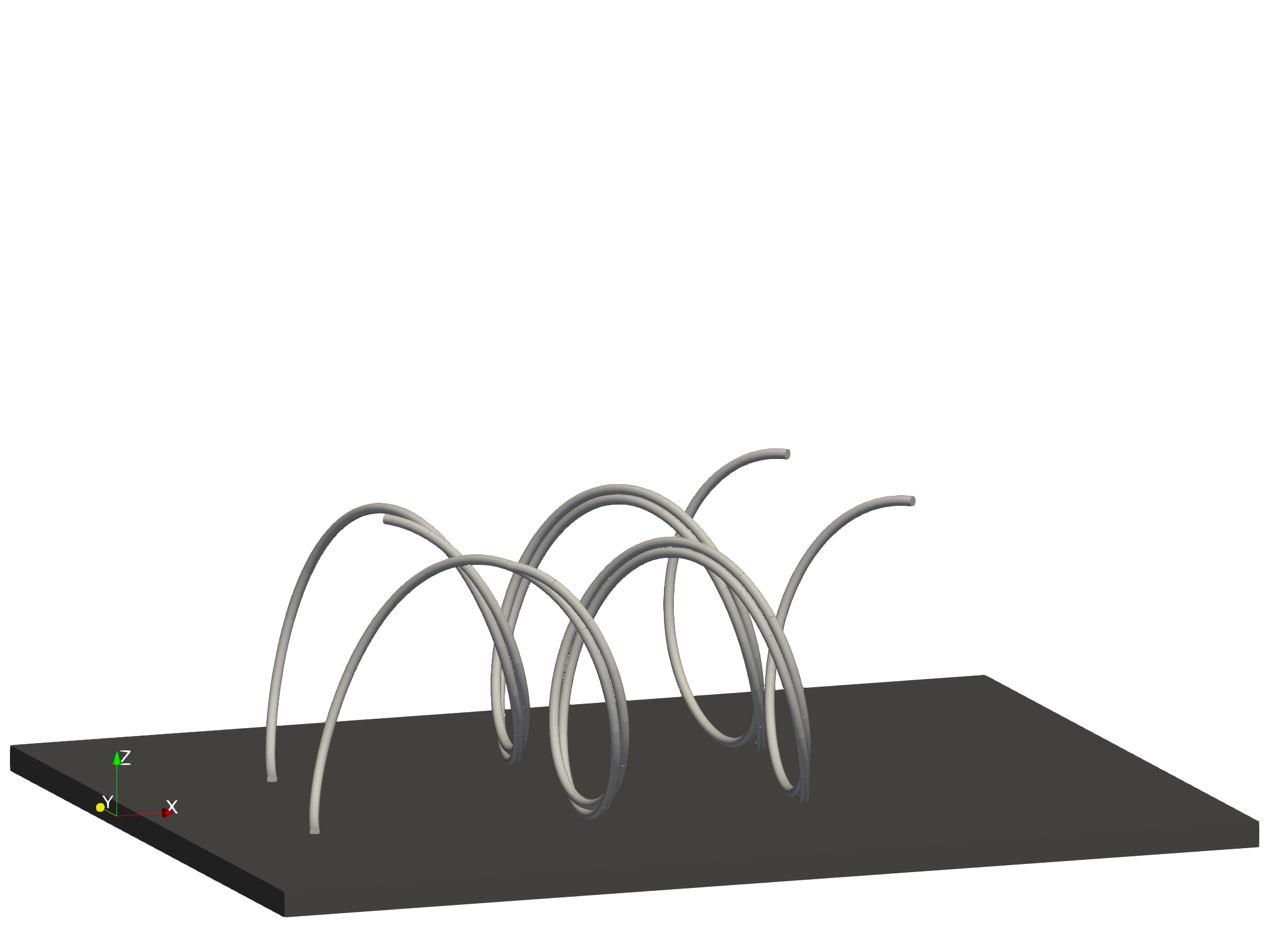}
    \label{fig::num_ex_adhesive_surfaces_simulation_snapshots_TwistAndPull_time2_5_step0868}
  }
  \vspace{-2.3cm}
  \subfigure[$t=3$, $\psi_z=37.5^\circ$: intermediate stage of twisting surfaces]{
    \includegraphics[width=0.4\textwidth]{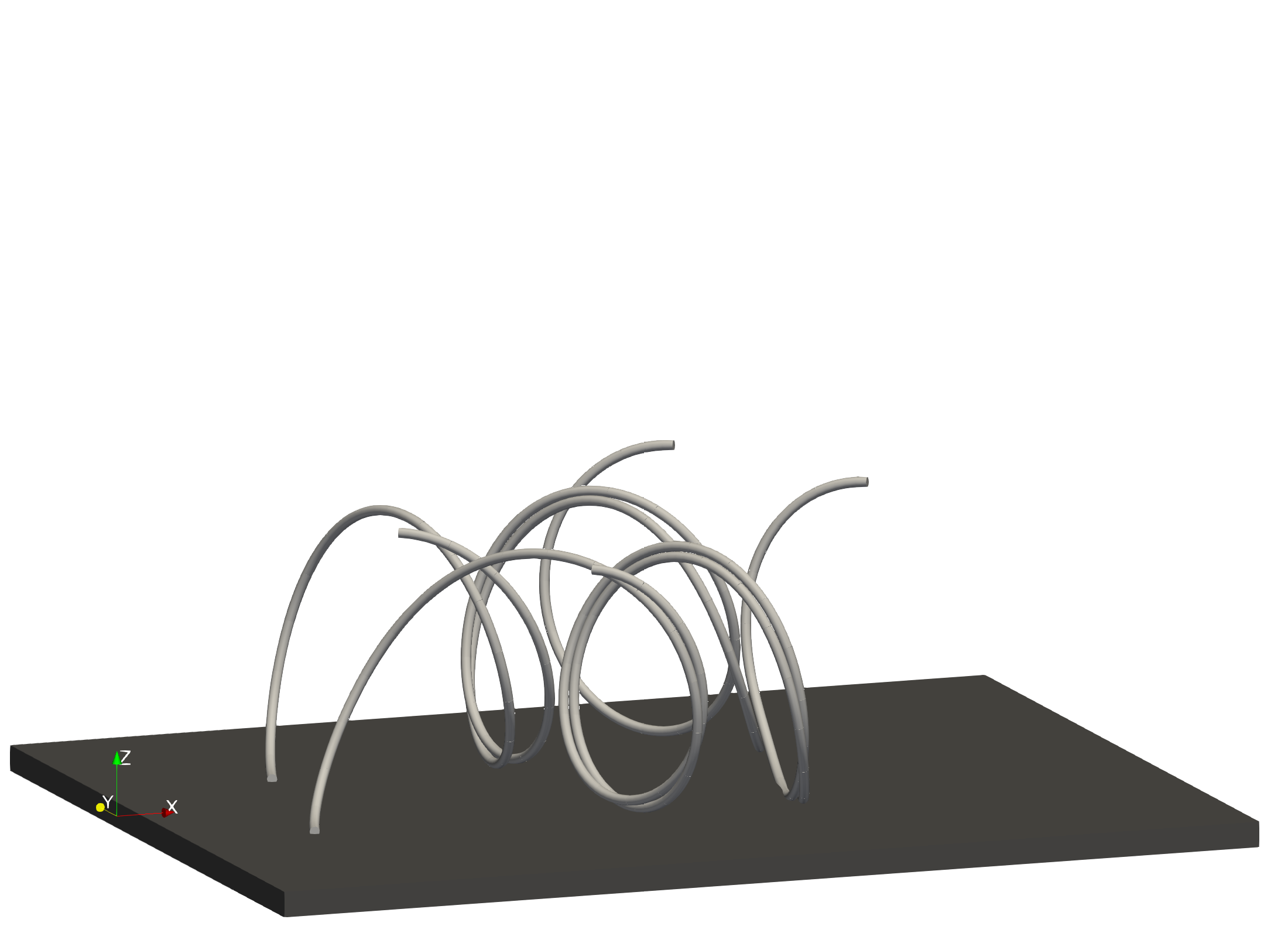}
    \label{fig::num_ex_adhesive_surfaces_simulation_snapshots_TwistAndPull_time3_step1092}
  }
  \vspace{-1cm}
  \subfigure[$t=3.5$, $\psi_z=56.25^\circ$: intermediate stage of twisting surfaces]{
    \includegraphics[width=0.4\textwidth]{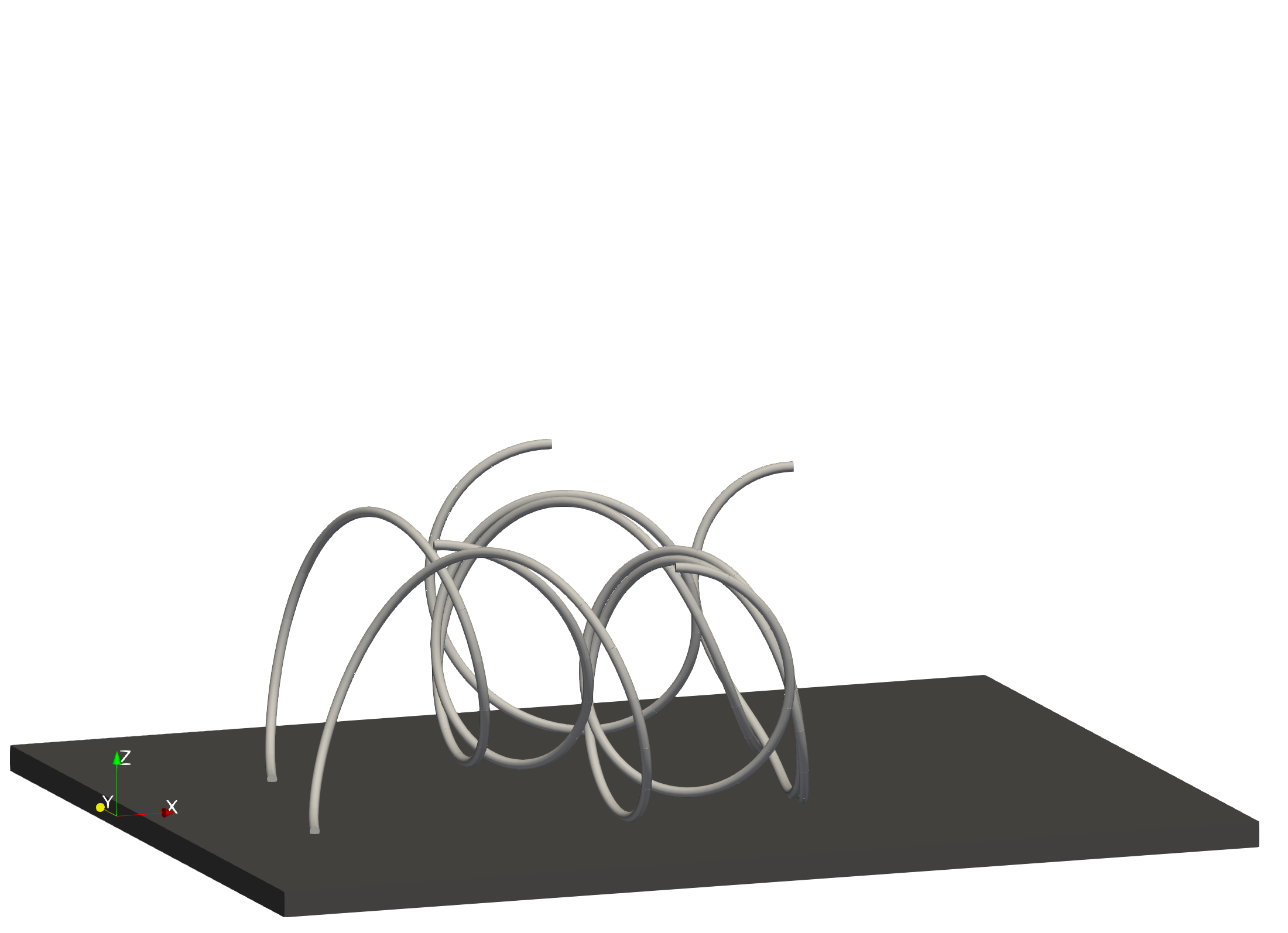}
    \label{fig::num_ex_adhesive_surfaces_simulation_snapshots_TwistAndPull_time3_5_step1355}
  }
  \subfigure[$t=4$, $\psi_z=75^\circ$: final twisted configuration]{
    \includegraphics[width=0.4\textwidth]{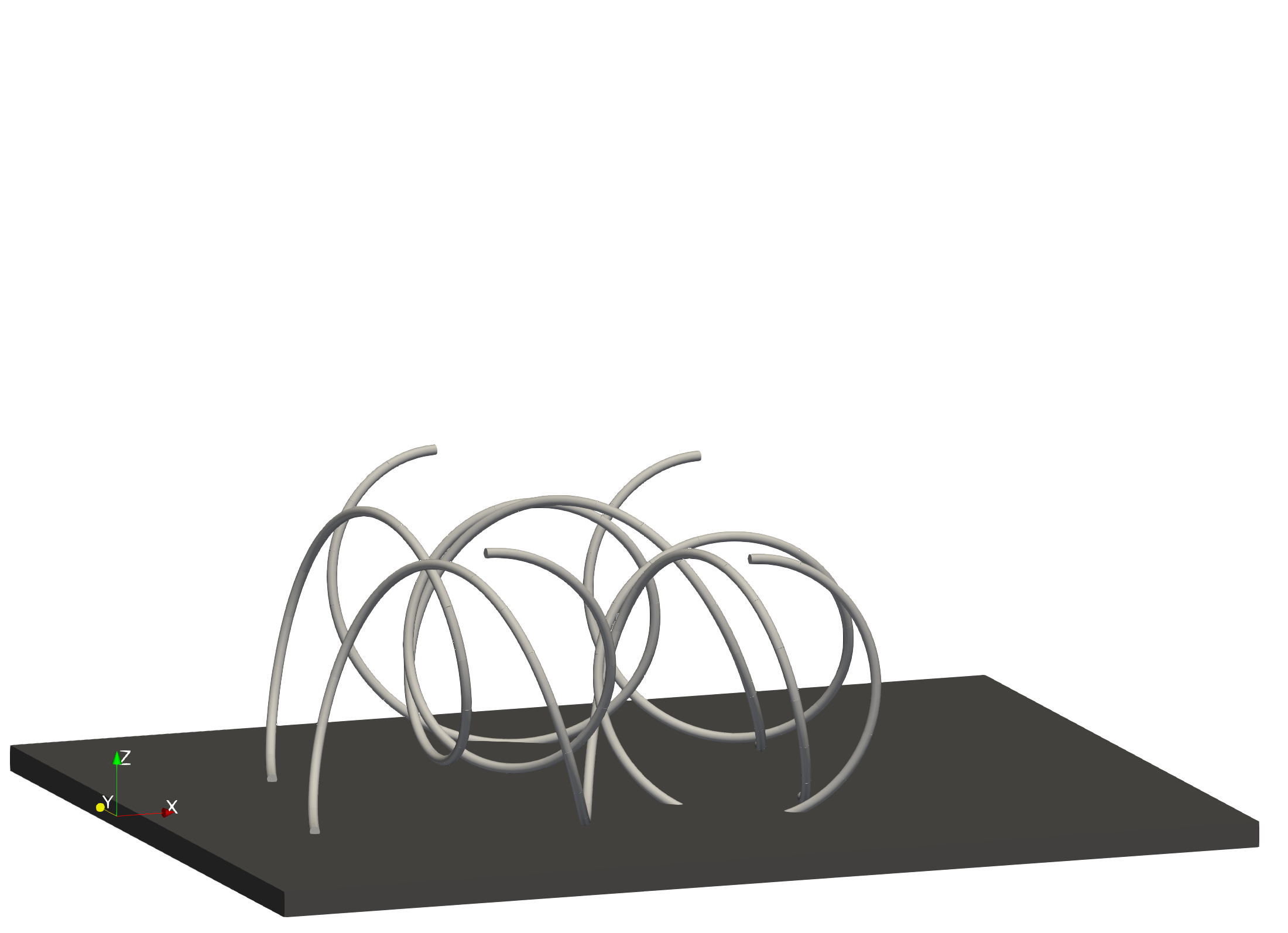}
    \label{fig::num_ex_adhesive_surfaces_simulation_snapshots_TwistAndPull_time4_step1672}
  }
  \subfigure[$t=5$, $u_z=0$: intermediate stage of pulling surfaces apart]{
    \includegraphics[width=0.4\textwidth]{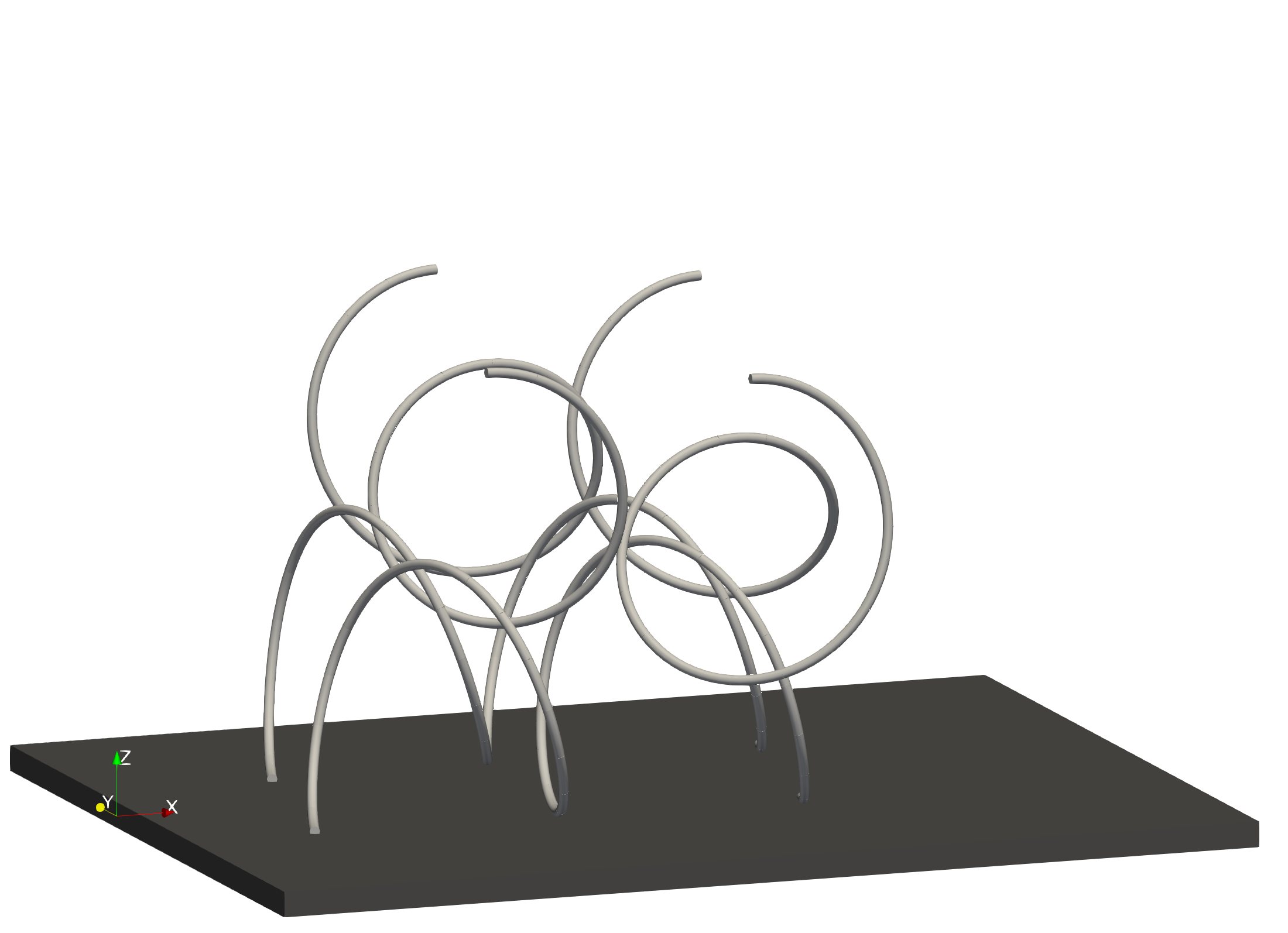}
    \label{fig::num_ex_adhesive_surfaces_simulation_snapshots_TwistAndPull_time5_step2007}
  }
  \subfigure[$t=5.72$, $u_z= 0.72 \, u_{z,\text{max}}$: detachment of the last fiber points]{
    \includegraphics[width=0.4\textwidth]{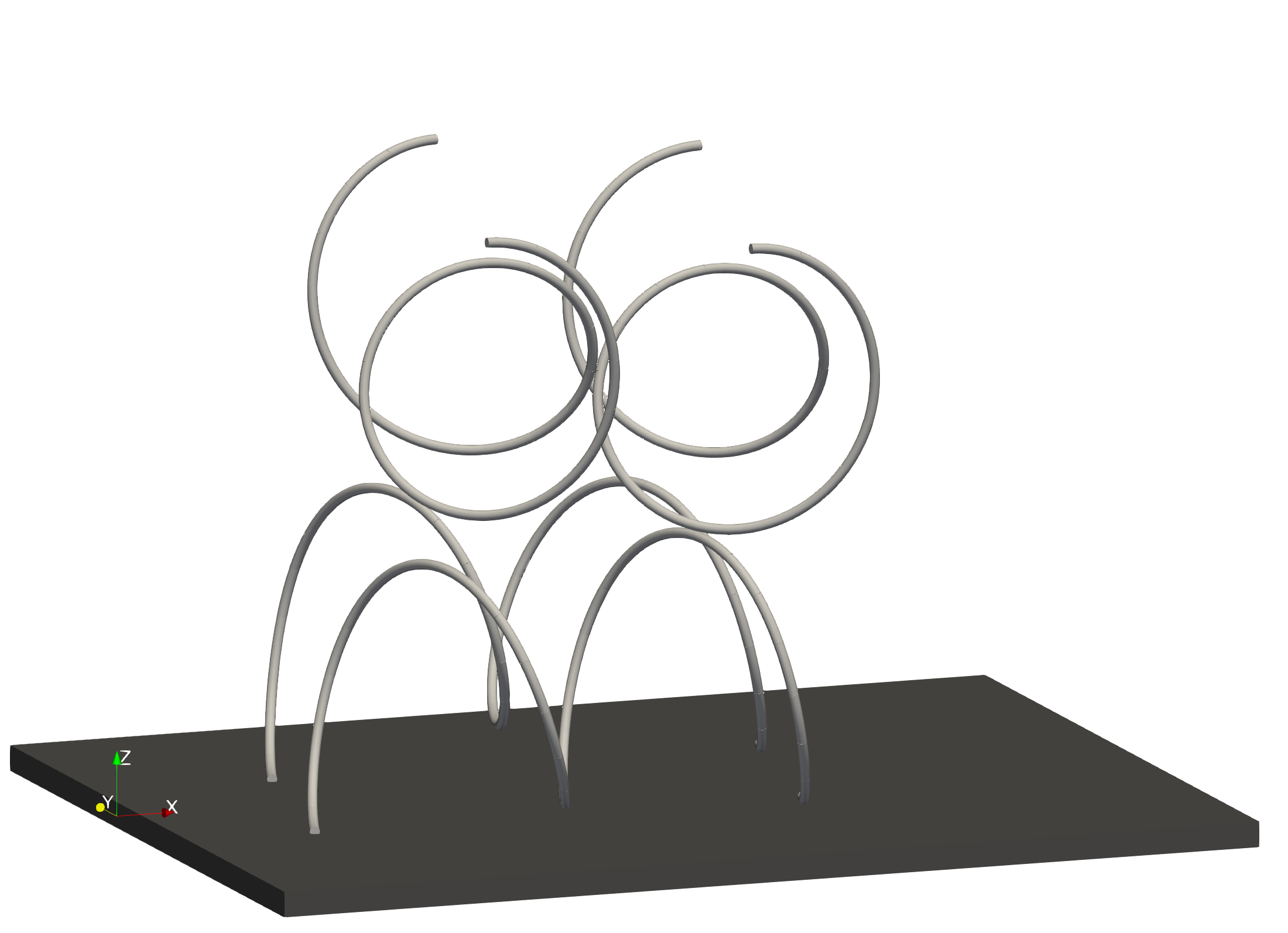}
    \label{fig::num_ex_adhesive_surfaces_simulation_snapshots_TwistAndPull_time5_72_step2262}
  }
  \caption{Sequence of simulation snapshots of the ``Twist \& Pull'' scenario. Top surface is hidden for better visibility of the fibers. Note that until time $t=2$, this scenario is identical to the ``Pull'' scenario shown in \figref{fig::num_ex_adhesive_surfaces_simulation_snapshots_2x2x2_Pull}.}
  \label{fig::num_ex_adhesive_surfaces_simulation_snapshots_2x2x2_Twistandpull}
\end{figure}

\figref{fig::num_ex_adhesive_surfaces_simulation_snapshots_2x2x2_Twistandpull} likewise shows selected simulation snapshots for the ``Twist \& Pull'' scenario, which differs from the ``Pull'' scenario in the phase~$t>2$.
\figref{fig::num_ex_adhesive_surfaces_simulation_snapshots_TwistAndPull_time2_5_step0868} - \ref{fig::num_ex_adhesive_surfaces_simulation_snapshots_TwistAndPull_time4_step1672} cover the twisting phase and show that twisting the surfaces by $75^\circ$ indeed separates the fibers almost entirely, which agrees with the idea of a subsequent, relatively easy separation of surfaces in normal direction.
A closer look reveals that the strength of adhesion suffices to keep the fibers in contact almost along the entire length up to approximately $37.5^\circ$, which leads to quite large and complex deformations of the fibers.
Note also that \figref{fig::num_ex_adhesive_surfaces_simulation_snapshots_TwistAndPull_time4_step1672} shows a limitation of our current setup in a way that the fibers can penetrate the surfaces, because the surfaces are not explicitly included in the simulation and only accounted for by means of Dirichlet boundary conditions on the grafted fiber endpoints as mentioned above.
This is however not expected to have a major influence on the pull-off forces to be discussed next, as it occurs rather rarely and from \figref{fig::num_ex_adhesive_surfaces_simulation_snapshots_TwistAndPull_time4_step1672} it can be estimated that a preclusion of the observed penetration would not change the overall deformed shape of the fibers in a significant manner.

\begin{figure}[htpb]%
  \centering
  \includegraphics[width=0.46\textwidth]{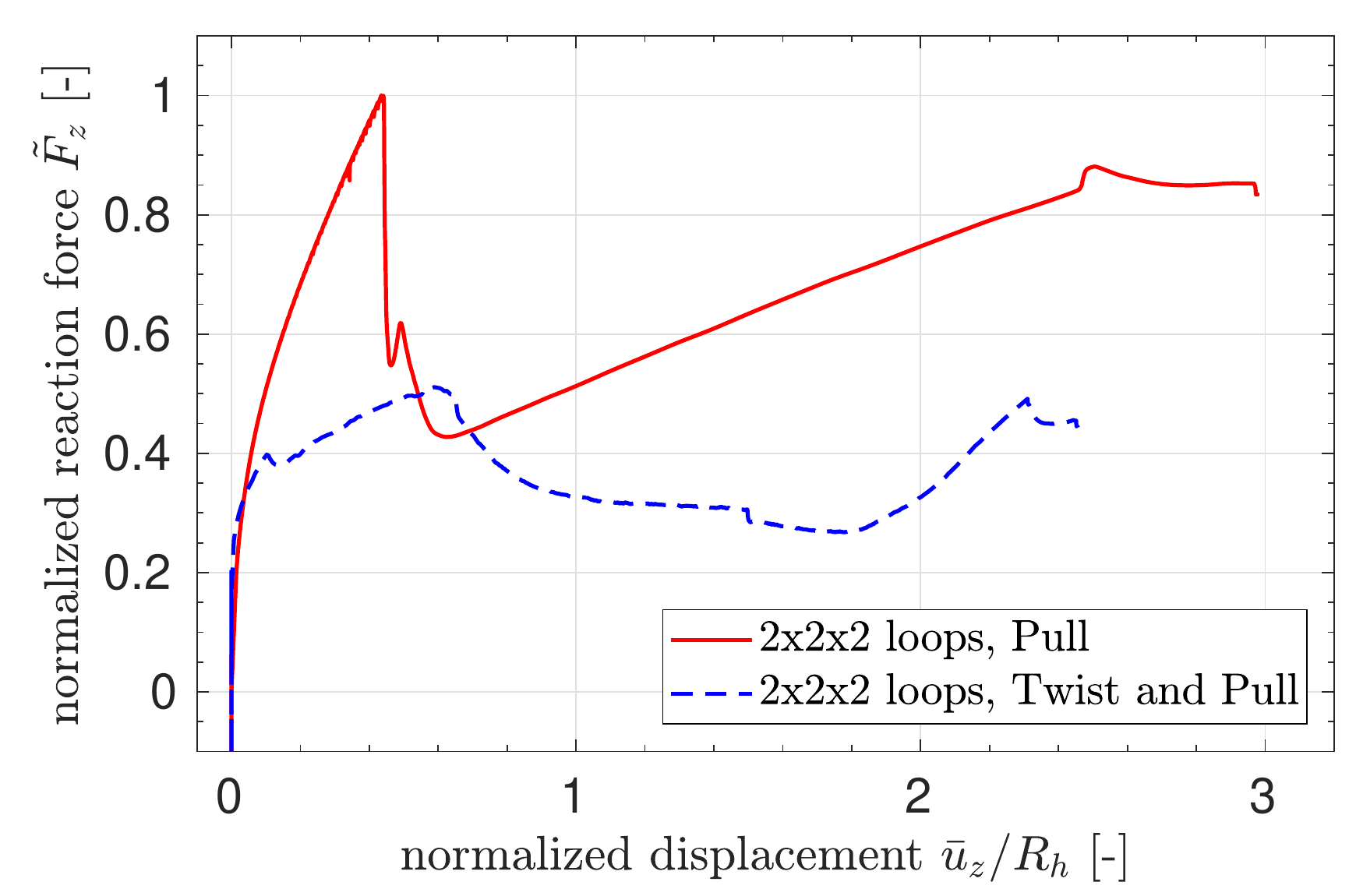}
  \caption{Force-displacement curves for the decisive phase, i.e., $2 \leq t < 5$ in the ``Pull'' scenario and $2 \leq t < 7$ in the ``Twist \& Pull'' scenario. Force values are normalized with respect to the maximal pull-off force~$F_{z,\text{max}}$.}
  \label{fig::num_ex_adhesive_surfaces_force_over_displacement}
\end{figure}

Finally, the force-displacement curves for both scenarios are shown in \figref{fig::num_ex_adhesive_surfaces_force_over_displacement}.
Specifically, the reaction force component~$F_z$ in the surface normal, i.e., $z$-direction, is summed over all nodes grafted onto the top surface and normalized by means of the maximum force~$F_{z,\text{max,2x2x2}} \approx 0.356$ that is observed for this 2x2x2 loop system.
As desired, the adhesive connection of surfaces is able to withstand approximately two times larger pull-off forces in the normal loading scenario (red line) as compared to the scenario meant to be used for the removal of surfaces (blue dashed line).
Generally, a rich and highly nonlinear system behavior can be observed from these force-displacement curves, which reflects the complex interplay of strongly adhesive LJ interactions and large and complex structural deformations in 3D.
Particularly for the ``Pull'' scenario, some characteristics of the fundamental problem of peeling two initially straight adhesive fibers (cf.~our previous article~\cite{GrillPeelingPulloff}) such as the sharp force peak, the extended pull-off phase and the sudden snap-free are however still recognizable.
Note also the considerable compressive forces that occur in the ``Twist \& Pull'' for $\bar u_z=0$, i.e., during the twisting phase~$2 \leq t < 7$.
In this context, it seems worth mentioning that the reaction forces in $x$- and $y$-direction are below~$0.1$ and thus more than ten times smaller than the maximal pull-off force observed here.

\subsubsection{Upscaling the computational experiment}\label{sec::num_ex_adhesive_surfaces_largescale}
Increasing the size of the system is important to judge the performance of the novel SBIP approach.
Therefore, a 2x8x8 and a 2x16x16 loop system will be considered now.
With respect to the original 2x2x2 loop system, this corresponds to a scaling factor of 16 and 64, respectively, which leads to a total of 2048 elements / 4112 nodes and 8192 elements / 16416 nodes, respectively.
Except for the system size, the setup and parameters stated in \secref{sec::num_ex_adhesive_surfaces_setup_params} remain unchanged.

\begin{figure}[htpb]%
  \centering
  \subfigure[$t=3.5$, $u_z=0.5 \, u_{z,\text{max}}$: maximum force]{
    \includegraphics[width=0.55\textwidth]{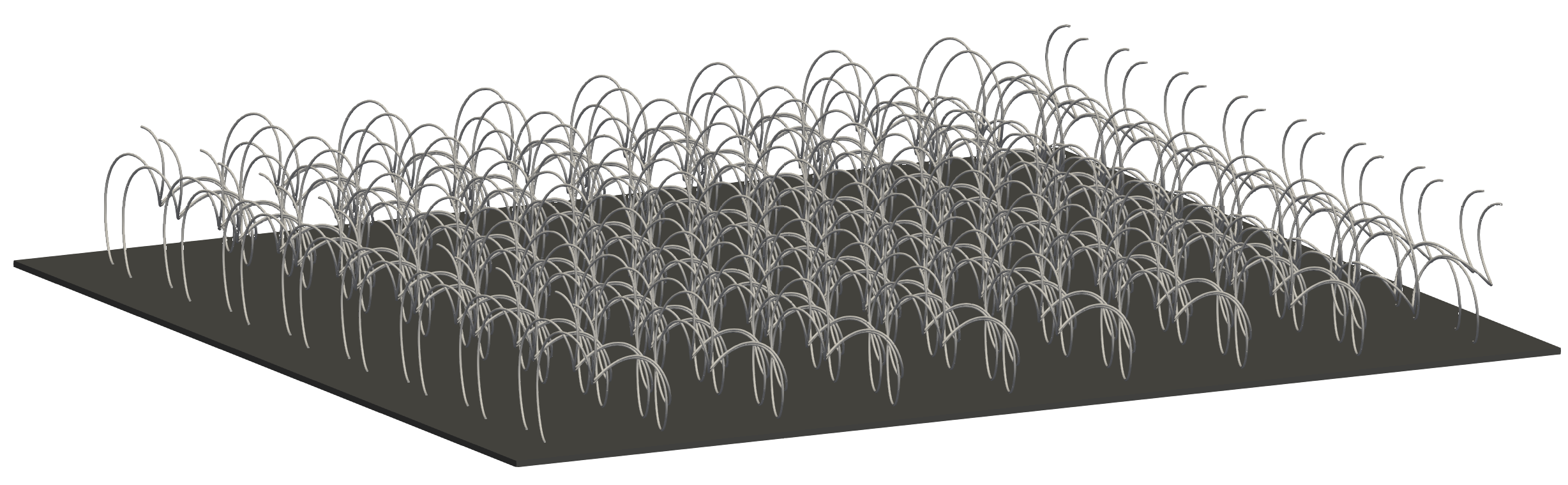}
    \label{fig::num_ex_adhesive_surfaces_simulation_snapshots_2x16x16loops_Pull_time3_5}
  }
  \subfigure[$t=3.5$: (view along $y$-axis)]{
    \includegraphics[width=0.4\textwidth]{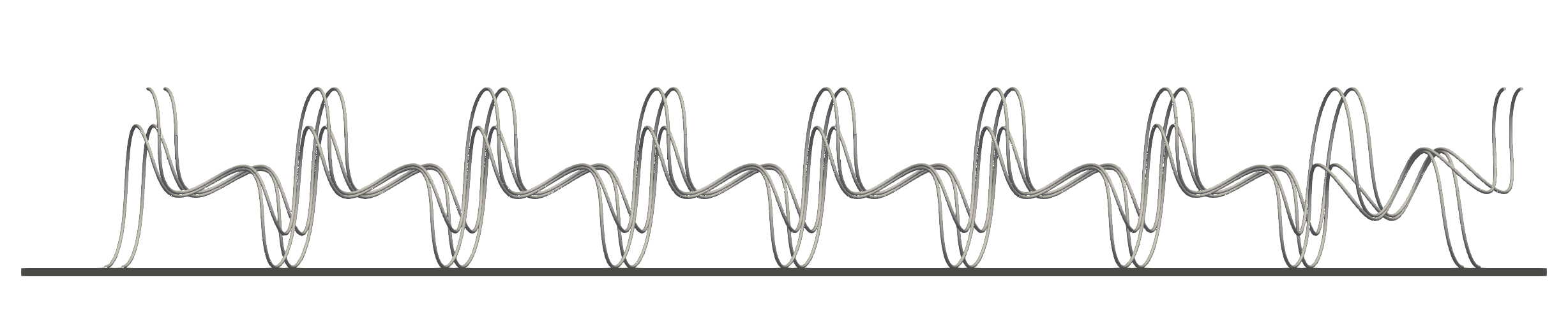}
    \label{fig::num_ex_adhesive_surfaces_simulation_snapshots_2x16x16loops_Pull_time3_5_Y-view}
  }
  \subfigure[$t=4.25$: detachment of the last fibers]{
    \includegraphics[width=0.55\textwidth]{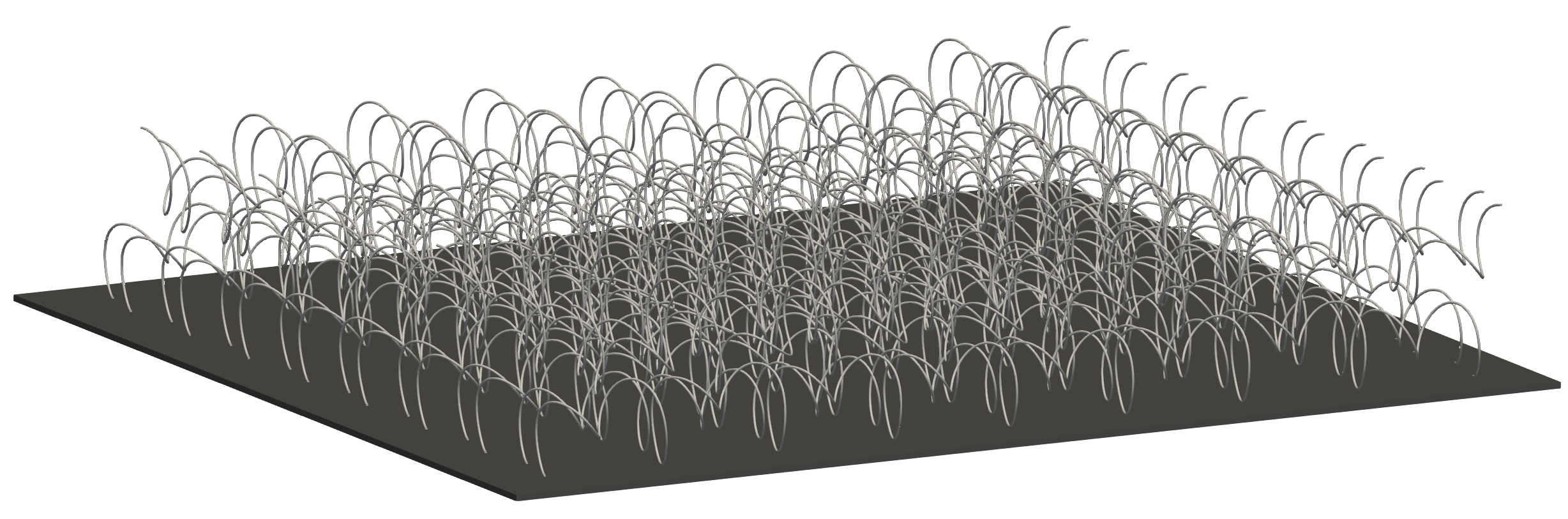}
    \label{fig::num_ex_adhesive_surfaces_simulation_snapshots_2x16x16loops_Pull_time4_25}
  }
  \subfigure[$t=4.25$ (view along $y$-axis)]{
    \includegraphics[width=0.4\textwidth]{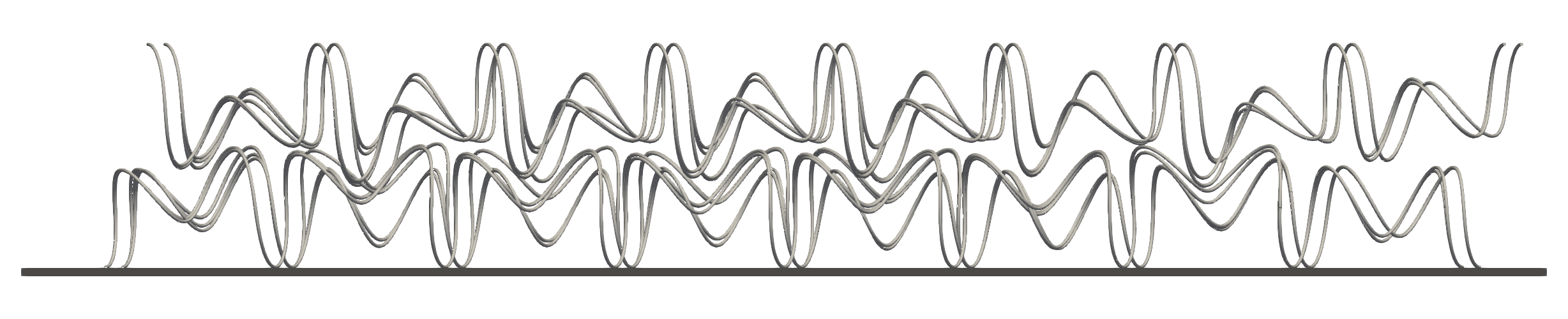}
    \label{fig::num_ex_adhesive_surfaces_simulation_snapshots_2x16x16loops_Pull_time4_25_Y-view}
  }
  \caption{Selected simulation snapshots of the ``Pull'' scenario with 2x16x16 loops. Top surface is hidden for better visibility of the fibers.}
  \label{fig::num_ex_adhesive_surfaces_simulation_snapshots_2x16x16_Pull}
\end{figure}

\figref{fig::num_ex_adhesive_surfaces_simulation_snapshots_2x16x16_Pull} shows selected simulation snapshots of the ``Pull'' scenario for the 2x16x16 loop system and confirms that the periodicity in the problem setup is also observable in the resulting deformed configurations.
Once again, refer to Ref.~\cite{GrillDiss} for a detailed investigation of the force-displacement curves obtained for these scaled system sizes.

\subsubsection{Entanglement of fibers}\label{sec::num_ex_adhesive_surfaces_entanglement}
\begin{figure}[htpb]%
  \centering
  \begin{minipage}{0.55\textwidth}
    \subfigure[$t=3$, $\psi_z=37.5^\circ$: intermediate stage of twisting surfaces]{
      \includegraphics[width=\textwidth]{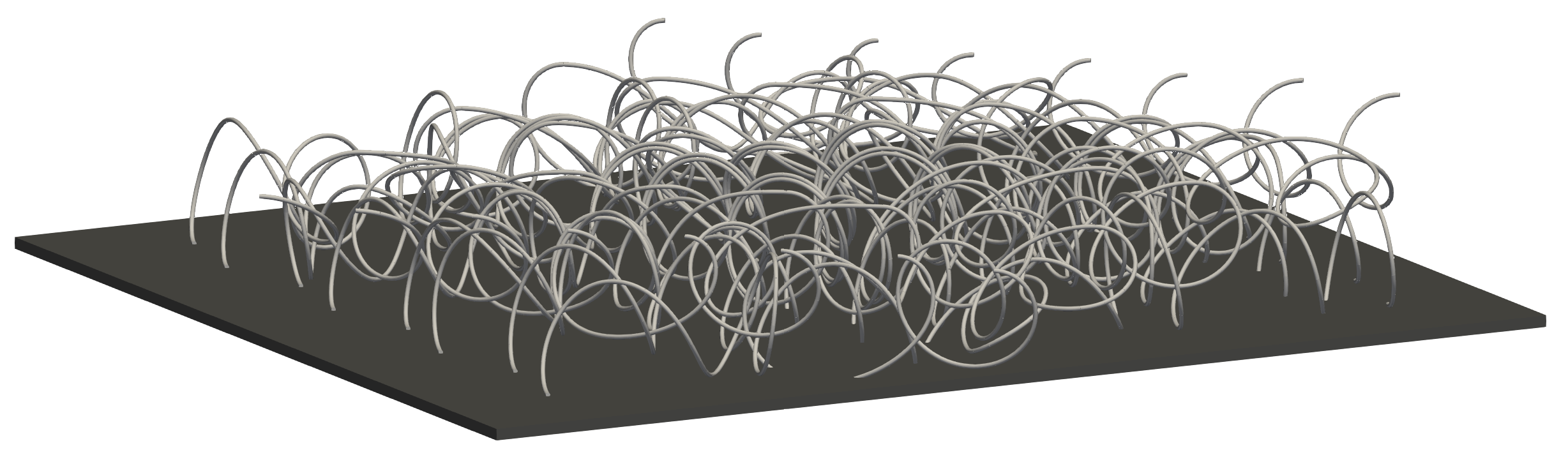}
      \label{fig::num_ex_adhesive_surfaces_simulation_snapshots_2x8x8_TwistAndPull_time3_step10754}
    }
    \subfigure[$t=4$, $\psi_z=75^\circ$: final twisted configuration]{
      \vspace{0.5cm}
      \includegraphics[width=\textwidth]{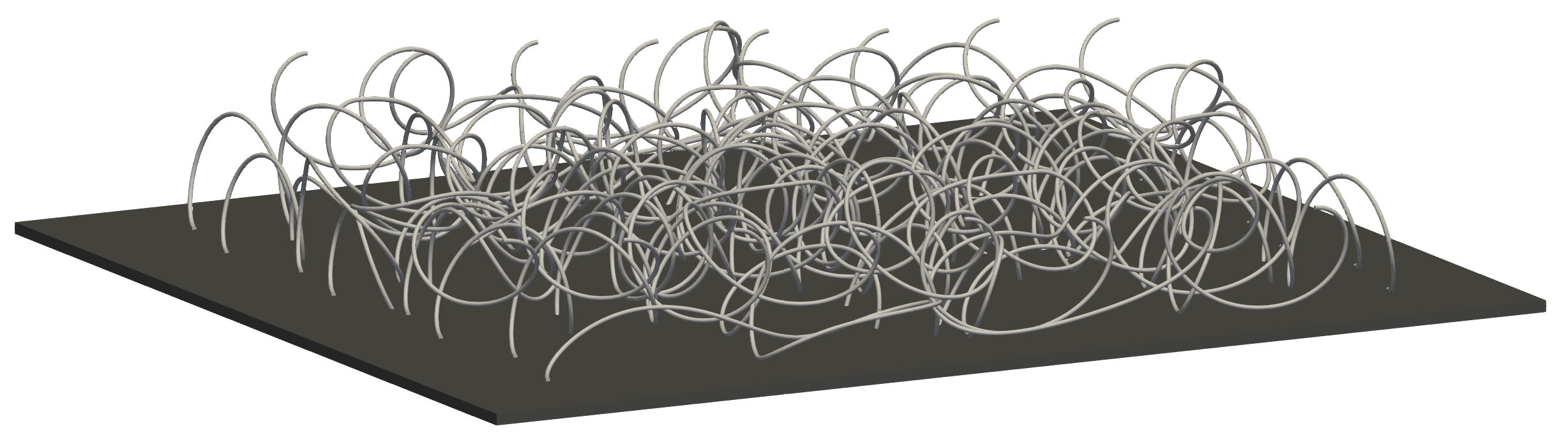}
      \label{fig::num_ex_adhesive_surfaces_simulation_snapshots_2x8x8_TwistAndPull_time4_step16969}
    }
    \subfigure[$t=5.5$: intermediate stage of pulling surfaces apart]{
      \vspace{0.5cm}
      \includegraphics[width=\textwidth]{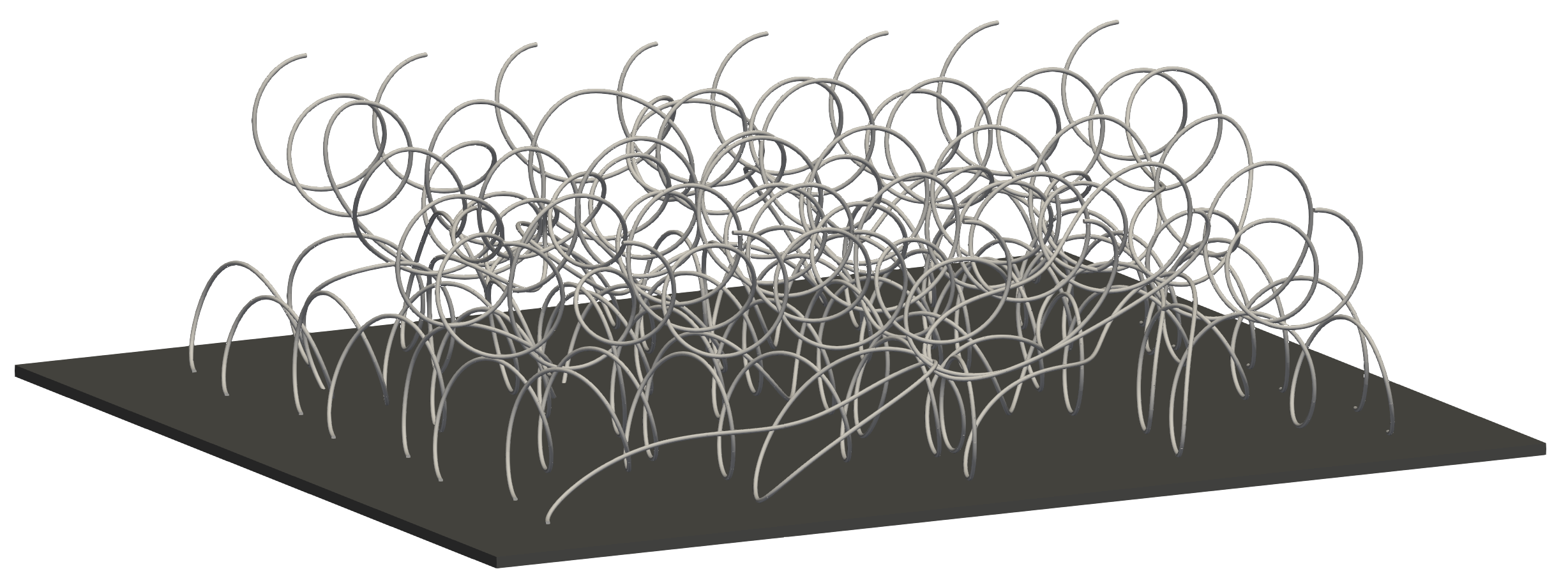}
      \label{fig::num_ex_adhesive_surfaces_simulation_snapshots_2x8x8_TwistAndPull_time5_5_step21949}
    }
    \subfigure[$t=7$: final configuration]{
      \includegraphics[width=\textwidth]{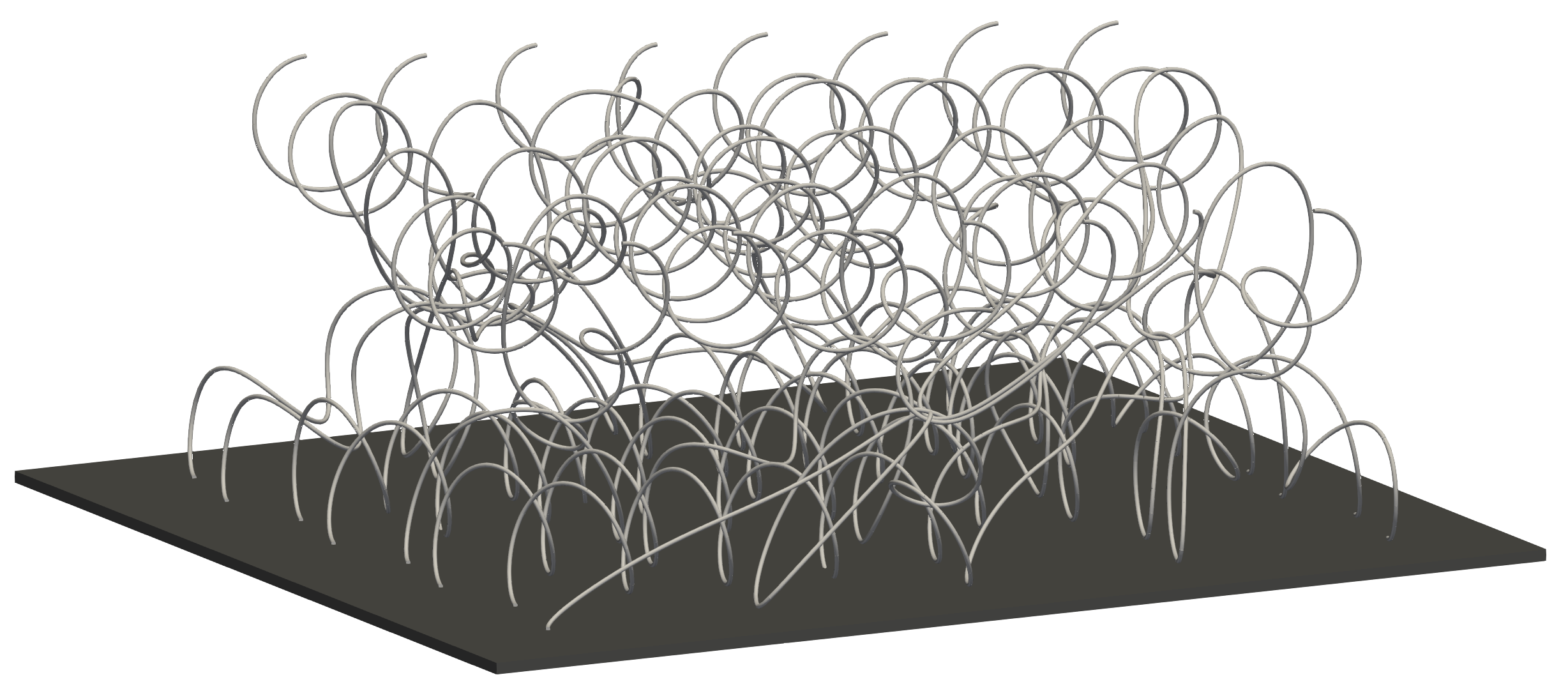}
      \label{fig::num_ex_adhesive_surfaces_simulation_snapshots_2x8x8_TwistAndPull_time7_step23516}
    }
  \end{minipage}
  \begin{minipage}{0.35\textwidth}
    \subfigure[$t=3$, $\psi_z=37.5^\circ$ (view from the top)]{
      \includegraphics[width=\textwidth]{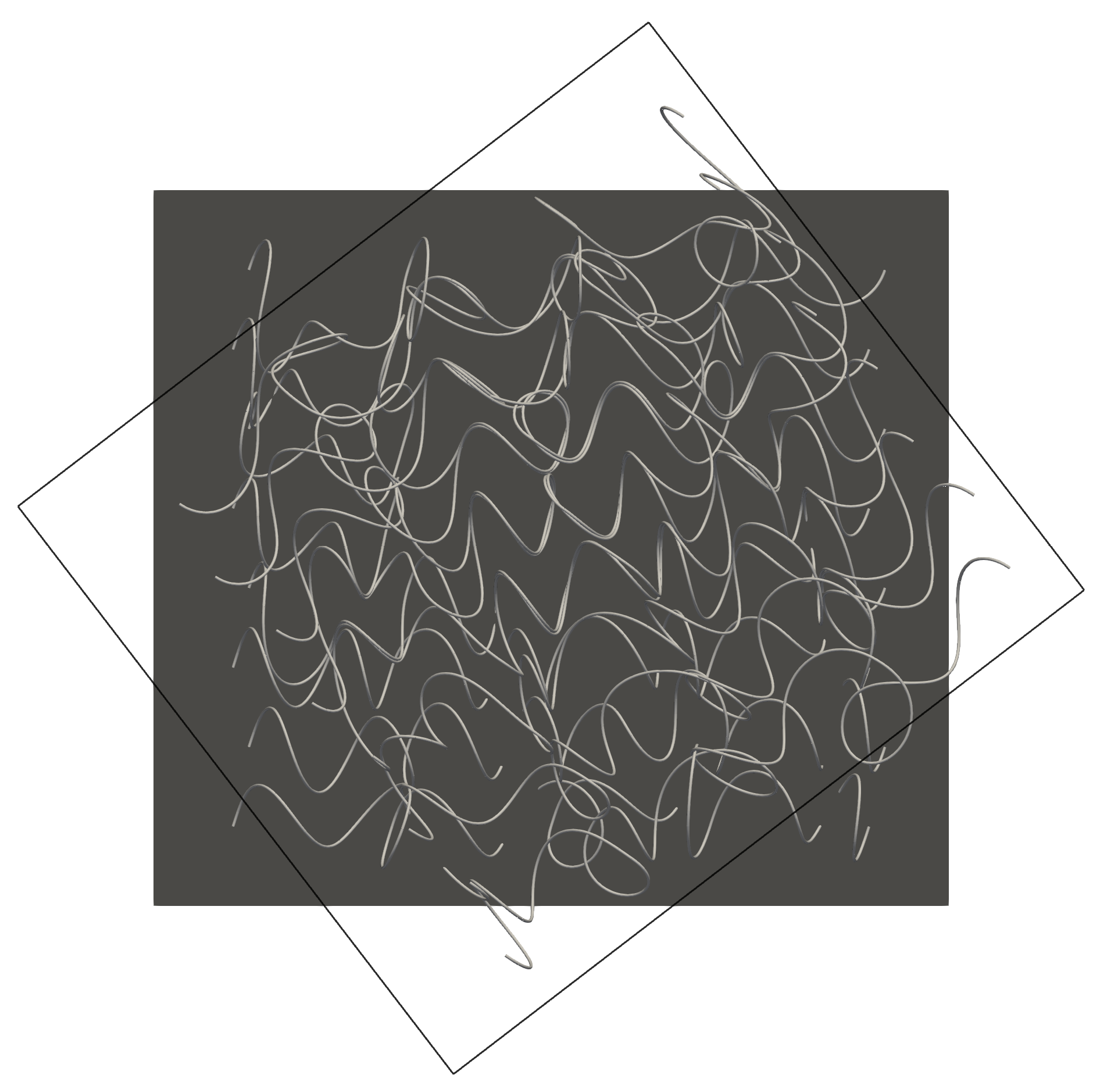}
      \label{fig::num_ex_adhesive_surfaces_simulation_snapshots_2x8x8_TwistAndPull_time3_step10754_Z-view}
    }
    \subfigure[$t=4$, $\psi_z=75^\circ$ (view from the top)]{
      \includegraphics[width=\textwidth]{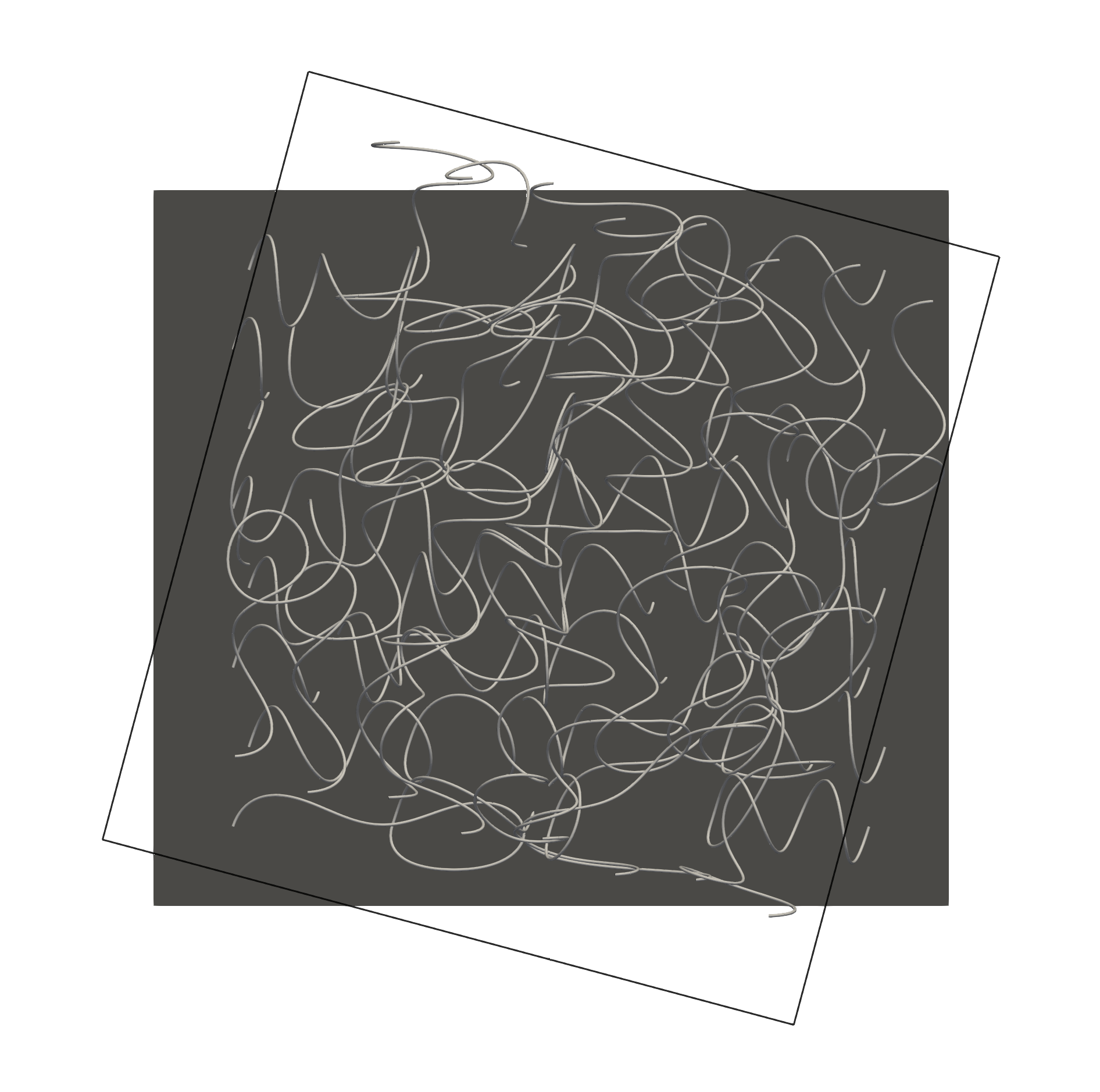}
      \label{fig::num_ex_adhesive_surfaces_simulation_snapshots_2x8x8_TwistAndPull_time4_step16969_Z-view}
    }
    \subfigure[$t=7$ (view along $y$-axis)]{
      \includegraphics[width=0.9\textwidth]{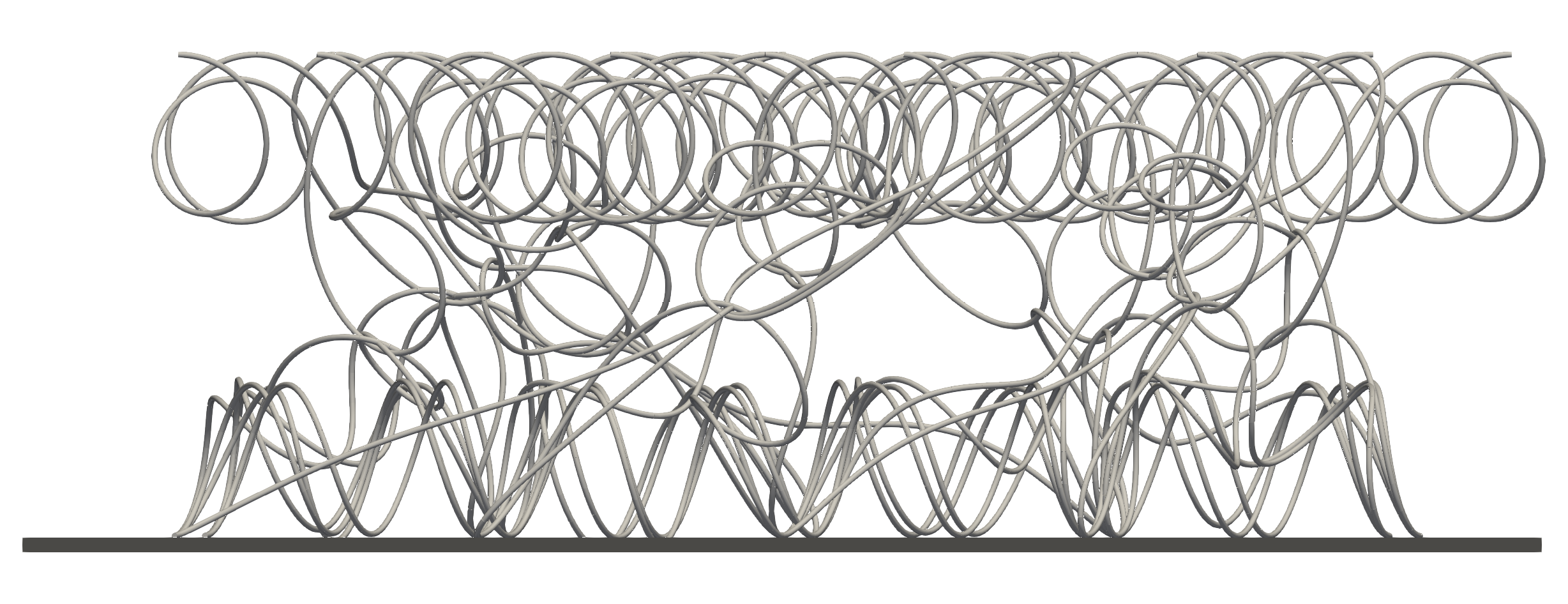}
      \label{fig::num_ex_adhesive_surfaces_simulation_snapshots_2x8x8_TwistAndPull_time7_step23516_Y-view}
    }
  \end{minipage}
  \caption{Sequence of simulation snapshots of the ``Twist \& Pull'' scenario with 2x8x8 loops. Top surface is hidden for better visibility of the fibers (and outlined by the black rectangle in the views from the top).}
  \label{fig::num_ex_adhesive_surfaces_simulation_snapshots_2x8x8_Twistandpull}
\end{figure}

Turning to the second considered scenario including the twisting of surfaces, an -- at first glance unintended -- entanglement of fibers has been observed, which also has a noticeable influence on the force-displacement curve.
This can be traced back to the (unphysical) behavior that during the twisting phase the outermost loop from the top surface dives under the fiber endpoint grafted onto the bottom surface and thus leaves the fibers entangled for the remaining course of the simulation, as can be seen from the simulation snapshots in \figref{fig::num_ex_adhesive_surfaces_simulation_snapshots_2x8x8_Twistandpull}.
While this is obviously neither physically correct nor desirable in the initial context of separating the adhesive fibers from each other, it is, however, the correct numerical solution given that no beam-solid contact has been considered between the fibers and the surfaces.
And, most importantly, it nicely demonstrates the effectiveness and robustness of the proposed SBIP approach even for highly complex, large 3D deformations in large-scale systems that include a broad variety of mutual orientations and separations of fibers.

\section{Summary, conclusions and outlook}
\label{sec::SBIP_conclusions_outlook}
Following up on the development of the first 3D beam-beam interaction formulation for molecular interactions -- the so-called section-section interaction potential (SSIP) approach~\cite{GrillSSIP} -- this article proposes an enhanced approach, which is specialized on the most challenging case of short-ranged interactions.
Examples for such short-ranged interactions include the van der Waals (vdW) adhesion and the steric repulsion of the Lennard-Jones (LJ) interaction.
Exploiting the characteristic rapid decay of the interaction potential with increasing distance allows to further reduce the dimensionality of the beam-beam interaction problem and thus achieve a significantly more efficient formulation than the more general SSIP approach.
The key idea is to approximate the second, arbitrarily deformed beam by a surrogate body with trivial geometry, which is located at the closest point from a given point on the first beam and oriented according to the centerline tangent vector of the second beam.
Mathematically, this surrogate body is the linear series expansion of the second, so-called master beam around the closest point.
In this manner, the interaction of the entire second beam with a given section of the first, so-called slave beam can be described by an analytical, closed-form interaction potential law, which replaces the numerical integration along the master beam by a single function evaluation.
This novel, specialized approach is therefore called the section-beam interaction potential (SBIP) approach.

In addition to being significantly more efficient, the novel formulation developed in this article is also significantly more accurate than the one used before.
This is due to the fact that the underlying reduced interaction law is superior to the previously used simple SSIP law, because it considers the relative rotation, i.e., the orientation of the interacting bodies in addition to their scalar separation.
The specific SBIP law used in the present work has been derived in the authors' recent contribution~\cite{Grilldiskcylpot}, considering the case of circular, homogeneous cross-sections and a generic inverse power law with exponent~$m \geq 6$ as fundamental point-pair interaction potential.
Most importantly for this work, the SBIP law ensures the correct asymptotic scaling behavior in the decisive regime of small separations, which has been identified as the most critical aspect when using the simple SSIP law from Eq.~(A12) of our previous contribution~\cite{GrillSSIP}, which neglects the mutual orientation of cross-sections.
This important property has been verified numerically by an exemplarily chosen test case considering two straight fibers at different relative orientations.
Moreover, using the numerical example of peeling two adhesive elastic fibers, it has been shown that only the SBIP law reproduces the quantitatively correct local adhesive force as well as equilibrium surface separation and therefore also the decisive global pull-off force.

These two novelties -- the general SBIP approach from \secref{sec::method_single_length_specific_integral} and the specific analytical interaction law from Ref.~\cite{Grilldiskcylpot} -- have been combined in \secref{sec::virtual_work_disk-cyl-pot_m} to obtain the resulting virtual work contribution.
The subsequent discretization by means of finite elements and its consistent linearization finally give rise to a novel beam-beam interaction formulation that can be seamlessly integrated in an existing nonlinear finite element framework for structural mechanics.
In this respect, it is important to point out that the approach does neither depend on a specific beam (element) formulation nor time integration scheme and is thus considered to be highly versatile.
Further methodological aspects such as a suitable numerical regularization of the characteristic singularity at zero separation in the reduced interaction law and a criterion to sort out beam element pairs with a larger separation than the cutoff radius before the actual evaluation to save computational resources have been presented in \secref{sec::discretization_algorithmic_aspects}.

By means of a numerical example studying the peeling and pull-off behavior of elastic adhesive fibers, the higher efficiency of the novel, specialized SBIP approach as compared to the previous, more general SSIP approach has been quantified.
In this example and even without using all savings the proposed formulation offers, the novel approach is approx.~4 times faster than the previous approach.
This is slightly less than the factor of 10, which would be expected from the comparison of algorithmic complexities, because the applied reduced SBIP interaction law (\eqref{eq::disk-cyl-pot_m}) is more complex due to the significantly enhanced accuracy if compared to the simple SSIP law (Eq.~(A12) of our previous contribution~\cite{GrillSSIP}).
It can be concluded that the combination of a better accuracy and higher efficiency leads to a drastically improved applicability -- for instance with respect to the maximum feasible system size and time scales -- and therefore opens the door to tackle a multitude of yet intractable problems in science and technology.

This is demonstrated by the comprehensive numerical example studying adhesive nanofiber-grafted surfaces, which confirms the effectiveness, efficiency and robustness of the novel formulation also for large-scale, complex systems with arbitrary mutual configurations and large deformations of the interacting fibers.
In addition, this example outlines the working principle of the proposed design of the nanofiber-grafted surfaces based on arrays of helical fibers by showing the desired large ratio of strong adhesive connection under load and easy removal of the surfaces if desired.
This showcases how suitable simulation tools such as the one developed in this article could contribute to the design and manufacturing of bioinspired artificial adhesives in the long run.

A future derivation of other SBIP laws for instance for the case of screened electrostatics or hydrophobic/-philic interactions would further extend the range of applications of the novel SBIP approach.
Given the importance of the mentioned interaction types in many biological systems on the nano- and microscale, this is considered a highly promising extension of the work presented here.
Moreover, the identification and calibration of the parameters of reduced interaction laws based on experimental results (such as the force-displacement curves for simple characteristic setups with two interacting fibers) would be an interesting avenue of future research that is expected to be of great benefit to the prediction quality of computational experiments.

\appendix

\section{Dimensionless key parameters of a system with adhesive elastic fibers}\label{sec::nondimensionalization}
This section aims to identify the dimensionless parameters that govern the fundamental behavior of a mechanical system with elastic, adhesive fibers.
Considering the space-continuous, static problem without external loads, the governing equation for the total potential energy (\eqref{eq::total_potential_energy_is_minimized}) simplifies to the internal energy contributions from elastic forces and moments~$\Pi_\text{int}$, which is well-known from beam theory, and the energy contribution from the interaction~$\Pi_\text{ia}$.
For the latter, we use the most general form (\eqref{eq::pot_fullvolint}) based on two nested volume integrals of the point-pair potential~$\Phi(r)$ and consider the LJ potential with its adhesive and repulsive component as an example.
The entire set of parameters carrying units is thus given by the length~$L$ and radius~$R$ of the fibers, its Young's modulus~$E$, the atom density~$\rho$, and finally the prefactors of the adhesive and repulsive part of the LJ law $k_6$ and $k_{12}$, respectively.
Note that instead of these two prefactors, an equivalent set of parameters~$r_\text{LJ,eq}$ and $\Phi_\text{LJ,eq}$ specifying the equilibrium distance and the corresponding minimal potential value of the LJ law is commonly used.
Nondimensionalization of this equation by means of normalization of the primary variables with suitable length measures allows to identify the following (non-unique) set of dimensionless parameters:
\begin{align}
  &\frac{L}{R} =: \zeta \quad &\text{slenderness ratio}\\
  &\frac{k_{12}}{\abs{k_6}\,L^6} \quad \text{or} \quad \frac{r_\text{LJ,eq}}{R} \quad &\text{normalized equilibrium distance}\\
  &\frac{\rho^2 \, \abs{k_6}}{E\, L^3} \quad \text{or} \quad \frac{\rho^2 \, \abs{\Phi_\text{LJ,eq}} \, L^3}{E} \quad &\text{adhesive compliance}
\end{align}
The slenderness ratio~$\zeta$ is known to be the only dimensionless parameter of the static elastic problem of beams and e.g.~determines the amount and thus relevance of shear deformation.
The second and third parameter are specific for the interactions between beams and given for the two alternative sets of LJ parameters mentioned above.
Using $r_\text{LJ,eq}$ and $\Phi_\text{LJ,eq}$, these parameters nicely disclose their meaning as equilibrium distance measure and measure for the strength of adhesion relative to the fibers' structural rigidity.
The latter is thus named ``adhesive compliance'', because it will determine how much the fibers will deform due to the exposure to adhesive interactions with other fibers.
As expected, this compliance will increase with increasing strength of adhesion, i.e., increasing atom density and increasing depth of the LJ potential minimum and decrease with increasing structural stiffness, i.e., increasing Young's modulus and decreasing fiber length.
Such a ratio has before been reported to crucially influence the peeling behavior of thin films from rigid surfaces~\cite{Sauer2011c} and likewise the peeling and pull-off behavior of two adhesive fibers (see \secref{sec::num_ex_peeling_pulloff_SBIP} and the preceding study in~\cite{GrillPeelingPulloff}).
An illustrative analogy for this adhesive compliance may be found as the ratio of fiber length~$L$ and persistence length~$l_\text{p}$, which is commonly used in (bio-)polymer physics to measure the amount of deformation of filaments to be expected due to thermal excitation in the context of Brownian dynamics and thus characterizes filaments to be either flexible, semi-flexible, or rather stiff (see e.g.~\cite{doi1988}).
Finally, note that these dimensionless parameters are not unique and other recombinations of the ones given above are possible.
Our approach of normalizing the governing equation as described above has been verified by means of the formalized dimensional analysis based on the Buckingham $\pi$ theorem, which in addition to the three dimensionless parameters suggests a fourth dimensional group~$\rho \, L^3$, which can be interpreted as the number of atoms in a cube with edge length~$L$.

\section{Linearization of the Virtual Work Contribution}\label{sec::linearization_SBIP}
In order to solve the nonlinear system of equations resulting from discretization of~\eqref{eq::total_virtual_work_is_zero} by means of Newton's method, the linearization, i.e., the tangent stiffness matrix, of the virtual work contribution from~\eqref{eq::virtual_work_SBIP_disk-cyl-pot} is required.
Its derivation is a mere application of differentiation rules and presented in the following.
Differentiation of the discrete residual vector resulting from \eqref{eq::virtual_work_SBIP_disk-cyl-pot} reads
\begin{align}\label{eq::linearization_SBIP_disk-cyl-pot}
  \Delta[ \delta \Pi_\text{ia} ] = \int \limits_0^{l_1} \Delta [ \diff{\tilde{\pi}}{{g_\text{ul}}} ] \, \delta g_\text{ul} +
                                    \diff{\tilde{\pi}}{{g_\text{ul}}} \, \Delta [ \delta g_\text{ul} ] +
                                    \Delta [ \diff{\tilde{\pi}}{{(\cos\alpha)}} ] \, \delta (\cos\alpha) +
                                    \diff{\tilde{\pi}}{{(\cos\alpha)}} \, \Delta [ \delta (\cos\alpha) ] \dd s_1,
\end{align}
where~$\Delta[ . ] := \diff{}{{\hat \vdd}}[ . ] \, \Delta \hat \vdd$ denotes the incremental change of a quantity expressed by an incremental change of the vector of degrees of freedom~$\hat \vdd$.
Refer to \eqref{eq::centerline_discretization} for the definition of this vector consisting of all those discrete degrees of freedom~$\hat \vdd = [\hat \vdd_1^T, \hat \vdd_2^T]^T$, which are required for the centerline discretization of both elements~$i=1,2$, and the definition of the correspondingly assembled matrix of shape functions~$\vdH$.
Therefore,
\begin{align}
  \Delta \vr_1 = [ \vdH_1, \vnull ] \, \Delta \hat \vdd, \,
  \Delta \vr^\shortmid_1 = [ \vdH^\shortmid_1, \vnull ] \, \Delta \hat \vdd, \,
  \Delta \vr_2 = [ \vnull, \vdH_2 ] \, \Delta \hat \vdd, \,
  \Delta \vr^\shortmid_2 = [ \vnull, \vdH_2^\shortmid ] \, \Delta \hat \vdd, \, \text{and} \,
  \Delta \vr^{\shortmid\shortmid}_2 = [ \vnull, \vdH_2^{\shortmid\shortmid} ] \, \Delta \hat \vdd,
\end{align}
and the four~$\Delta [.]$ terms from~\eqref{eq::linearization_SBIP_disk-cyl-pot} will now be expressed by means of these basic quantities.
To begin with,
\begin{align}
  \Delta [ \diff{\tilde{\pi}}{{g_\text{ul}}} ] = \diff{ {\tilde \pi} }{ {g_\text{ul}^2} } \, \Delta g_\text{ul} +
                                                 \diff{ {\tilde \pi} }{ {g_\text{ul}}{(\cos\alpha)} } \, \Delta [ \cos\alpha ],
\end{align}
where
\begin{align}
  \Delta g_\text{ul} = \vn_\text{ul}^T (\Delta \vr_1 - \Delta \vr_2)
\end{align}
and
\begin{align}
  \Delta [ \cos\alpha ] = \left( \vv_{\alpha1}^T \, \Delta \vr_1^{\shortmid} + \vv_{\alpha2}^T \, \Delta \vr_2^{\shortmid} + (\vr_2^{\shortmid\shortmid T} \vv_{\alpha2}) \, \Delta \xi_\text{2c} \right) \, \sgn(\vt_1^T \vt_2)
\end{align}
are analogous to the variations of these quantities given in \eqref{eq::var_gap_ul} and \eqref{eq::var_cosalpha}.
The required second derivatives of the disk-cylinder interaction potential~$\tilde \pi$ from \eqref{eq::disk-cyl-pot_m} can -- like the first derivatives in \eqref{eq::disk-cyl-pot_m_firstderiv_gap_ul} and \eqref{eq::disk-cyl-pot_m_firstderiv_cos_alpha} -- be conveniently expressed in recursive manner as
\begin{align}
  \diff{ {\tilde \pi} }{ {g_\text{ul}^2} } &= (-m+\frac{7}{2}) \frac{1}{g_\text{ul}} \diff{{\tilde \pi}}{ {g_\text{ul}} } \quad \text{and}\\\
  \diff{ {\tilde \pi} }{ {g_\text{ul}}{(\cos\alpha)} } &= (-m+\frac{9}{2}) \, \frac{1}{g_\text{ul}} \, \diff{ {\tilde \pi} }{ {(\cos\alpha)} }.
\end{align}
Furthermore,
\begin{align}
  \Delta [ \diff{\tilde{\pi}}{{(\cos\alpha)}} ] = \diff{ {\tilde \pi} }{ {g_\text{ul}}{(\cos\alpha)} } \, \Delta g_\text{ul} +
                                                  \diff{ {\tilde \pi} }{ {(\cos\alpha)^2} } \, \Delta [ \cos\alpha ],
\end{align}
and the missing second derivative of~$\tilde \pi$ with respect to~$\cos\alpha$:
\begin{align}
  \diff{ {\tilde \pi} }{ {(\cos\alpha)^2} } &= \left( - \frac{ R_1 }{R_1 \cos\alpha + R_2} + \frac{3 \, R_1^2 \cos^2\alpha}{ (R_1 \cos\alpha + R_2)^2 } \right) \, \tilde \pi
\end{align}
The linearization of the variation of the unilateral gap reads
\begin{align}
  \Delta [ \delta g_\text{ul} ] = \left( \delta \vr_1^T - \delta \vr_2^T \right) \Delta \vn_\text{ul}
                                - \delta \vr_2^{\shortmid T} \vn_\text{ul} \otimes \Delta \xi_\text{2c},
\end{align}
where in turn the linearization of the unilateral unit normal vector
\begin{align}
  \Delta \vn_\text{ul} = \frac{1}{d_\text{ul}} \left( \vI_{3\times3} - \vn_\text{ul} \otimes \vn_\text{ul} \right) \Delta \vd_\text{ul}
\end{align}
and unilateral distance vector
\begin{align}
  \Delta \vd_\text{ul} = \Delta \vr_1 - \Delta \vr_2 - \vr_2^\shortmid \otimes \Delta \xi_\text{2c}
\end{align}
are required.
In these equations, the linearization of the (closest-point) arc-length coordinate on the master side~$\Delta \xi_\text{2c}$ appears, which is a result from the fact that~$\xi_\text{2c}$ is determined via a closest-point-to-curve projection and thus depends on the current configuration, i.e., primary degrees of freedom.
For this reason, we need
\begin{align}\label{eq::lin_total_pos_master}
  \Delta [ \vr_2(\xi_{2\text{c}}) ] &= \Delta \vr_2 (\xi_{2\text{c}}) + \vr_2^\shortmid (\xi_{2\text{c}}) \, \Delta \xi_\text{2c}\\
  \Delta [ \vr_2^\shortmid(\xi_{2\text{c}}) ] &= \Delta \vr_2^\shortmid (\xi_{2\text{c}}) + \vr_2^{\shortmid\shortmid} (\xi_{2\text{c}}) \, \Delta \xi_\text{2c} \label{eq::lin_total_tangent_master}
\end{align}
with
\begin{align}
  \Delta \xi_\text{2c} = \frac{1}{p_{2,\xi_2}} \left( - \vr_2^{\shortmid T} \Delta \vr_1 + \vr_2^{\shortmid T} \Delta \vr_2 - \vd_\text{ul}^T \Delta \vr_2^{\shortmid} \right).
\end{align}
Note the analogy to the corresponding variation~$\delta \xi_\text{2c}$ from \eqref{eq::var_total_pos_master} and the definition of~$p_{2,\xi_2}$ given in \eqref{eq::orthogonalitycondition_master_deriv}.\\

\noindent\textit{Remark.}
Up to this point, the number and complexity of the expressions is comparable to the ones required in the macroscopic line contact formulation proposed in~\cite{meier2016} and the combination, i.e., blending%
\footnote{
on contact force level
}
thereof with a point contact formulation in~\cite{Meier2017a}.
Indeed, only the expressions that depend on the applied disk-cylinder interaction potential law deviate from the line contact case, where most commonly a quadratic penalty potential~$\tilde \pi = 0.5 \, \varepsilon_\parallel \, g_\text{ul}^2$ is used.\\

\noindent Here, the additional expression for the linearization of the variation of the cosine of the mutual angle
\begin{align}
  \Delta [ \delta (\cos\alpha) ] = \Big(&\delta \vr_1^{\shortmid T} \, \Delta \vv_{\alpha1}
                                        + \delta \vr_2^{\shortmid T} \, \Delta \vv_{\alpha2}
                                        + \delta \xi_\text{2c} \left( \vr_2^{\shortmid\shortmid T} \Delta \vv_{\alpha2} + \vv_{\alpha2}^T \Delta[ \vr_2^{\shortmid\shortmid}(\xi_\text{2c}) ] \right)\\
                                        &+ \vv_{\alpha2}^T \Delta [ \delta \vr_2^\shortmid ] + ( \vr_2^{\shortmid\shortmid T} \vv_{\alpha2} ) \Delta [ \delta \xi_\text{2c} ]
                                   \Big) \, \sgn(\vt_1^T \vt_2)\nonumber
\end{align}
is required for a consistent linearization and in fact most of the complexity comes from this contribution or, to be more precise, from the linearization of the variation of the element parameter of the closest-point on the master side
\begin{align}
  \Delta [ \delta \xi_\text{2c} ] &= \frac{1}{p_{2,\xi_2}^2} \left( \delta \vr_1^T \vr_2^\shortmid
                                                                  - \delta \vr_2^T \vr_2^\shortmid
                                                                  + \delta \vr_2^{\shortmid T} \vd_\text{ul}
                                                             \right) \Delta [ p_{2,\xi_2} ]\\
                                   &+\frac{1}{p_{2,\xi_2}} \Big( -\delta \vr_1^T \Delta [ \vr_2^\shortmid(\xi_\text{2c})]
                                                                 + \delta \vr_2^T \Delta [ \vr_2^\shortmid(\xi_\text{2c})]
                                                                 - \delta \vr_2^{\shortmid T} \Delta \vd_\text{ul}
                                                                 + \vr_2^{\shortmid T} \Delta [ \delta \vr_2 ]
                                                                 - \vd_\text{ul}^T \Delta [ \delta \vr_2^\shortmid ]
                                                           \Big).\nonumber
\end{align}
In addition, the following quantities have been introduced in these last two equations.
\begin{align}
  \Delta [ \vv_{\alpha1} ] &= - \frac{1}{\norm{\vr_1^\shortmid}^2} \left( \left( (\vI_{3\times3} - \vt_1 \otimes \vt_1^T) \vt_2 \right) \otimes \vr_1^{\shortmid T} \right) \Delta \vr_1^\shortmid\\
                           &+ \frac{1}{\norm{\vr_1^\shortmid}} \left( \vI_{3\times3} - \vt_1 \otimes \vt_1^T \right) \Delta \vt_2
                            - \frac{1}{\norm{\vr_1^\shortmid}} \left( (\vt_1^T \vt_2) \vI_{3\times3} + \vt_1 \otimes \vt_2^T \right) \Delta \vt_1\nonumber
\end{align}
\begin{align}
  \Delta [ \vv_{\alpha2} ] &= - \frac{1}{\norm{\vr_2^\shortmid}^2} \left( \left( (\vI_{3\times3} - \vt_2 \otimes \vt_2^T) \vt_1 \right) \otimes \vr_2^{\shortmid T} \right) \Delta [\vr_2^\shortmid(\xi_\text{2c})]\\
                           &+ \frac{1}{\norm{\vr_2^\shortmid}} \left( \vI_{3\times3} - \vt_2 \otimes \vt_2^T \right) \Delta \vt_1
                            - \frac{1}{\norm{\vr_2^\shortmid}} \left( (\vt_1^T \vt_2) \vI_{3\times3} + \vt_2 \otimes \vt_1^T \right) \Delta \vt_2\nonumber
\end{align}
\begin{align}
  \Delta [ \vr_2^{\shortmid\shortmid}(\xi_{2\text{c}}) ] &= \Delta \vr_2^{\shortmid\shortmid} (\xi_{2\text{c}}) + \vr_2^{\shortmid\shortmid\shortmid} (\xi_{2\text{c}}) \, \Delta \xi_\text{2c}
\end{align}
\begin{align}
  \Delta [ \delta \vr_2^\shortmid ] = \delta \vr_2^{\shortmid\shortmid} \otimes \Delta \xi_\text{2c}
\end{align}
\begin{align}
  \Delta \vt_i = \frac{1}{\norm{ \vr_i^\shortmid }} \left( \vI_{3\times3} - \vt_i \otimes \vt_i^T \right) \Delta [ \vr_i^\shortmid(\xi_i) ], \quad \text{with } i=1,2
\end{align}
Note that the correctness of the presented and implemented analytical linearization has been verified by means of automatic differentiation~\cite{Trilinos2012} of the virtual work expression from \eqref{eq::virtual_work_SBIP_disk-cyl-pot}.
Due to the high complexity of~$\Delta [ \delta \xi_\text{2c} ]$, it would be interesting to investigate how a neglect of this contribution would influence the computational cost on the one hand and the performance of the nonlinear solver on the other hand.
In the context of this work, however, always the fully consistent linearization has been used.

\section{Additionally required expressions for the regularization of the reduced disk-cylinder interaction law}\label{sec::regularization_SBIP_expressions}
The equation describing the proposed quadratic extrapolation below~$g_\text{ul,reg}$ for a general section-beam interaction potential law~$\tilde\pi$ is given as follows:
\begin{align}
  \tilde \pi_\text{reg} &=
      \begin{dcases}
        \tilde \pi(g_\text{ul,reg}) + \diff{{\tilde\pi}}{{g_\text{ul}}}\biggr\rvert_{g_\text{ul,reg}} \, (g_\text{ul} - g_\text{ul,reg})
            + \frac{1}{2} \, \diff{{\tilde\pi}}{{g_\text{ul}^2}}\biggr\rvert_{g_\text{ul,reg}} \, (g_\text{ul} - g_\text{ul,reg})^2, \quad & g_\text{ul} < g_\text{ul,reg}\\
        \tilde \pi(g_\text{ul}), \quad & g_\text{ul} \geq g_\text{ul,reg}
      \end{dcases}%
\end{align}
Note that the required derivatives of this regularized potential law~$\tilde\pi_\text{reg}$ accordingly change to
\begin{align}
  \diff{{\tilde\pi_\text{reg}}}{{g_\text{ul}}} &=
      \begin{dcases}
        \diff{{\tilde\pi}}{{g_\text{ul}}}\biggr\rvert_{g_\text{ul,reg}}
            + \diff{{\tilde\pi}}{{g_\text{ul}^2}}\biggr\rvert_{g_\text{ul,reg}} \, (g_\text{ul} - g_\text{ul,reg}), \quad & g_\text{ul} < g_\text{ul,reg}\\
        \diff{{\tilde\pi}}{{g_\text{ul}}}(g_\text{ul}), \quad & g_\text{ul} \geq g_\text{ul,reg}
      \end{dcases}\\&\nonumber\\
  \diff{{\tilde\pi_\text{reg}}}{{g_\text{ul}^2}} &=
      \begin{dcases}
        \diff{{\tilde\pi}}{{g_\text{ul}^2}}\biggr\rvert_{g_\text{ul,reg}}, \quad & g_\text{ul} < g_\text{ul,reg}\\
        \diff{{\tilde\pi}}{{g_\text{ul,reg}}}(g_\text{ul}), \quad & g_\text{ul} \geq g_\text{ul,reg}
      \end{dcases}\\&\nonumber\\
  \diff{{\tilde\pi_\text{reg}}}{{(\cos\alpha)}}&=
      \begin{dcases}
        \diff{{\tilde\pi}}{{(\cos\alpha)}}\biggr\rvert_{g_\text{ul,reg}}
            + \diff{{\tilde\pi}}{{(\cos\alpha)}{g_\text{ul}}}\biggr\rvert_{g_\text{ul,reg}} \, (g_\text{ul} - g_\text{ul,reg}) \ldots\\
            \qquad + \frac{1}{2} \, \diff{{\tilde\pi}}{{(\cos\alpha)}{g_\text{ul}^2}}\biggr\rvert_{g_\text{ul,reg}} \, (g_\text{ul} - g_\text{ul,reg})^2, \quad & g_\text{ul} < g_\text{ul,reg}\\
        \diff{{\tilde\pi}}{{(\cos\alpha)}}(g_\text{ul}), \quad & g_\text{ul} \geq g_\text{ul,reg}
      \end{dcases}
\end{align}
\begin{align}
  \diff{{\tilde\pi_\text{reg}}}{{(\cos\alpha)}{g_\text{ul}}}&=
      \begin{dcases}
        \diff{{\tilde\pi}}{{(\cos\alpha)}{g_\text{ul}}}\biggr\rvert_{g_\text{ul,reg}}
            + \diff{{\tilde\pi}}{{(\cos\alpha)}{g_\text{ul}^2}}\biggr\rvert_{g_\text{ul,reg}} \, (g_\text{ul} - g_\text{ul,reg}), \quad & g_\text{ul} < g_\text{ul,reg}\\
        \diff{{\tilde\pi}}{{(\cos\alpha)}{g_\text{ul}}}(g_\text{ul}), \quad & g_\text{ul} \geq g_\text{ul,reg}
      \end{dcases}\\&\nonumber\\
  \diff{{\tilde\pi_\text{reg}}}{{(\cos\alpha)^2}}&=
      \begin{dcases}
        \diff{{\tilde\pi}}{{(\cos\alpha)^2}}\biggr\rvert_{g_\text{ul,reg}}
            + \diff{{\tilde\pi}}{{(\cos\alpha)^2}{g_\text{ul}}}\biggr\rvert_{g_\text{ul,reg}} \, (g_\text{ul} - g_\text{ul,reg}) \ldots\\
            \qquad + \frac{1}{2} \, \diff{{\tilde\pi}}{{(\cos\alpha)^2}{g_\text{ul}^2}}\biggr\rvert_{g_\text{ul,reg}} \, (g_\text{ul} - g_\text{ul,reg})^2, \quad & g_\text{ul} < g_\text{ul,reg}\\
        \diff{{\tilde\pi}}{{(\cos\alpha)^2}}(g_\text{ul}), \quad & g_\text{ul} \geq g_\text{ul,reg}.
      \end{dcases}
\end{align}
These expressions for the regularized general interaction potential~$\tilde \pi$ and its first and second derivatives replace the original expressions presented in the context of Sections~\ref{sec::ia_pot_single_length_specific_evaluation_vdW} and \ref{sec::virtual_work_disk-cyl-pot_m} and~\appref{sec::linearization_SBIP}, respectively.
For the specific disk-cylinder interaction potential from~\eqref{eq::disk-cyl-pot_m}, most of the derivatives required on the right hand side of these equations have already been presented in~\secref{sec::virtual_work_disk-cyl-pot_m} and~\appref{sec::linearization_SBIP}, and the additionally required third and fourth derivative follow (recursively) as
\begin{align}
  \diff{{\tilde\pi_\text{m,disk-cyl}}}{{(\cos\alpha)^2}{g_\text{ul}}} &= (-m+\frac{9}{2}) \, \frac{1}{g_\text{ul}} \,  \diff{{\tilde\pi_\text{m,disk-cyl}}}{{(\cos\alpha)^2}},\\
  \diff{{\tilde\pi_\text{m,disk-cyl}}}{{(\cos\alpha)}{g_\text{ul}^2}} &= (-m+\frac{9}{2})(-m+\frac{7}{2}) \, \frac{1}{g_\text{ul}^2} \,  \diff{{\tilde\pi_\text{m,disk-cyl}}}{{(\cos\alpha)}}
\end{align}
and
\begin{align}
  \diff{{\tilde\pi_\text{m,disk-cyl}}}{{(\cos\alpha)^2}{g_\text{ul}^2}} = (-m+\frac{9}{2})(-m+\frac{7}{2}) \, \frac{1}{g_\text{ul}^2} \,  \diff{{\tilde\pi_\text{m,disk-cyl}}}{{(\cos\alpha)^2}}.
\end{align}
Note that this regularization needs to be applied to both the adhesive~$m=6$ and the repulsive~$m=12$ part of the LJ potential law.

\bibliography{SBIP}

\begin{thebibliography}{10}
\providecommand \doibase [0]{http://dx.doi.org/}%

\bibitem{lindstrom2010biopolymer}
Lindstr{\"o}m SB, Vader DA, Kulachenko A, Weitz DA. Biopolymer network
  geometries: Characterization, regeneration, and elastic properties. {\it
  Physical Review E} 2010\string; 82(5)\string: 051905.

\bibitem{Castro2011}
Castro CE, Kilchherr F, Kim DN, et al. {A primer to scaffolded DNA origami}.
  {\it Nature Methods} 2011\string; 8(3)\string: 221--229.

\bibitem{Gautieri2012}
Gautieri A, Pate MI, Vesentini S, Redaelli A, Buehler MJ. {Hydration and
  distance dependence of intermolecular shearing between collagen molecules in
  a model microfibril}. {\it Journal of Biomechanics} 2012\string;
  45(12)\string: 2079--2083.

\bibitem{Sauer2009}
Sauer RA. {Multiscale modelling and simulation of the deformation and adhesion
  of a single gecko seta}. {\it Computer Methods in Biomechanics and Biomedical
  Engineering} 2009\string; 12(6)\string: 627--640.

\bibitem{Mueller2014rheology}
M{\"{u}}ller KW, Bruinsma RF, Lieleg O, Bausch AR, Wall WA, Levine AJ.
  {Rheology of Semiflexible Bundle Networks with Transient Linkers}. {\it
  Physical Review Letters} 2014\string; 112(23)\string: 238102.

\bibitem{Mueller2015interpolatedcrosslinks}
M{\"{u}}ller KW, Meier C, Wall WA. {Resolution of sub-element length scales in
  Brownian dynamics simulations of biopolymer networks with geometrically exact
  beam finite elements}. {\it Journal of Computational Physics} 2015\string;
  303\string: 185--202.

\bibitem{Negi2018}
Negi V, Picu RC. {Mechanical behavior of cross-linked random fiber networks
  with inter-fiber adhesion}. {\it Journal of the Mechanics and Physics of
  Solids} 2018\string; 122\string: 418--434.

\bibitem{Goodrich2018}
Goodrich CP, Brenner MP, Ribbeck K. {Enhanced diffusion by binding to the
  crosslinks of a polymer gel}. {\it Nature Communications} 2018\string;
  9(1)\string: 4348.

\bibitem{GrillParticleMobilityHydrogels}
Grill MJ, Eichinger JF, Koban J, Meier C, Lieleg O, Wall WA. {A novel modelling
  and simulation approach for the hindered mobility of charged particles in
  biological hydrogels}. {\it Proceedings of the Royal Society A: Mathematical,
  Physical and Engineering Sciences} 2021\string; 477(2249)\string: 20210039.

\bibitem{Eichinger2021}
Eichinger JF, Grill MJ, Kermani ID, et al. {A computational framework for
  modeling cell-matrix interactions in soft biological tissues}. {\it
  Biomechanics and Modeling in Mechanobiology} 2021.

\bibitem{BundlesPNAS}
Slepukhin VM, Grill MJ, Hu Q, Botvinick EL, Wall WA, Levine AJ. {Topological
  defects produce kinks in biopolymer filament bundles}. {\it Proceedings of
  the National Academy of Sciences} 2021\string; 118(15)\string: e2024362118.

\bibitem{durville2010}
Durville D. {Simulation of the mechanical behaviour of woven fabrics at the
  scale of fibers}. {\it International Journal of Material Forming}
  2010\string; 3(2)\string: 1241--1251.

\bibitem{Kulachenko2012}
Kulachenko A, Uesaka T. {Direct simulations of fiber network deformation and
  failure}. {\it Mechanics of Materials} 2012\string; 51\string: 1--14.

\bibitem{Durville2015}
Wielhorski Y, Durville D. Finite element simulation of a 3D woven fabric:
  Determination of the initial configuration and characterization of the
  mechanical behavior. {\it Texcomp-12 Conference} 2015\string; 2(May)\string:
  26--29.

\bibitem{Weeger2016}
Weeger O, Kang YSB, Yeung SK, Dunn ML. {Optimal Design and Manufacture of
  Active Rod Structures with Spatially Variable Materials}. {\it 3D Printing
  and Additive Manufacturing} 2016\string; 3(4)\string: 204--215.

\bibitem{Meier2017b}
Meier C, Grill MJ, Wall WA, Popp A. {Geometrically exact beam elements and
  smooth contact schemes for the modeling of fiber-based materials and
  structures}. {\it International Journal of Solids and Structures}
  2018\string; 154\string: 124--146.

\bibitem{pattinson2019additive}
Pattinson SW, Huber ME, Kim S, et al. Additive manufacturing of biomechanically
  tailored meshes for compliant wearable and implantable devices. {\it Advanced
  Functional Materials} 2019\string; 29(32)\string: 1901815.

\bibitem{steinbrecher2020mortar}
Steinbrecher I, Mayr M, Grill MJ, Kremheller J, Meier C, Popp A. A mortar-type
  finite element approach for embedding 1D beams into 3D solid volumes. {\it
  Computational Mechanics} 2020\string; 66(6)\string: 1377--1398.

\bibitem{Khristenko2021}
Khristenko U, Schu{\ss} S, Kr{\"u}ger M, Schmidt F, Wohlmuth B, Hesch C.
  Multidimensional coupling: A variationally consistent approach to
  fiber-reinforced materials. {\it Computer Methods in Applied Mechanics and
  Engineering} 2021\string; 382\string: 113869.

\bibitem{steinbrecher2021}
Steinbrecher I, Popp A, Meier C. Consistent coupling of positions and rotations
  for embedding 1D Cosserat beams into 3D solid volumes. {\it Computational
  Mechanics} 2021\string: 1--32.

\bibitem{Argento1997}
Argento C, Jagota A, Carter WC. {Surface formulation for molecular interactions
  of macroscopic bodies}. {\it Journal of the Mechanics and Physics of Solids}
  1997\string; 45(7)\string: 1161--1183.

\bibitem{Sauer2007a}
Sauer RA, Li S. {A contact mechanics model for quasi-continua}. {\it
  International Journal for Numerical Methods in Engineering} 2007\string;
  71(8)\string: 931--962.

\bibitem{Sauer2009a}
Sauer RA, Wriggers P. {Formulation and analysis of a three-dimensional finite
  element implementation for adhesive contact at the nanoscale}. {\it Computer
  Methods in Applied Mechanics and Engineering} 2009\string; 198(49)\string:
  3871--3883.

\bibitem{Sauer2013}
Sauer RA, {De Lorenzis} L. {A computational contact formulation based on
  surface potentials}. {\it Computer Methods in Applied Mechanics and
  Engineering} 2013\string; 253\string: 369--395.

\bibitem{Fan2015}
Fan H, Li S. {A three-dimensional surface stress tensor formulation for
  simulation of adhesive contact in finite deformation}. {\it International
  Journal for Numerical Methods in Engineering} 2016\string; 107(3)\string:
  252--270.

\bibitem{Du2019}
Du S, {Ben Dhia} H. {An asymptotic numerical method to solve compliant
  Lennard-Jones-based contact problems involving adhesive instabilities}. {\it
  Computational Mechanics} 2019\string; 63(6)\string: 1261--1281.

\bibitem{Mergel2019}
Mergel JC, Sahli R, Scheibert J, Sauer RA. {Continuum contact models for
  coupled adhesion and friction}. {\it The Journal of Adhesion} 2019\string;
  95(12)\string: 1101--1133.

\bibitem{wriggers1997}
Wriggers P, Zavarise G. {On contact between three-dimensional beams undergoing
  large deflections}. {\it Communications in Numerical Methods in Engineering}
  1997\string; 13(6)\string: 429--438.

\bibitem{litewka2005}
Litewka P. {The penalty and Lagrange multiplier methods in the frictional 3d
  beam-to-beam contact problem}. {\it Civil and Environmental Engineering
  Reports} 2005\string; 1\string: 189--207.

\bibitem{Chamekh2014}
Chamekh M, Mani-Aouadi S, Moakher M. {Stability of elastic rods with
  self-contact}. {\it Computer Methods in Applied Mechanics and Engineering}
  2014\string; 279\string: 227--246.

\bibitem{GayNeto2016a}
{Gay Neto} A, Pimenta PM, Wriggers P. {A master-surface to master-surface
  formulation for beam to beam contact. Part I: Frictionless interaction}. {\it
  Computer Methods in Applied Mechanics and Engineering} 2016\string;
  303\string: 400--429.

\bibitem{Konyukhov2016}
Konyukhov A, Mrenes O, Schweizerhof K. {Consistent Development of a
  Beam-To-Beam Contact Algorithm via the Curve-to-Solid Beam Contact - Analysis
  for the Nonfrictional Case}. {\it International Journal for Numerical Methods
  in Engineering} 2018\string; 113(7)\string: 1108--1144.

\bibitem{Weeger2017}
Weeger O, Narayanan B, {De Lorenzis} L, Kiendl J, Dunn ML. {An isogeometric
  collocation method for frictionless contact of Cosserat rods}. {\it Computer
  Methods in Applied Mechanics and Engineering} 2017\string; 321\string:
  361--382.

\bibitem{meier2016}
Meier C, Popp A, Wall WA. {A finite element approach for the line-to-line
  contact interaction of thin beams with arbitrary orientation}. {\it Computer
  Methods in Applied Mechanics and Engineering} 2016\string; 308\string:
  377--413.

\bibitem{Meier2017a}
Meier C, Wall WA, Popp A. {A unified approach for beam-to-beam contact}. {\it
  Computer Methods in Applied Mechanics and Engineering} 2017\string;
  315\string: 972--1010.

\bibitem{bosten2022mortar}
Bosten A, Cosimo A, Linn J, Br{\"u}ls O. A mortar formulation for frictionless
  line-to-line beam contact. {\it Multibody System Dynamics} 2022\string;
  54(1)\string: 31--52.

\bibitem{Sauer2014}
Sauer RA, Mergel JC. {A geometrically exact finite beam element formulation for
  thin film adhesion and debonding}. {\it Finite Elements in Analysis and
  Design} 2014\string; 86\string: 120--135.

\bibitem{Schmidt2015}
Schmidt MG, Ismail AE, Sauer RA. {A continuum mechanical surrogate model for
  atomic beam structures}. {\it International Journal for Multiscale
  Computational Engineering} 2015\string; 13(5)\string: 413--442.

\bibitem{Meier2021CosseratPotential}
Meier C, Grill MJ, Wall WA. {Generalized section-section interaction potentials
  in the geometrically exact beam theory: Modeling of Intermolecular Forces,
  Asymptotic Limit as Strain-Energy Function, and Formulation of Rotational
  Constraints}. {\it submitted for publication, arXiv preprint
  arXiv:2105.10032}.

\bibitem{GrillSSIP}
Grill MJ, Wall WA, Meier C. {A computational model for molecular interactions
  between curved slender fibers undergoing large 3D deformations with a focus
  on electrostatic, van der Waals, and repulsive steric forces}. {\it
  International Journal for Numerical Methods in Engineering} 2020\string;
  121(10)\string: 2285--2330.

\bibitem{Grilldiskcylpot}
Grill MJ, Wall WA, Meier C. {Analytical disk-cylinder interaction potential
  laws for the computational modeling of adhesive, deformable (nano)fibers}.
  {\it submitted for publication, arXiv preprint arXiv:2208.03074}.

\bibitem{israel2011}
Israelachvili JN. {\it {Intermolecular and surface forces}}.
\newblock Oxford: Academic press.
\newblock 3rd~ed. 2011.

\bibitem{parsegian2005}
Parsegian VA. {\it {Van der Waals forces: a handbook for biologists, chemists,
  engineers, and physicists}}.
\newblock Cambridge, UK: Cambridge University Press .
\newblock 2005.

\bibitem{brenner1974}
Brenner SL, Parsegian VA. {A physical method for deriving the electrostatic
  interaction between rod-like polyions at all mutual angles}. {\it Biophysical
  Journal} 1974\string; 14(4)\string: 327--334.

\bibitem{Langbein1974}
Langbein D. {Theory of Van der Waals attraction}. In: Springer, Berlin,
  Heidelberg.  1974 (pp. 1--139).

\bibitem{Rajter2007}
Rajter RF, Podgornik R, Parsegian VA, French RH, Ching WY. {Van der
  Waals-London dispersion interactions for optically anisotropic cylinders:
  Metallic and semiconducting single-wall carbon nanotubes}. {\it Physical
  Review B - Condensed Matter and Materials Physics} 2007\string; 76(4)\string:
  1--16.

\bibitem{Dobson2012}
Dobson JF, Gould T. {Calculation of dispersion energies}. {\it Journal of
  Physics Condensed Matter} 2012\string; 24(7).

\bibitem{Meier2017c}
Meier C, Popp A, Wall WA. {Geometrically Exact Finite Element Formulations for
  Slender Beams: Kirchhoff-Love Theory Versus Simo-Reissner Theory}. {\it
  Archives of Computational Methods in Engineering} 2019\string; 26(1)\string:
  163--243.

\bibitem{jelenic1999}
Jeleni{\'{c}} G, Crisfield MA. {Geometrically exact 3D beam theory:
  implementation of a strain-invariant finite element for statics and
  dynamics}. {\it Computer Methods in Applied Mechanics and Engineering}
  1999\string; 171(1--2)\string: 141--171.

\bibitem{crisfield1999}
Crisfield MA, Jeleni{\'{c}} G. {Objectivity of strain measures in the
  geometrically exact three-dimensional beam theory and its finite-element
  implementation}. {\it Proceedings of the Royal Society of London. Series A:
  Mathematical, Physical and Engineering Sciences} 1999\string; 455\string:
  1125--1147.

\bibitem{Sauer2011}
Sauer RA. {Enriched contact finite elements for stable peeling computations}.
  {\it International Journal for Numerical Methods in Engineering} 2011\string;
  87(6)\string: 593--616.

\bibitem{Papadopoulos1995}
Papadopoulos P, Jones RE, Solberg JM. {A novel finite element formulation for
  frictionless contact problems}. {\it International Journal for Numerical
  Methods in Engineering} 1995\string; 38(15)\string: 2603--2617.

\bibitem{montgomery2000}
Montgomery SW, Franchek MA, Goldschmidt VW. {Analytical Dispersion Force
  Calculations for Nontraditional Geometries.}. {\it Journal of colloid and
  interface science} 2000\string; 227(2)\string: 567--584.

\bibitem{Trilinos2012}
Heroux MA, Willenbring JM. {A New Overview of the Trilinos Project}. {\it
  Scientific Programming} 2012\string; 20(2)\string: 83--88.

\bibitem{GrillPeelingPulloff}
Grill MJ, Meier C, Wall WA. {Investigation of the peeling and pull-off behavior
  of adhesive elastic fibers via a novel computational beam interaction model}.
  {\it The Journal of Adhesion} 2019.

\bibitem{durville2012}
Durville D. {Contact-friction modeling within elastic beam assemblies: an
  application to knot tightening}. {\it Computational Mechanics} 2012\string;
  49(6)\string: 687--707.

\bibitem{Wriggers2006}
Wriggers P. {\it {Computational Contact Mechanics}}.
\newblock Springer.
\newblock 2nd~ed. 2006.

\bibitem{BACI2020}
{BACI: A Comprehensive Multi-Physics Simulation Framework} .
  {https://baci.pages.gitlab.lrz.de/website}.;  2020.

\bibitem{MATLAB2017b}
{The MathWorks Inc} . {MATLAB R2017b}.;  2017.

\bibitem{Brodoceanu2016}
Brodoceanu D, Bauer CT, Kroner E, Arzt E, Kraus T. {Hierarchical bioinspired
  adhesive surfaces - A review}. {\it Bioinspiration and Biomimetics}
  2016\string; 11(5)\string: 051001.

\bibitem{Autumn2000}
Autumn K, Liang YA, Hsieh ST, et al. {Adhesive force of a single gecko
  foot-hair}. {\it Nature} 2000\string; 405(6787)\string: 681--685.

\bibitem{Kesel2003}
Kesel AB, Martin A, Seidl T. {Adhesion measurements on the attachment devices
  of the jumping spider Evarcha arcuata}. {\it Journal of Experimental Biology}
  2003\string; 206(16)\string: 2733 -- 2738.

\bibitem{Lee2007}
Lee H, Lee BP, Messersmith PB. {A reversible wet/dry adhesive inspired by
  mussels and geckos}. {\it Nature} 2007\string; 448(7151)\string: 338--341.

\bibitem{Autumn2002}
Autumn K, Sitti M, Liang YA, et al. {Evidence for van der Waals adhesion in
  gecko setae}. {\it Proceedings of the National Academy of Sciences}
  2002\string; 99(19)\string: 12252 -- 12256.

\bibitem{Gao2005}
Gao H, Wang X, Yao H, Gorb S, Arzt E. {Mechanics of hierarchical adhesion
  structures of geckos}. {\it Mechanics of Materials} 2005\string;
  37(2)\string: 275--285.

\bibitem{GrillDiss}
Grill MJ. {\it {Computational Models and Methods for Molecular Interactions of
  Deformable Fibers in Complex Biophysical Systems}}. Dissertation. Technical
  University of Munich, http://mediatum.ub.tum.de/?id=1537775;  2020.

\bibitem{Sauer2011c}
Sauer RA. {The peeling behavior of thin films with finite bending stiffness and
  the implications on gecko adhesion}. {\it Journal of Adhesion} 2011\string;
  87(7-8)\string: 624--643.

\bibitem{doi1988}
Doi M, Edwards SF. {\it {The theory of polymer dynamics}}.
\newblock Oxford university press .
\newblock 1988.

\end{thebibliography}

\newpage

\end{document}